\documentclass{aastex62}
\usepackage{times}
\usepackage[utf8]{inputenc}
\usepackage{amsmath}

\usepackage{scalerel}
\usepackage{tabularx}
\usepackage{threeparttable}

\usepackage{graphicx}
\usepackage{epstopdf}
\usepackage{natbib}
\usepackage{float}
\usepackage{textcomp}
\newcommand{\comment}[1]{}
\newcommand{\tess}{\textit{TESS}}
\newcommand{\kepler}{\textit{Kepler}}
\DeclareMathOperator{\sinc}{sinc}
\begin{document}

\title{\LARGE Precise Radial Velocities of Cool Low Mass Stars With iSHELL}
\correspondingauthor{Bryson Cale}
\email{bcale@masonlive.gmu.edu}

\author[0000-0001-6279-0595]{Bryson Cale}
\affiliation{George Mason University,
4400 University Drive,
Fairfax, VA 22030}

\author[0000-0002-8864-1667]{Peter Plavchan}
\affiliation{George Mason University,
4400 University Drive,
Fairfax, VA 22030}

\author{Danny LeBrun}
\affiliation{George Mason University,
4400 University Drive,
Fairfax, VA 22030}

\author[0000-0002-2592-9612]{Jonathan Gagn\'{e}}
\affiliation{Universit\'{e} de Montr\'{e}al,
2900 Edouard Montpetit Blvd.,
Montreal, QC H3T 1J4, Canada}

\author[0000-0002-8518-9601]{Peter Gao}
\affiliation{University of California, Berkeley,
501 Campbell Hall, MC 3411,
Berkeley, CA 94720-3411}

\author[0000-0002-2903-2140]{Angelle Tanner}
\affiliation{Mississippi State University,
75 B. S. Hood Road,
Mississippi State, MS 39762}

\author[0000-0002-5627-5471]{Charles Beichman}
\affiliation{NASA Jet Propulsion Laboratory,
4800 Oak Grove Dr.,
Pasadena, CA 91109}

%\author[0000-0001-6792-8637]{Bernie Walp}
%\affiliation{NASA Infrared Telescope Facility,
%Mauna Kea Access Rd.,
%Hilo, HI 96720}

\author[0000-0002-6937-9034]{Sharon Xeusong-Wang}
\affiliation{Carnegie Department of Terrestrial Magnetism
5241 Broad Branch Road, N.W.
Washington, DC 20015-1305}

\author[0000-0002-5258-6846]{Eric Gaidos}
\affiliation{University of Hawaii,
2500 Campus Rd.,
Honolulu, HI 96822}

\author{Johanna Teske}
\affiliation{Carnegie Department of Terrestrial Magnetism,
5241 Broad Branch Road, N.W.,
Washington, DC 20015-1305}

%\author[0000-0002-9061-2865]{Todd Henry}
%\affiliation{Georgia State University,
%P.O. Box 3965,
%Atlanta, GA 30302-3965}

%\author[0000-0001-5313-7498]{Russel White}
%\affiliation{Georgia State University,
%P.O. Box 3965,
%Atlanta, GA 30302-3965}

%\author{John Johnson}
%\affiliation{Harvard University,
%60 Garden Street,
%Cambridge, MA 02138}

\author[0000-0002-5741-3047]{David Ciardi}
\affiliation{California Institute of Technology,
1200 E California Blvd.,
Pasadena, CA 91125}

\author[0000-0002-1871-6264]{Gautam Vasisht}
\affiliation{NASA Jet Propulsion Laboratory,
4800 Oak Grove Dr.,
Pasadena, CA 91109}
 
\author[0000-0002-7084-0529]{Stephen R. Kane}
\affiliation{University of California, Riverside,
900 University Ave,
Riverside, CA 92521}

\author[0000-0002-5823-4630]{Kaspar von Braun}
\affiliation{Lowell Observatory,
1400 W Mars Hill Rd.,
Flagstaff, AZ 86001}

\begin{abstract}
The coolest dwarf stars are intrinsically faint at visible wavelengths and exhibit rotationally modulated stellar activity from spots and plages. It is advantageous to observe these stars at near infrared (NIR) wavelengths (1-2.5 \micron) where they emit the bulk of their bolometric luminosity and are most quiescent. In this work we describe our methodology and results in obtaining precise radial velocity (RV) measurements of low mass stars using K-band spectra taken with the R\texttildelow80,000 iSHELL spectrograph and the NASA Infrared Telescope Facility (IRTF) using a methane isotopologue gas cell in the calibration unit. Our novel analysis pipeline extracts RVs by minimizing the RMS of the residuals between the observed spectrum and a forward model. The model accounts for the gas cell, tellurics, blaze function, multiple sources of quasi-sinusoidal fringing, and line spread function of the spectrograph (LSF). The stellar template is derived iteratively using the target observations themselves through averaging barycenter-shifted residuals. We have demonstrated 5 ms$^{-1}$ precision over one-year timescales for the M4 dwarf Barnard's Star and K dwarf 61 Cygni A, and 3 ms$^{-1}$ over a month for the M2 dwarf GJ 15 A. This work demonstrates the potential for iSHELL to determine dynamical masses for candidate exoplanets discovered with the NASA \tess\ mission, and to search for exoplanets orbiting moderately active and/or young K \& M dwarfs.
\end{abstract}


%%%% INTRODUCTION %%%%%%%%%%%%%%%%%%%%%%%%%%%%%%%%%%%%%%%%%%%%%%%%%%%%%%%%%%%%%%

% Section 1
\section{Introduction} \label{sec:intro}
The radial velocity (RV) technique has been successfully applied to reveal hundreds of systems around solar-type (FGK dwarf) stars since the discovery of 51 Pegasi b in 1995 \citep{1995Natur.378..355M}. Observations are usually performed at visible wavelengths where these stars are brightest and telluric lines are relatively scarce \citep{2015arXiv150301770P, 2016PASP..128f6001F}. However, 75\% of the stars in the solar neighborhood are cooler, less luminous M dwarfs \citep{recons_survey}. The ``habitable zone" around M dwarfs is much closer to the star due to their lower luminosities, significantly shortening the timescales needed to detect habitable zone worlds. The lower stellar mass also means a larger reflex velocity for a given planet mass. These advantages have been recently exploited in the discovery of the planet in the habitable zone of the M dwarf Proxima Cen \citep{2016Natur.536..437A}.

% Spectral features such as the Calcium II triplet at \texttildelow 8500nm and the potassium doublet at \texttildelow 766nm are found to correlate with the canonical optical activity tracer H$\alpha$ for M dwarfs \citep{2016ApJ...832..112R}, although neither are captured by most NIR RV spectrographs operating past Y-band.

RV observations of M dwarfs, especially mid- to late-type M dwarfs (M4+), are more challenging than those of solar-type stars. M Dwarfs are intrinsically faint and require long integration times per epoch to acquire sufficient SNR, especially for spectroscopic observations where light is dispersed. Late M dwarfs are more magnetically active \citep{2004AJ....128..426W,2010ApJ...713L.155B}. Star spots and other sources of activity can introduce RV noise and spurious signals at periods corresponding to the stellar rotation period $P_{\textrm{rot}}$ and harmonics $P_{\textrm{rot}}/n$ where \textit{n} is a small integer \citep{2015ApJ...805L..22R}. A further realization for M1-M4 dwarf stars is the stellar rotation period overlaps with the range in periods of habitable zone worlds \citep{2016ApJ...821L..19N, 2016MNRAS.459.3565V}.

Due to the difficulties of RV observations of M dwarfs in the optical, interest has grown in the last decade in developing near-infrared (NIR) spectrographs for these observations. M dwarfs are brightest at near-infrared (NIR) wavelengths, and the flux contrast between star spots and the surrounding chromosphere is reduced \citep{2010ASPC..430..181M, 2011ApJ...736..123M,2012ApJ...761..164C, 2013AN....334..184A, 2015ApJ...798...63M}. To first order in wavelength, the flux contrast (and thus RV signal) from activity is expected to follow a $\lambda^{-1}$ relationship, although additional challenges arise from the wavelength dependence of limb-darkening and convective blue-shift \citep{2010ApJ...710..432R}. NIR RV efforts have made rapid progress in precision (and thus mass detection) capabilities. NIRSPEC on Keck (K-band, R\texttildelow25000) obtained 45 ms$^{-1}$ precision observing late M dwarfs \citep{2012ApJS..203...10T}. CSHELL (K-band, R\texttildelow46,000) on IRTF obtained 35 ms$^{-1}$ observing GJ 15 A \citep{2016PASP..128j4501G}. NIR RV efforts have gained traction both with absorption gas cells \citep[e.g.][]{2010ApJ...713..410B} and the use of fiber-feeds to stabilized environments for instruments. As a recent example, the Habitable Zone Planet Finder (HPF) on the 10 meter Hobby-Eberly Telescope (R\texttildelow50,000, Y- and J-bands) has shown $<$ 3 ms$^{-1}$ precision on Barnard's Star, sufficient to detect rocky worlds in the habitable zone of M dwarfs \citep{2019arXiv190202817M,2019arXiv190200500M}. 

Most NIR RV instruments exploited the $Y$, $J$, and $H$ regions of the spectrum, although RV information content in the Y- and J-band is found to be lower than expected from synthetic spectra \citep{2018A&A...612A..49R}. The K-band spectra of M dwarfs also contains deep sharp lines of CO at $\lambda > 2.29$ \micron\ suitable for RV measurements. However, observations in this wavelength region are also plagued with telluric lines of water and methane as well as other trace molecules that complicate the data analysis \citep{2010A&A...524A..11S}.

In this paper, we describe our methodology in obtaining precise NIR RVs from observations of cool low mass stars using the iSHELL spectrograph \citep{2016SPIE.9908E..84R} on the 3 meter NASA Infrared Telescope Facility (IRTF) at K-band wavelengths (2.18-2.47 $\micron$). We present our novel pipeline extending the work of \cite{2016PASP..128j4501G} with CSHELL in Sections \ref{sec:data_reduction} and \ref{sec:rv_pipe}. We test our pipeline with observations of Barnard's Star (GJ 699), GJ 15 A, and 61 Cygni A, all previously used in RV searches and suitable as RV standards, and present the results in Section \ref{sec:results}. We analyze the fidelity of our stellar template retrieval in Section \ref{sec:template_2}, and forward model parameter distributions in Section \ref{sec:model_params}. In Section \ref{sec:discussion} we discuss our particular choice of forward model and how our obtained iSHELL RV precision compares to other precise NIR spectrographs, as well as prospects for planet confirmation. A summary of this work is provided in Section \ref{sec:summary}.

%%%% OBSERVATIONS %%%%%%%%%%%%%%%%%%%%%%%%%%%%%%%%%%%%%%%%%%%%%%%%%%%%%%%%%%%%%%
% Section 2
\section{Observations} \label{sec:obs}

\begin{figure}[H]
    \center
    \includegraphics[width=0.85\textwidth]{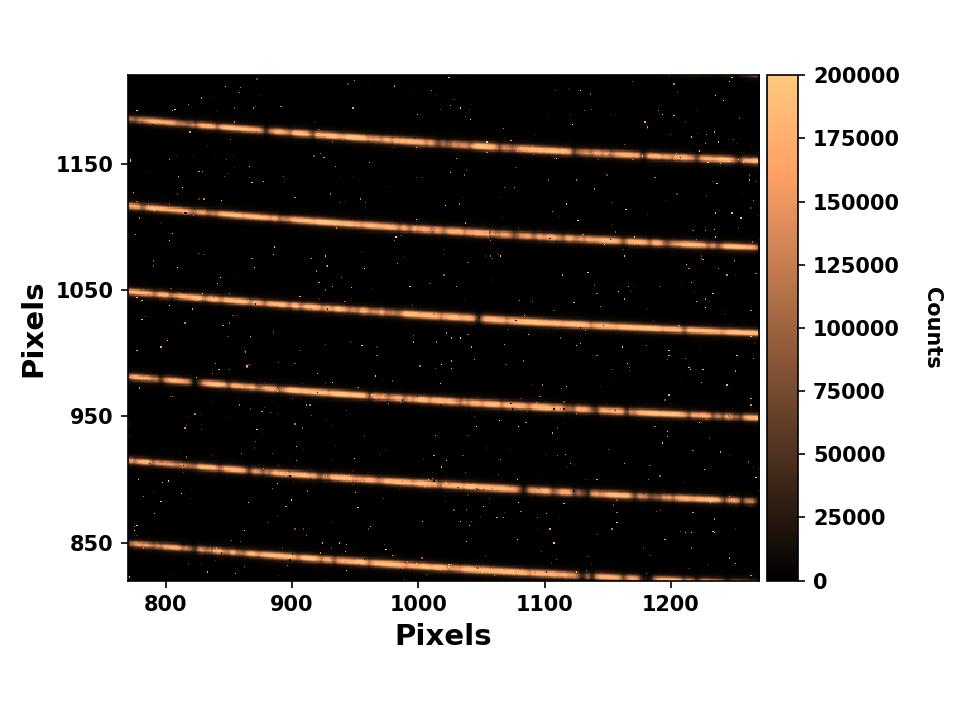}
    \caption{A sub-frame of an unprocessed multi-order spectral image of 61 Cygni A from Oct 16. The full frame contains 29 orders with dimensions 2048 x 2048. Numerous ``hot" pixels appearing on the stellar trace as dark values, and between traces as bright values, are flagged during data reduction, but can be missed if the SNR is not sufficient to properly identify them as outliers.}
    \label{fig:raw_data}
\end{figure}

We obtained spectra with the iSHELL spectrograph on the 3 meter NASA Infrared Telescope Facility (IRTF) between October 2016 and October 2017, with the majority of observations taking place during the first half of this period. Table \ref{tab:obs} provides the estimated SNR (per detector pixel) of each observation as well as the number of observations (N$_{\textrm{obs}}$) obtained each night. Spectra are recorded using the iSHELL spectrograph in \textit{KGAS} mode (2.18-2.47 $\micron$) with a 0.375'' slit at R\texttildelow80,000. A Hawaii 2RG array records 29 cross-dispersed echelle orders ($m=212-240$) spanning this spectral range. A methane isotopologue ($^{13}\textrm{CH}_{4}$) gas cell in the calibration unit with 90\% continuum throughput is used to provide a common optical path wavelength reference and to constrain the variable line spread function (\textit{LSF}) of the spectrograph \citep{2012PASP..124..586A,2013SPIE.8864E..1JP}. To minimize errors in the barycenter correction and telluric optical depths of individual spectra, integration times are limited to 5 minutes. The exposure midpoint (opposed to flux-weighted) is used to determine the barycentric correction since iSHELL does not have an exposure meter as is common in visible precise RV instruments \citep{2014PASP..126..838W}. Unfortunately, the error in barycenter correction scales with square of the exposure time \citep{tronsgaard}, so doubling an exposure time will quadruple our barycenter velocity uncertainty. Various factors can further contribute to a non-constant photon rate, particularly changes in airmass and atmospheric transparency. The telluric content (particularly water) can be variable on time-scales of an hour or less along with the changing airmass of our target \citep{2018PASP..130g4502S}, so we limit our maximum exposure time to avoid any errors that could potentially be introduced in the telluric modeling, although we do not characterize this. We further limit integration times to avoid the nonlinear detector regime of the detector for brighter targets. After every target is observed, a set of five flat-fields is obtained before slewing the telescope to the next target. An example of a raw unprocessed 2-dimensional fits image is shown in Fig. \ref{fig:raw_data}. All data are available online at the NASA/IPAC Infrared Science Archive.\footnote{Available at \url{https://irsa.ipac.caltech.edu/applications/irtf/}. All IRTF data has a proprietary period of 18 months starting at the end of that observing semester.}

We choose three bright RV standards to evaluate the RV precision obtainable with iSHELL. These targets have already been observed to show RV stability at or below the expected iSHELL precision, and are summarized in Table \ref{tab:targets}. Barnard's Star (GJ 699) shows no RV variations down to 2 ms$^{-1}$ over several years using data sets from Keck/HIRES \& Lick/Hamilton after accounting for the observed secular acceleration of 4.515 ms$^{-1}$yr$^{-1}$ \citep{2012AAS...21924504C}. More recent efforts have found a low-amplitude periodic signal at 233 days and $K_{\textrm{amp}}=1.2$ ms$^{-1}$ \citep{2018Natur.563..365R}. 61 Cygni is a binary system of two comparatively bright K dwarfs with an orbital period of 653 years, neither of which are known to host any planets. GJ 15 A is suspected to host a single planet with $K_{\textrm{amp}}=2.9$ ms$^{-1}$ and a period of 11.44 days \citep{2014ApJ...794...51H}.

% Basic Table about targets

\begin{table}[H]
\caption{Summary of Observed Targets.}
\begin{center}
    \begin{tabular}{| m{20mm} | m{18mm} | m{10mm} | m{8mm} | m{45mm} | m{35mm} |}
    \hline
    Star & R.A./Decl. & Spec. Type & $\mathrm{K_{\textrm{mag}}}$ & Planets & Reference \\
    \hline \hline
    GJ 699 & 17:57:48.5 \newline +04:41:36.1 & M4V & 4.52 & b: $K_{\textrm{amp}}=1.2$ ms$^{-1}$, P=233 days & \cite{2009ApJ...704..975J} \newline \cite{2018Natur.563..365R}\\
    \hline
    GJ 15 A & 00:18:22.9 \newline +44:01:22.6 & M2V & 4.02 & b: $K_{\textrm{amp}}=2.9$ ms$^{-1}$, P=11.4 days & \cite{2009ApJ...704..975J} \newline \cite{2014ApJ...794...51H} \\
    \hline
    61 Cygni A & 21:06:53.9 \newline +38:44:57.9 & K5V & 2.68 & No known planets & \cite{2009ApJ...694.1085V} \\
    \hline
    \end{tabular}
\end{center}
\label{tab:targets}
\end{table}

% Observations Table
% For the Gaidos Observing, take SNR_i to be 41 per each 20 Section exposure for GJ 699

\begin{table}[H]
\centering
\begin{threeparttable}
\caption{The obtained SNR per detector pixel for all observations.}\label{tab:obs}
    \begin{tabular}{| m{25mm} | m{20mm} | m{20mm} | m{20mm} | m{20mm} |}
    \hline
    UT Date & $\mathrm{N_{obs}}$ & Int. Time & $\textrm{SNR}_{i}$\tnote{a} & $\textrm{SNR}_{\textrm{tot}}$\tnote{b} \\
    \hline \hline
    \multicolumn{5}{|c|}{|Barnard's Star|} \\
    \hline
    2016 Oct. 16 & 4 & 5 min & 139 & 277 \\
    2016 Oct. 23 & 7 & 5 min & 159 & 420 \\
    2016 Nov. 06 & 8 & 5 min & 146 & 413 \\
    2016 Nov. 07 & 4 & 5 min & 201 & 402 \\
    2017 Apr. 06 & 9 & 5 min & 132 & 397 \\
    2017 Jun. 18 & 6 & 5 min & 164 & 402 \\
    2017 Jun. 26 & 6 & 5 min & 160 & 392 \\
    2017 Jul. 05 & 7 & 5 min & 152 & 402 \\
    2017 Jul. 29 & 10 & 5 min & 133 & 420 \\
    2017 Oct. 20 & 16 & 20 sec & 41\tnote{c} & 164\tnote{c} \\
    2017 Oct. 21 & 16 & 20 sec & 41\tnote{c} & 164\tnote{c} \\
    2017 Oct. 22 & 16 & 20 sec & 41\tnote{c} & 164\tnote{c} \\
    2017 Oct. 23 & 16 & 20 sec & 41\tnote{c} & 164\tnote{c} \\
    \hline
    
    \multicolumn{5}{|c|}{|GJ 15 A|} \\
    \hline
    2016 Oct. 16 & 7 & 2.5 min & 198 & 525 \\
    2016 Oct. 17 & 12 & 2.5 min & 150 & 521 \\
    2016 Oct. 22 & 8 & 2.5 min & 179 & 505 \\
    2016 Oct. 23 & 17 & 2.5 min & 122 & 503 \\
    2016 Nov. 06 & 16 & 1.5 min & 127 & 506 \\
    2016 Nov. 07 & 11 & 2.5 min & 158 & 524 \\
    \hline
    
    \multicolumn{5}{|c|}{|61 Cygni A|} \\
    \hline
    2016 Oct. 16 & 10 & 30 sec & 170 & 537 \\
    2016 Oct. 17 & 12 & 1 min & 152 & 525 \\
    2016 Oct. 22 & 23 & 15 sec & 105 & 502 \\
    2016 Oct. 23 & 10 & 1 min & 162 & 513 \\
    2016 Nov. 06 & 6 & 1 min & 127 & 506 \\
    2016 Nov. 07 & 16 & 15 sec & 126 & 502 \\
    2017 Apr. 06 & 7 & 1.5 min & 194 & 514 \\
    2017 Apr. 12 & 6 & 1.5 min & 210 & 515 \\
    2017 Jun. 18 & 5 & 1.5 min & 258 & 577 \\
    2017 Jun. 26 & 6 & 1.5 min & 223 & 546 \\
    2017 Jul. 05 & 11 & 1.5 min & 154 & 510 \\
    \hline
    \end{tabular}
\begin{tablenotes}
    \item [a] $\textrm{SNR}_{i}$ represents the SNR per spectral pixel for an individual spectrum measured by summing the area of a Gaussian PSF model to the data near the blaze peak with no spectral features with the iSHELL observing gui. Sky noise is not used in this estimation.
    \item [b] $\textrm{SNR}_{\textrm{tot}}$ represents the total (co-added) SNR per spectral pixel for a consecutive series of observations, with $\textrm{SNR}_{\textrm{tot}}=\sqrt{\textrm{N}_{\textrm{Obs}}}\textrm{SNR}_{i}$.
    \item [c] SNR was not recorded during the last four observations for Barnard's Star and are an estimation from the previous nights assuming a relationship of SNR $\propto$ $\sqrt{t_{\textrm{exp}}}$.
    \end{tablenotes}
    \end{threeparttable}
\end{table}

%%%% DATA REDUCTION %%%%%%%%%%%%%%%%%%%%%%%%%%%%%%%%%%%%%%%%%%%%%%%%%%%%%%%%%%%%%%
% Section 3
\pagebreak
\section{Data Reduction} \label{sec:data_reduction}

We reduce iSHELL data with custom Interactive Data Language (IDL) software routines\footnote{Available at \url{https://github.com/jgagneastro/ishell_reduction}; version 1.1 and commit number \texttt{dc5367c93d4ee37ae72286cf338388a2a9727155} were used.}. For each nightly target, we locate the spectral trace using the corresponding flat-fields. We generate a master flat field which is free of sinusoidal fringing through various smoothing techniques. We then divide the corresponding science data by this master flat to correct for the slit illumination and pixel-to-pixel response. We finally iteratively extract the 1-dimensional spectral orders through an iterative process where we obtain a better estimation of the spatial point spread function (PSF), spectral flux density, and identification of bad pixels with each iteration.

\begin{figure}[H]
    \center
    \includegraphics[width=0.45\textwidth]{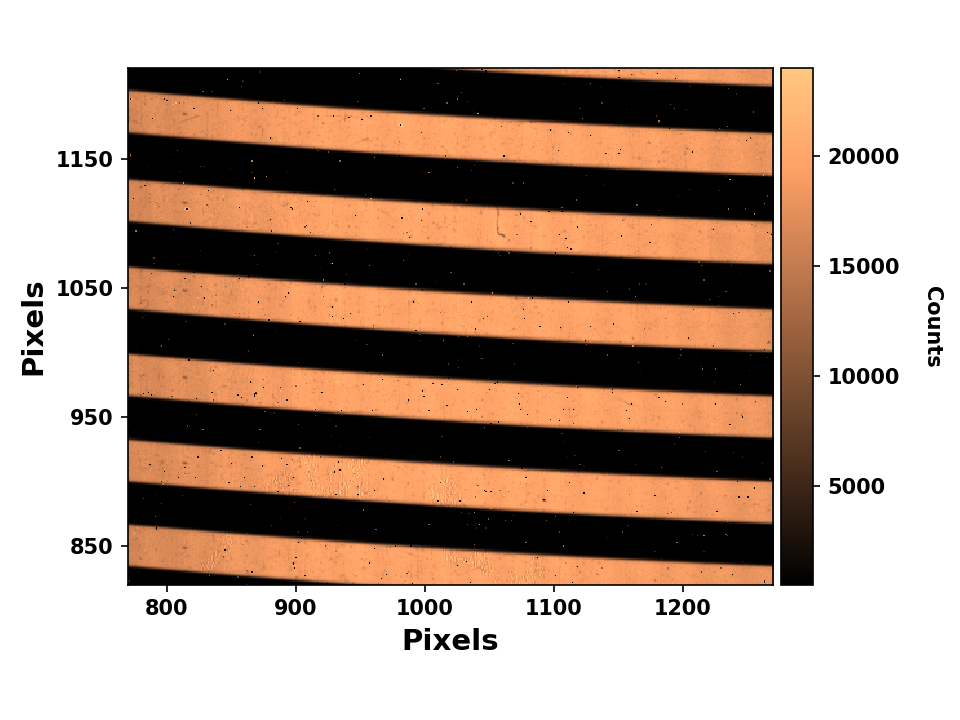}
    \includegraphics[width=0.45\textwidth]{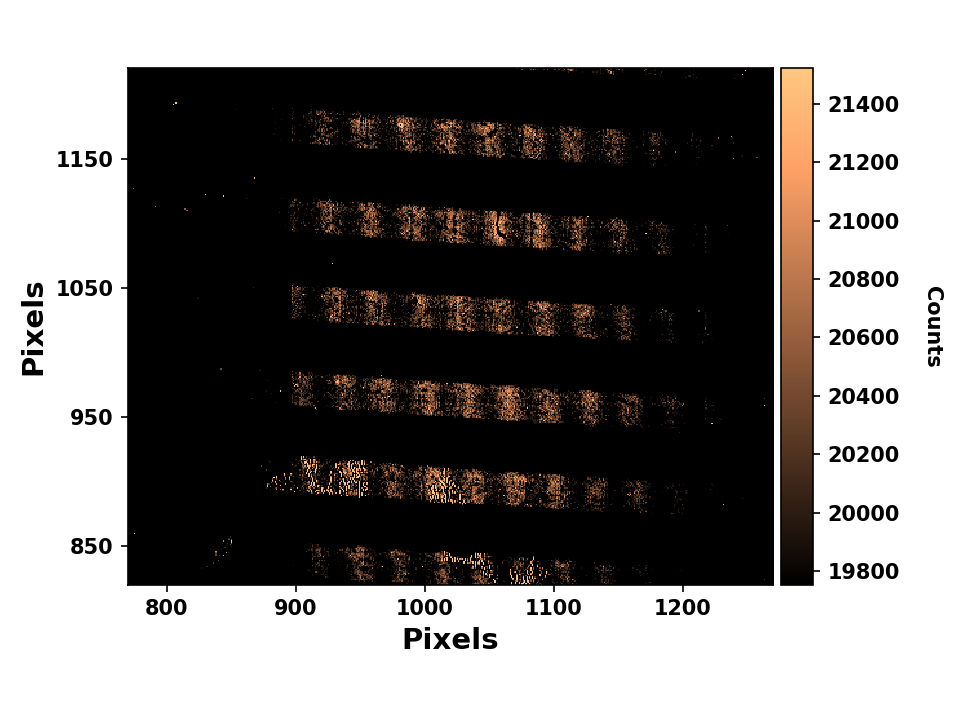}
    \includegraphics[width=0.45\textwidth]{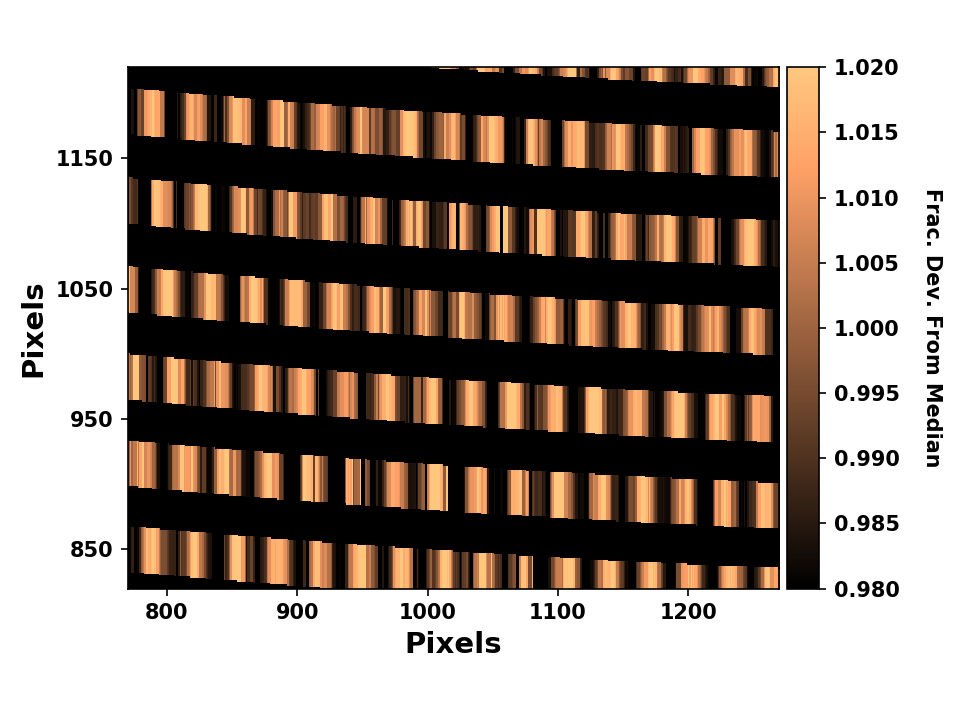}
    \includegraphics[width=0.45\textwidth]{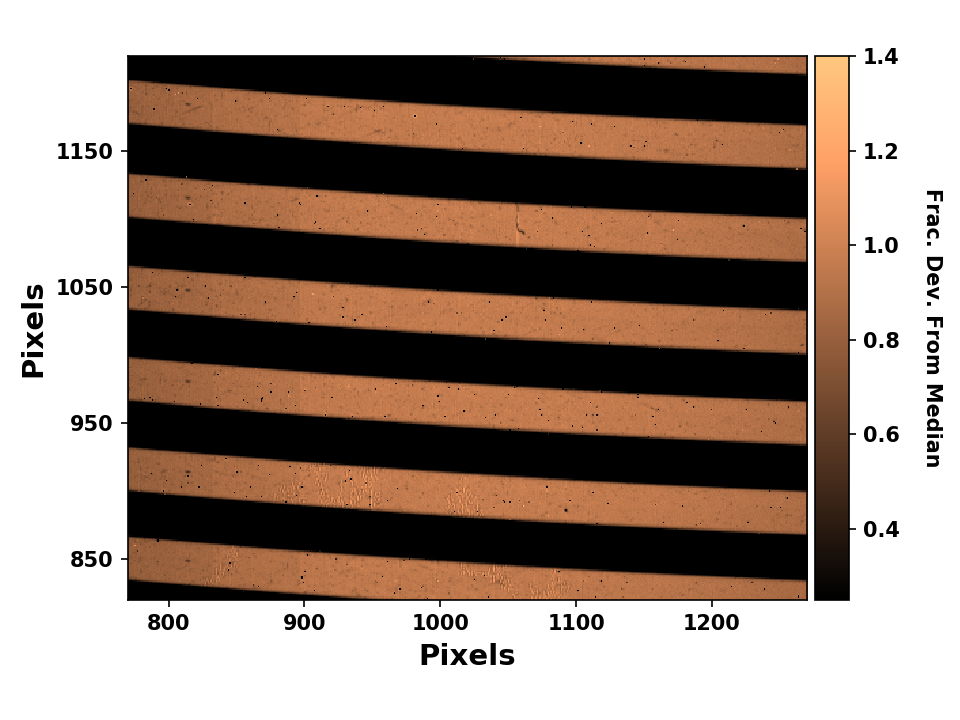}
    \caption{\textit{Top Left}: A sub-frame of a raw flat-field image. \textit{Top Right}: The same image and region as the top left but with a high resolution color scaling (narrow range in counts) centered around the fringing signal. \textit{Bottom Left}: The fringing present in the flat-fields isolated through a rolling median with a window comparable to the dominant period in wavelength of $\Delta \lambda \sim 0.3$nm. iSHELL fringing is further discussed in Section \ref{sec:fringing}. \textit{Bottom Right}: A median combined flat-field to be used to correct the science data. Fringing shown in panels 2 and 3 are removed from this image.}
    \label{fig:fringing}
\end{figure}

First, we median combine and use the flat-fields to detect the 29 echelle order traces. We model the edges and center of each order with independent second-degree polynomials. In each flat-field order, we straighten the order into a rectangular array by linear interpolation in the spatial direction. We smooth in the spectral direction using a 45-pixel rolling median. We then divide the straightened flat by the spectrally-smoothed flat to isolate instrumental fringing in the flat-field (See Section \ref{sec:fringing}). A one-dimensional version of the fringing is then obtained by taking the vertical median in the spatial direction, and smoothing the resulting 1-dimensional array with a 3-pixel rolling median. We divide the fringing pattern in each order to finally remove the fringing present in the flat-fields (Fig. \ref{fig:fringing}).

Next we divide the corrected flat-field into all individual science exposures to correct the slit illumination and pixel-to-pixel response function without injecting an additional fringing pattern in the science exposures. The fringing in iSHELL is not stable in phase and amplitude, nor is the slit-illumination function identical between flat-field (quartz lamp) and stellar (point) sources, and thus does not perfectly divide out if left in the flat-fields.

To extract the spectra, we use a multi-step iterative process. First, we straighten the individual spectral orders of each science exposure with the corresponding order’s central position polynomial (from the flat-fields). For each order, an initial (spatial) point spread function is constructed using a median filter in the dispersion (wavelength) direction. A first spectral extraction is performed by using the estimated point spread function as an extraction weight on the rectified order. A cross-correlation of the estimated profile with the straightened data is next performed at each spectral position where the estimated spectral flux density is above half of a cutoff value, set at the 80\% quartile value of the spectral flux density. This results in a more precise trace position of the data within each order, which is next modeled with a second-degree polynomial.

We next create a curved two-dimensional spectral profile from the spatial point spread function and refined second-degree polynomial of the science trace position. We obtain a better estimate of the spectral flux density and avoid interpolation by using this two-dimensional profile as an extraction weight on the non-straightened science order. Significant outliers in the resulting spectral flux density are also masked iteratively by flagging large deviations taking place within less than three spectral pixels.

Next, this cleaned up version of the spectral flux density is used in combination with the two-dimensional trace to build a clean version of the two-dimensional science trace. Dividing the science trace by this resulting image allows us to flag bad pixels directly in the two-dimensional data by looking at significant outliers in flux deviations that happen within three pixels. This allows us to mask the deviant pixels directly in the two-dimensional image and to refine our best estimate of the spectral flux density by performing an optimal (maximum SNR) extraction \citep{1986PASP...98..609H} using the masked two-dimensional spectral trace and the curved two-dimensional point spread function.

\begin{figure}[H]
    \centering
    \includegraphics[width=0.85\textwidth]{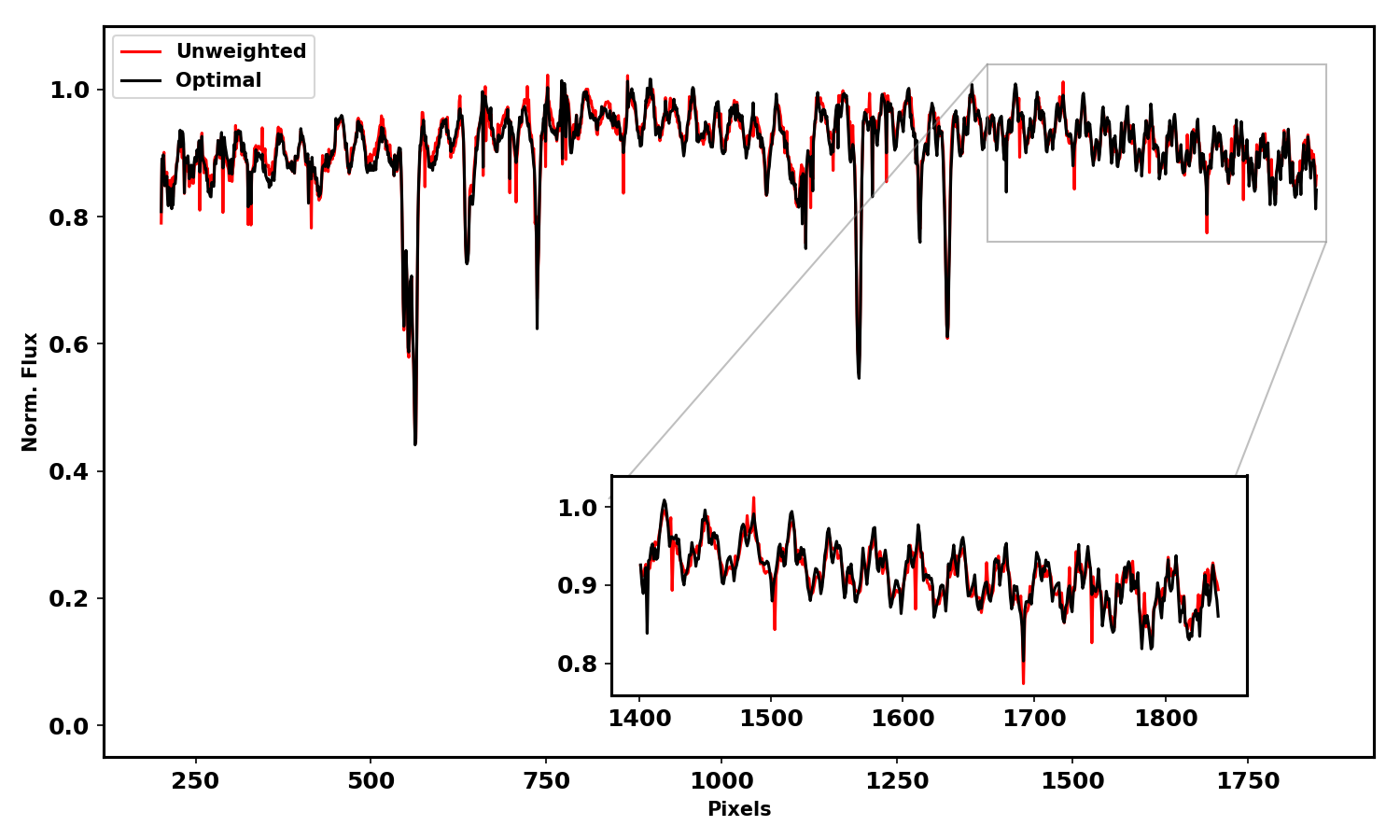}
    \caption{A reduced spectrum as a function of pixels (blue to red in wavelength) for 61 Cygni A from Oct. 16, 2016 for order 28 ($m=239$, $\lambda=2.18-2.194\ \micron$). The optimally (weighted) extracted spectrum used in RV calculations is shown in black, and the unweighted is shown in red. Inversely weighing pixels by their distance from the center of the trace mitigates sky noise resulting in fewer outliers and an overall smoother spectrum. Order 28 is relatively free of tellurics, gas cell, and stellar lines, so the OS and AR fringing components (see Section \ref{sec:fringing}) are clearly seen with overall peak-to-peak amplitudes of \texttildelow10\%.}
    \label{fig:data_reduced}
\end{figure}

\begin{figure}[H]
    \centering
    \includegraphics[width=0.85\textwidth]{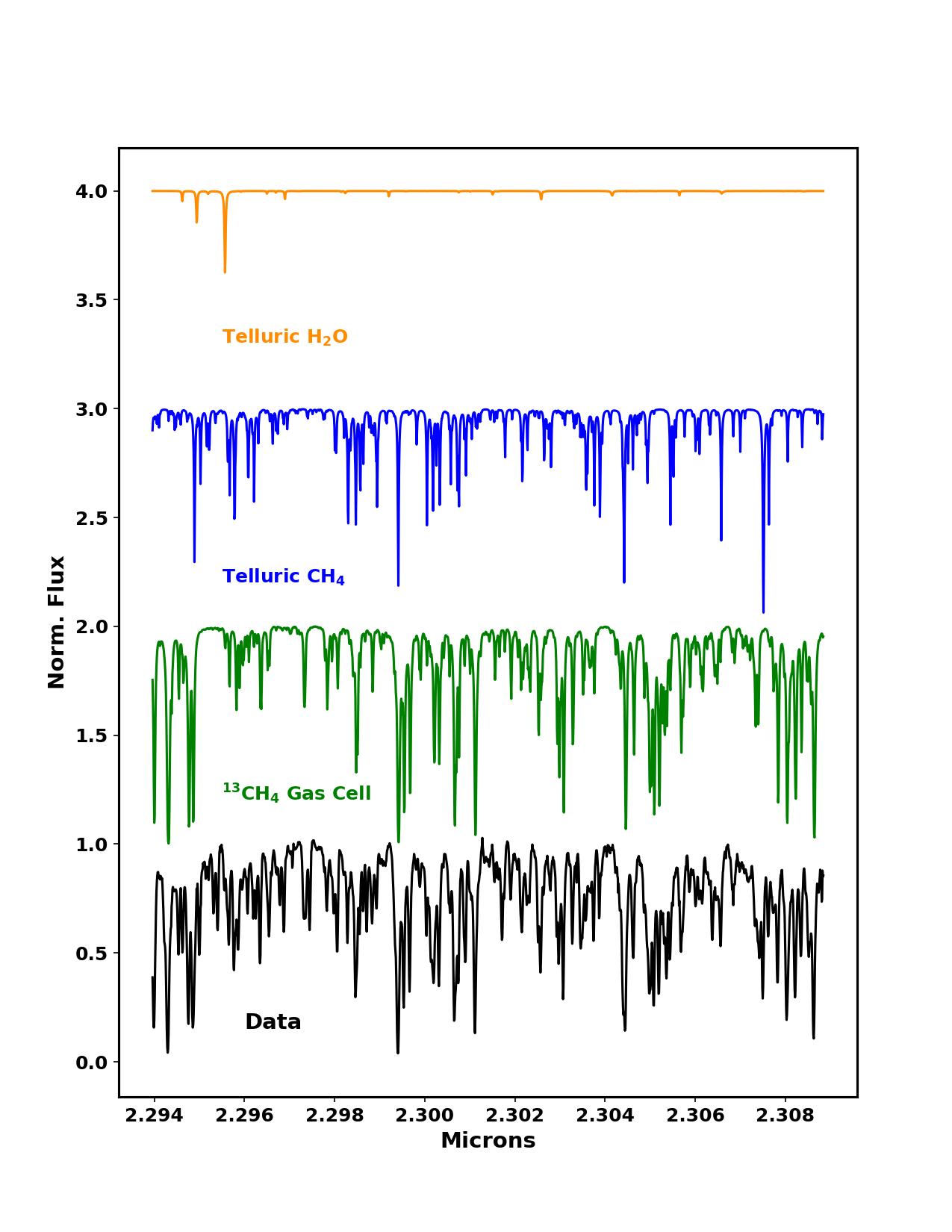}
    \caption{A reduced spectrum for 61 Cygni A from Oct. 16, 2016 for order 16 ($m=227$). The wavelength grid was generated with the initial guess parameters to the RV pipeline (see Section \ref{sec:rv_pipe}). The unmodified input templates for the methane gas cell, telluric water, \& telluric methane for this order are also shown.}
    \label{fig:templates}
\end{figure}

As a final step to refine the spectral flux density, we allow the width of the spatial line profile to vary linearly in the spectral direction within each order. To do this required modeling the spatial point spread function; we found that a Gaussian profile represents the data adequately. A Gaussian profile is fit at each spectral position of the masked two-dimensional science trace, and the resulting Gaussian profile width versus spectral pixel position is fit with a first-degree polynomial. This is used to build a final version of the curved two-dimensional extraction profile, and a last optimal extraction is performed with this profile to obtain our final extracted spectral flux density. Examples of reduced spectra are shown in Figs. \ref{fig:data_reduced} \& \ref{fig:templates}.

\pagebreak
%%%% RV PIPELINE %%%%%%%%%%%%%%%%%%%%%%%%%%%%%%%%%%%%%%%%%%%%%%%%%%%%%%%%%%%%%%

\section{Radial Velocity Pipeline} \label{sec:rv_pipe}
In this section, we describe the methods used to extract RVs by forward modeling single-order extracted (one-dimensional) science spectra (\ref{sec:forward_model}--\ref{sec:template_1}), then compute nightly RV measurements and finally optimize multi-order RVs in Section \ref{sec:rv_calc}. We adapt the RV pipeline for CSHELL spectra described in \cite{2016PASP..128j4501G} to iSHELL data. We have rewritten the CSHELL code in a \textit{Python} script \textit{PySHELL}\footnote{Available upon request.} taking into account iSHELL's larger spectral grasp with multiple orders. Due to variability in the blaze function and due to the lower SNR, we choose to ignore the first and last 200 pixels at the edges of the 2048-element extracted spectra. Utilizing the remaining pixels is a subject of future work. Our radial velocity pipeline represents a significant departure from traditional analyses with iodine cell data. Rather than splitting orders into smaller chunks and introducing discontinuities at the boundaries, we model entire orders as a single ``chunk''. This necessitates a more complex forward model than is used with traditional iodine cells (e.g. \cite{1996PASP..108..500B}).

% Section 4.1
\subsection{Choice of Numerical Solver} \label{sec:numerical_solver}

To fit a model to the extracted one-dimensional spectra, we have implemented a custom downhill Nelder-Mead algorithm that performs simplex calls for the entire parameter space followed by consecutive two-dimensional subspace calls for all neighboring pairs of parameters to better handle the large dimensional space, as standard Nelder-Mead algorithms fail to converge. A similar approach that we did not test would be to: first use \textit{SciPy's optimize} routine with \textit{method=Nelder-Mead} a single time for the entire parameter space; second use \textit{optimize} for each consecutive pair of parameters keeping others constant $((1, 2), (2, 3), ..., (N_{\textrm{pars}}-1, N_{\textrm{pars}}), (N_{\textrm{pars}}, 1))$; and finally third, repeat the first two steps for the number of parameters in the model. The minimum RMS is continuously improved with each call to \textit{optimize} as the parameters converge. Our algorithm is therefore dependent on the parameter ordering, and we do not explore the impact of parameter ordering in this work. However, the RMS typically converges before $\sim N_{\textrm{pars}}/2$ iterations of the algorithm. Our specific Nelder-Mead algorithm in \textit{Python} for a given simplex is based on that used for CSHELL given by \cite{matlab_simps} but with stricter convergence requirements. Specifically, the largest fractional difference in the RMS for the current simplex must be less than $10^{-5}$ three times in a row for the solver to be considered successfully converged.

\subsection{Spectral Forward Model} \label{sec:forward_model}
For a given echelle order, we define the forward model intensity as

\begin{gather}
    I_{M}(\lambda) = B(\lambda) F_{\scaleto{\textrm{AR}}{4pt}}(\lambda) LSF(\lambda) * [I_{\star}(\lambda_{\star})\ T_{g}^{\tau_{g}}(\lambda)\ T_{t}^{\tau_{t}}(\lambda_{t})\ F_{\textrm{\scaleto{\textrm{OS}}{4pt}}}(\lambda)] \label{eq:model}
\end{gather}

\noindent where * represents a convolution. We describe each of the forward model terms in turn. $I_{\star}(\lambda_{\star})$ is the Doppler shifted stellar spectrum derived iteratively and described in detail in Section \ref{sec:template_1}. $T_{g}$ is our gas cell spectrum, obtained with a Fourier Transform Spectrometer (FTS) at the NASA Jet Propulsion Laboratory (JPL) at R\texttildelow500,000 \citep{2012PASP..124..586A,2013SPIE.8864E..1JP}. Like \cite{2016PASP..128j4501G}, we find the gas cell optical depth $\tau_{g}=0.97$ (vs. unity) because of the off-axis angle the gas cell was placed in the FTS, as opposed to CSHELL \& iSHELL where the path length is minimized. $T_{t}(\lambda_{t})$ corresponds to the Doppler shifted telluric absorption spectrum with optical depth $\tau_{t}$. For \textit{KGAS} mode, the relevant telluric components are water ($\textrm{H}_{2}\textrm{O}$), methane ($\textrm{CH}_{4}$), nitrous oxide ($\textrm{N}_{2}\textrm{O}$), and carbon dioxide ($\textrm{CO}_{2}$). Each component is obtained from \textit{Transmissions of the AtmosPhere for AStromomical data (TAPAS)} \citep{2014A&A...564A..46B}. We use realistic temperature-pressure profiles for Maunakea corresponding to the zenith and date of April 12, 2018 at midnight (arbitrarily chosen). The telluric shift is common to all species, but each can have different optical depths to account for variable atmospheric content. The stellar and telluric shifts are computed on a logarithmic grid keeping $\Delta \ln\lambda = v/c=\textrm{constant}.$ If a telluric component has no absorption features $> 1\%$ prior to convolution, it is excluded from the fit for that order. The effective resolution of the gas cell and telluric templates in our spectral model are 5 and 15 times that of the data, respectively.

$B(\lambda)$ is the residual blaze function left over after the flat division in data reduction. The residual blaze is relatively consistent across orders for a sequence of observations, and is approximately quadratic. While the deviations from the quadratic are not well-modeled with an analytic function, they are relatively small in flux ($<$ 10\%). We first model the blaze with a quadratic to approximate the general curvature of the continuum, then use 14 cubic splines as an additive correction. A wavelength grid for the blaze function spline correction is generated by first starting from an initial guess for the wavelength solution. The corresponding $\lambda_{j}$ grid for each spline point $bs_{j}$ is then generated using a linearly spaced array with endpoints corresponding to the estimated wavelengths of pixels 200 and 1848 (the cropped data) with an extra padding of 0.1nm to account for the error in the initial wavelength solution, ensuring that no spline points are outside of the cropped data.

\textit{LSF} represents the line spread function (line profile) of the spectrograph and is constructed using a sum of Gaussians with Hermite polynomial coefficients \citep{lsf_hermite}. These are derived iteratively using

\begin{gather}
    \psi_{k}(x)=\sqrt{\frac{2}{k}} \bigg(x\psi_{k-1}(x)-\sqrt{\frac{k-1}{2}}\psi_{k-2}(x)\bigg) \\[20pt]
    \text{with}\ \ \psi_{0}(x)=\pi^{-\frac{1}{4}}e^{-\frac{1}{2}x^{2}}\ \ \text{and} \ \ \psi_{1}(x)=\sqrt{2}x\psi_{0}(x)
\end{gather}

\noindent where $x=\lambda / a_{0}$ with $a_{0}$ being the Gaussian width of $\psi_{0}$. The \textit{LSF} is then constructed by summing over $\psi_{k}$,

\begin{gather}
    LSF(x) = \psi_{0}(x) + \sum_{k=1}^{N_{\textrm{H}}}a_{k}\psi_{k}(x)
\end{gather}

\noindent where $N_{\textrm{H}}$ is the highest order of the Hermite function series. We use $N_{\textrm{H}}=6$ (up to $a_{5}$), and explore other \textit{LSF} models in Section \ref{sec:lsf_model}. The \textit{LSF} is normalized as a final step. Further, the convolution is only performed within a window of $\pm 0.17$ nm for the model pixel as convolution is computationally expensive.

Like the residual blaze function, we compute the wavelength grid of the data, $\lambda(P_i)$ for pixels $\{P_i\}$, via a main quadratic component, plus a cubic spline correction for small local deviations. Unlike the blaze, a need for splines here is not initially obvious. As discussed further in Section \ref{sec:wave_sol}, however, inclusion of a spline correction in the wavelength solution improves the resulting RVs. To obtain the main quadratic component for the wavelength solution, pixel $P_{i}\in \{1, 1024.5, 2048\}$ (from blue to red) is mapped to a window $\lambda_{i} \pm 0.05$nm. An initial guess for the zero points $\lambda_{i}$ are predetermined from modeling several nights of Vega data with no stellar lines. From here, the polynomial coefficients are determined through a matrix inversion and a quadratic wavelength solution is obtained for all pixels. While $\lambda_{i}$ are not orthogonal parameters, we find that polynomial coefficients yield similar RV precision, and opt to use these ``Lagrange points'' for their simple behavior. Wavelength splines are placed on top of the quadratic by first choosing evenly spaced pixels (for the cropped data) equal to the number of splines plus one. Each pixel gets mapped to the range $\pm 0.0125$nm and are interpolated onto the data pixel grid using cubic spline interpolation. The sum of the quadratic and spline correction yields the final wavelength solution for a given spectrum.

% Section 4.1.2
\subsubsection{Fringing} \label{sec:fringing}

$F_{\scaleto{\textrm{OS}}{4pt}}(\lambda)$ and $F_{\scaleto{\textrm{AR}}{4pt}}(\lambda)$ represent the two sources of fringing present in iSHELL data. The first is introduced by the order selection (OS) filter, before the light is diffracted at the echelle grating\footnote{Upgrades planned for late 2019 will replace the order selection filter with a wedged version to eliminate this source of fringing in iSHELL spectra (Rayner, private comm.).}. The OS filter behaves as a Fabry-Per\'{o}t cavity, introducing a sinusoidal-like pattern with an amplitude of a few percent, and is modeled by:

\begin{gather}
    F(\lambda)=1-A\bigg[\frac{2}{\mathcal{F}}\bigg(\frac{1+\mathcal{F}}{1+\mathcal{F}\sin^{2}(\delta/2)}-1\bigg)-1\bigg], \\[15pt]
    \text{where } \delta_{\scaleto{\scaleto{\textrm{OS}}{4pt}}{4pt}}=\frac{2\pi D_{\scaleto{\textrm{OS}}{4pt}}}{\lambda} \text{ and } \mathcal{F}=\frac{4R}{(1-R)^{2}}.
\end{gather}

\noindent $A$ is the amplitude of the signal and $D_{\scaleto{\textrm{OS}}{4pt}}$ traces the optical path length through the cavity. $\mathcal{F}$ corresponds to the finesse of the cavity, where R is the reflectance \citep{fb_cavity}. A large finesse manifests as sharper downward cavity absorption spikes for the sinusoid, but we don't see significant evidence for a large cavity finesse for $F_{\scaleto{\textrm{OS}}{4pt}}(\lambda)$ in our data. Varying $\mathcal{F}_{\scaleto{\textrm{OS}}{4pt}}$ reveals no obvious preference for any particular value and solutions settle at both upper and lower bounds (0.1, 2) and shows no significant improvement in RVs, so we force $\mathcal{F}_{\scaleto{\textrm{OS}}{4pt}}=1$.

The second source of fringing is introduced by the anti-reflective (AR) coating of the silicon immersion grating surface on the echelle as the light exits the grating. It is modeled with a similar Fabry-Per\'{o}t absorption equation, but in a more complex form because of the wavelength-dependent angle of the incidence on the cavity:

\begin{gather}
    \delta_{\scaleto{\textrm{AR}}{4pt}}=\frac{D_{\scaleto{\textrm{AR}}{4pt}}}{\lambda}\bigg[\cos\bigg(\beta_{0} - \arcsin\bigg(\frac{\lambda}{\lambda_{\scaleto{\textrm{AR0}}{4pt}}} (\sin\beta_{0} + \sin\theta_{b}) - \sin\beta_0\bigg)\bigg) - 1\bigg] + \phi. \label{eq:fringing_ar}
\end{gather}

\noindent $\beta_{0}$ and $\theta_{b}$ are geometric properties of the immersion grating and assumed constant. $D_{\scaleto{\textrm{AR}}{4pt}}$ again traces the optical thickness of the cavity, and $\phi$ allows for a phase shift. $\lambda_{\scaleto{\textrm{AR0}}{4pt}}$ corresponds to the wavelength with the shortest orthogonal path through the AR cavity with $\delta_{\scaleto{\textrm{AR}}{4pt}}=\phi$. We further opt to replace $D_{\scaleto{\textrm{AR}}{4pt}}$ with a second wavelength ``set point'' $\lambda_{\scaleto{\textrm{AR2}}{4pt}}$ corresponding to an overall phase of $\delta_{\scaleto{\textrm{AR}}{4pt}}=\phi-68\pi$, where 68 is arbitrarily chosen to span a significant fraction of the order. $\lambda_{\scaleto{\textrm{AR0}}{4pt}}$, $\lambda_{\scaleto{\textrm{AR2}}{4pt}}$ and $\phi$ are free parameters, along with the amplitude $A_{\scaleto{\textrm{AR}}{4pt}}$ and finesse $\mathcal{F}_{\scaleto{\textrm{AR}}{4pt}}$. A detailed derivation of this equation will be presented in Cale et al. (in prep).

\begin{figure}[H]
    \center
    \includegraphics[width=0.99\textwidth]{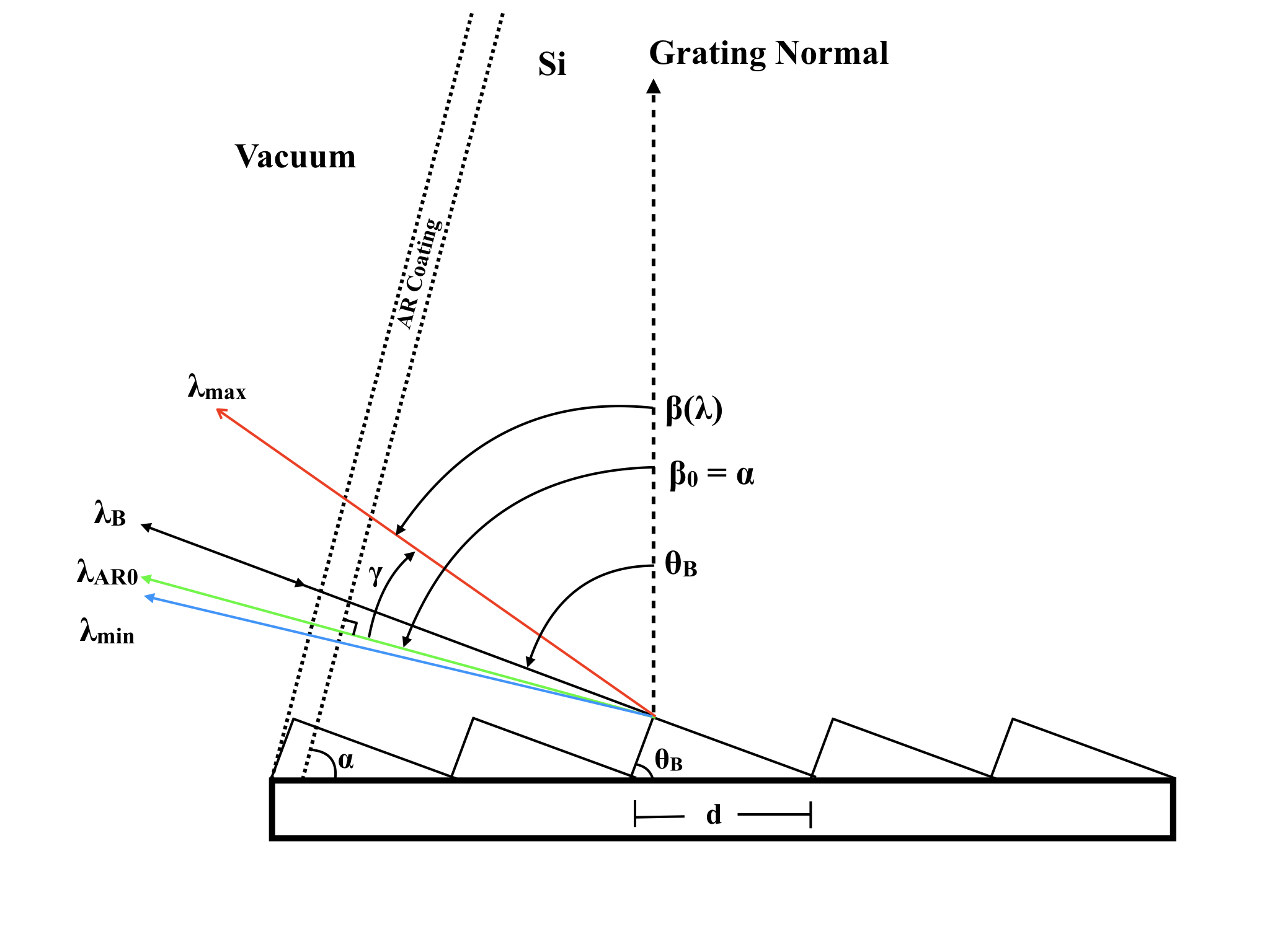}
    \caption{A diagram of the silicon immersion grating. A single echelle order spans from $\lambda_{\textrm{min}}$ to $\lambda_{\textrm{max}}$. After the incident beam (along the the blaze wavelength $\lambda_{B}$) is diffracted at grating, the light is spatially separated and travels through the AR coating with different path lengths, resulting in a unique $D$ for each wavelength. $\lambda_{\scaleto{\textrm{AR0}}{4pt}}$ is the wavelength with the shortest path through the AR coating from Eq. \ref{eq:fringing_ar}. $\theta_{B}$ and $\beta_{0}$ are the blaze angle and grating tilt angle with values given in Table \ref{tab:pars}.}
    \label{fig:fringing_diagram}
\end{figure}

Eq. \ref{eq:fringing_ar} deviates from the standard fringing equation because the incident light at the AR exit face has already been diffracted at the echelle and is therefore spatially separated as a function of $\lambda$, resulting in different resonant cavity lengths for different wavelengths as they enter the AR cavity at different angles (Fig. \ref{fig:fringing_diagram}). The AR fringing component has noticeably sharp downward cavity absorption features, indicating a large finesse, so we fit for $\mathcal{F_{\scaleto{\textrm{AR}}{4pt}}}$. Lastly, we multiply the model by $F_{\scaleto{\textrm{AR}}{4pt}}$ after convolution because it's introduced post-diffraction. One signal we do not observe in our data is fringing corresponding to interference when light enters the AR cavity before getting diffracted at the echelle. We expect this would induce a signal similar to that of the OS filter with a corresponding amplitude and value for $D$. A summary of all forward model parameters is given in Table \ref{tab:pars}.

\begin{table}[H]
\caption{Forward Model Parameters}
\begin{center}
    \begin{tabular}{ | m{8mm} | m{75mm} | m{14mm} | m{40mm} | }
    \hline
    Num. & Description [units] & Symbol & Value/Bounds \\
    \hline
    1 & Stellar Doppler Shift [ms$^{-1}$] & $v_{\star}$ & unbounded \\
    2 & Gas Cell Optical Depth & $\tau_{g}$ & 0.97 \\
    3 & Telluric Doppler Shift [ms$^{-1}$] & $v_{t}$ & (-200, 200) \\
    4 & $\textrm{H}_{2}\textrm{O}$ Optical Depth & $\tau_{t1}$ & (0.02, 4.0) \\
    5 & $\textrm{CH}_{4}$ Optical Depth & $\tau_{t2}$ & (0.1, 3.0) \\
    6 & $\textrm{N}_{2}\textrm{O}$ Optical Depth & $\tau_{t3}$ & (0.05, 3.0) \\
    7 & $\textrm{CO}_{2}$ Optical Depth & $\tau_{t4}$ & (0.05, 3.0) \\
    8 & OS Filter Fringing Amplitude & $A_{\scaleto{\textrm{OS}}{4pt}}$ & (0.015, 0.043) \\
    9 & OS Filter Fringing Cavity Length Scale [nm] & $D_{\scaleto{\textrm{OS}}{4pt}}$ & ($1.8390 \times 10^7$, $1.8393 \times 10^7$) \\
    10 & OS Filter Fringing Finesse & $\mathcal{F}_{\scaleto{\textrm{AR}}{4pt}}$ & 1.0 \\
    11 & AR Fringing Amplitude &  $A_{\scaleto{\textrm{AR}}{4pt}}$ & (0, 0.03) \\
    12 & AR Fringing Reflection Point [nm] & $\lambda_{\scaleto{\textrm{AR0}}{4pt}}$ & $\pm$ 0.075 \\
    13 & AR Fringing Set Point [nm] & $\lambda_{\scaleto{\textrm{AR2}}{4pt}}$ & $\pm$ 0.05 \\
    14 & AR Fringing Phase & $\phi$ & (0, 2$\pi$) \\
    15 & AR Fringing Finesse & $\mathcal{F}_{\scaleto{\textrm{AR}}{4pt}}$ & (1, 4) \\
    16 & Immersion Grating Tilt Angle & $\beta_{0}$ & 71.57097\textdegree \\
    17 & Blaze Angle & $\theta_{b}$ & 71.56795\textdegree \\
    18-20 & Wavelength Solution Lagrange Points (3 total) [nm] & $\lambda_{i}$ & $\pm$ 0.05 \\
    21 & Blaze Function Quadratic Term & $b_{2}$ & ($-5 \times 10^{-5}$, $1 \times 10^{-8}$) \\
    22 & Blaze Function Linear Term & $b_{1}$ & ($-5 \times 10^{-4}$, $5 \times 10^{-4}$) \\
    23 & Blaze Function $0^{th}$ Order Term & $b_{0}$ & (0.98, 1.08) \\
    24 & \textit{LSF} Width [Model pixels] & $a_{0}$ & (5.5, 12) \\
    25-30 & \textit{LSF} Hermite Terms (6 total) & $a_{j}$ & $\pm$ 0.4 \\
    31-45 & Blaze Spline Lagrange Point (15 total) & $bs_{j}$ & $\pm$ 0.135 \\
    46-52 & Wavelength Solution Spline Lagrange Points (7 total) [nm] & $ws_{j}$ & $\pm$ 0.0125 \\
    \hline
    \end{tabular}
\end{center}
\label{tab:pars}
\end{table}

% Section 4.2
\subsection{Stellar Template Retrieval} \label{sec:template_1}

The derivation of the unconvolved stellar spectrum $I_{\star}$ has consistently proven to be a difficult step in forward modeling spectra, particularly in the NIR \citep{2010ApJ...713..410B}.  One approach is to use synthetic model spectra instead.  Models of stellar atmospheres can produce synthetic stellar spectra given the effective temperature, metallicity, and surface gravity \citep{2010ApJ...723..684B, 2011ApJ...735...78C, 2012ApJ...749...16B, 2012ApJS..203...10T}. Due to their lower effective temperatures, atmospheres of late M dwarfs (and brown dwarfs) contain molecular ro-vibrational transitions which can which can significantly contribute to the opacity and affect the emitted spectrum at certain wavelengths \citep{2003IAUS..211..325A}. While the addition and refinement of molecular opacities and full 3D radiation transfer in newer models (such as the BT-Settl PHEONIX models) are providing a better match with observations \citep{2011ASPC..448...91A}, there are still some deficiencies.

A second approach is to deconvolve spectra of A or B stars with little to no stellar spectral features observed through an absorption gas cell (e.g iodine). Spectral lines from the gas cell (and tellurics) provide a means of obtaining the line profile (\textit{LSF}) of the spectrograph, and this can be used to deconvolve the spectrum of a science target taken just before or soon after at a similar airmass and direction in the sky. However this approach presumes the \textit{LSF} remains stable between observations. This may be true for instruments relying on stabilization, but may not be the case for iSHELL as it slews with the telescope at the Cassegrain focus.

We therefore choose to rely on the target observations themselves to extract the stellar spectrum using an iterative deconvolution method described in \cite{2002PASJ...54..873S}. If $I_{\star}$ is the only unknown variable in the model, then the residuals from a model using an imperfect stellar template correspond to the missing (or extra) features of the stellar template, up to a convolution and Doppler shift. Furthermore, by averaging together many spectra, the coadded signal-to-noise is much higher than in individual spectra. In the limit of iteratively adding the residuals back to the stellar template, the template approaches the unconvolved spectrum. This iterative deconvolution method does have its own limitations. First, sufficient sampling at multiple barycenter velocities with high combined SNR are necessary (e.g. two RV data points are not enough). Second, residual correlated noise can gradually get repeatedly added into the stellar template from missed bad pixels, or from non-stellar spectral features that are not well fit.

In our work, we start with a flat guess for $I_{\star}$ and forward model all spectra. We choose a forward model wavelength grid resolution that is about 8 times the data spectral resolution to oversample the data and \textit{LSF}. Higher resolution models yield similar RV precision and RMS values. To compare the model to the observed spectrum (compute an RMS), the high resolution model is linearly interpolated onto the lower-resolution data grid. We shift each set of residuals to a pseudo rest frame of the star according to the barycenter corrections ($v_{\scaleto{\textrm{BC}}{4pt}}$) obtained from \textit{barycentric\_vel.pro}\footnote{Available at http://astroutils.astronomy.ohio-state.edu/exofast/} \citep{2014PASP..126..838W}, decoupling stellar features from any coherent features in the rest frame of the gas cell. We interpolate residuals onto the high resolution model wavelength grid using cubic splines and then median combine across spectra, weighted by RMS$^{-2}$ of the residuals from the forward model fit. We add the median values to the previous template, and re-fit the spectra. We repeat this process until the RVs stabilize, which happens anywhere between 5-40 iterations for orders low in RV content, but typically at later iterations for orders high in RV content. We run all targets through 41 iterations to assess convergence and RV precision.

\begin{figure}[H]
    \center
    \includegraphics[width=0.99\textwidth]{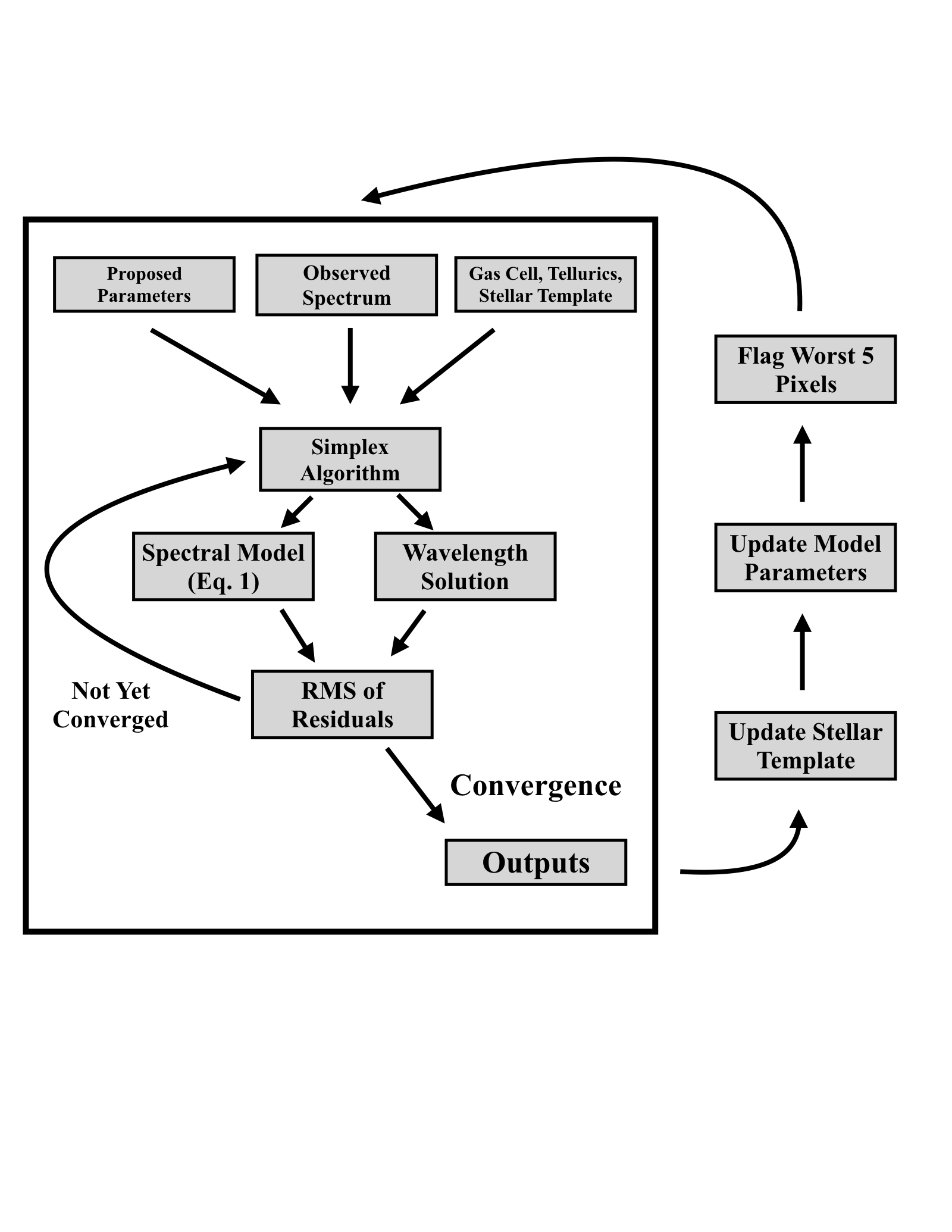}
    \caption{A schematic of the RV pipeline. The proposed parameters for the first iteration are given in Table \ref{tab:pars}. After each iteration, a new stellar template is generated by co-adding the barycenter shifted residuals, and the new proposed parameters are set to the previous iteration's converged values. The worst 5 pixels are only flagged on iteration 10 and each iteration thereafter.}
    \label{fig:code_outline}
\end{figure}

Furthermore, we run the flat template twice on the ``first'' iteration, where we attempt to minimize the effect of the deep stellar CO lines on the solver by masking values deeper than $4\sigma$ in the residuals of the first attempt. The blaze function splines and AR fringing are not well-constrained in the presence of poorly fit stellar lines, and are not included in the first iteration. We do not assess the impact of the initial error in the blaze on the RVs and stellar template generation at later iterations. However, with the variation in phase and decoupling from the star, any fringing that survives the weighted median combination in the pseudo rest frame of the star is minimized. We also force max\{$I_{\star}$\} $\leq$ 1, as the continuum is not well-constrained in early iterations. This requirement may be loosened at later iterations, although this is not explored in this work. Lastly, on iteration 10 and each iteration thereafter, we flag the worst 5 pixels in each set of residuals (see Fig. \ref{fig:data_model}). After 41 iterations, this flags nearly 10\% of all originally used pixels (150 of \texttildelow1648), but improves RV precision at later iterations. Each iteration produces a Doppler shift $v_{\star}$ for each individual spectrum. To calculate an individual relative RV, we ``subtract'' off the barycenter correction from the full Doppler shift, $RV_{\star}=v_{\star} + v_{\scaleto{\textrm{BC}}{4pt}}$. An outline of the forward model is given in Fig. \ref{fig:code_outline}.

After each iteration for a given order, we output text files and corresponding figures for the:

\begin{itemize}
        \item Best fit model to the data and corresponding parameters.
        \item Wavelength solution to the data.
        \item Flagged (ignored) pixels.
        \item The stellar template used for this iteration.
        \item Individual and co-added nightly RVs.
        \item Residuals between the data and models with flagged pixels marked as zeros and the corresponding RMS values and number of target function calls.
  \end{itemize}

To forward model our spectra in a timely manner, we use the ARGO cluster provided by the Office of Research Computing at George Mason University, VA, which can designate 280 cores to a single user, and the exo cluster at George Mason University, currently with 84 cores. Forward modeling a single-order spectrum takes 5-15 minutes at early iterations per core, but only 1-5 minutes at later iterations as parameters have already converged from their updated initial guess.

%%%% RV CALCULATIONS %%%%%%%%%%%%%%%%%%%%%%%%%%%%%%%%%%%%%%%%%%%%%%%%%%%%%%%%%%%%%%
% INDICES:
% i=night
% k=pixels
% j=individual observations
% m=Orders
% Section 4.3
\subsection{RV Calculations} \label{sec:rv_calc}
We explore two methods for computing one radial velocity measurement for each night averaged across echelle orders. The first extends on \citet{2016PASP..128j4501G} utilizing a series of weighted statistical formulas (Section \ref{sec:rv_calc1}). We also explore a second approach which numerically solves for the relative ``zero-points'' for each echelle order (Section \ref{sec:rv_calc2}).

\subsection{Weighted Statistics} \label{sec:rv_calc1}
In the equations that follow, $i,\ j,\ k,$ \& $m$ correspond to the $i^{\textrm{th}}$ night, $j^{\textrm{th}}$ individual observation, $k^{\textrm{th}}$ data pixel, and $m^{\textrm{th}}$ echelle order, respectively. In order to minimize our RV error per epoch, individual observations at $\textrm{SNR}_{i}$ are co-added to obtain a measurement at $\textrm{SNR}_{\textrm{tot}}$ (see Table \ref{tab:obs}), weighted by RMS$^{-2}$ from the forward model fit:

\begin{gather}
    RV_{i,m} = \frac{\sum\limits_{j}^{N_{\scaleto{Obs}{3pt}}^{i}}RV_{m,j}w_{m,j}}{\sum\limits_{j}^{N_{\scaleto{Obs}{3pt}}^{i}} w_{m,j}},  \label{eq:rv1} \\[15pt]
    \textrm{where } w_{m,j} = \frac{N_{\scaleto{\textrm{pix}}{4pt}}^{m,j}}{\sum\limits_{k}^{N_{\scaleto{\textrm{pix}}{4pt}}^{m,j}} [I_{\textrm{Obs}}(\lambda_{k}) - I_{M}(\lambda_{k})]^{2}} \label{eq:rv2}
\end{gather}

\noindent where $RV_{m,j}$ and $w_{m,j}$ are the $j^{\textrm{th}}$ individual RVs and weights for order m, respectively. $N_{\textrm{Obs}}^{i}$ corresponds to the number of observations for the $i^{\textrm{th}}$ night. $I_{\textrm{Obs}}$ and $I_{M}$ are the observed and model spectra, respectively, computed at the $k^{\textrm{th}}$ data pixel. $N_{\scaleto{\textrm{pix}}{4pt}}^{m,j}$ is the number of used pixels for the $j^{\textrm{th}}$ observation for order $m$ (e.g. $N_{\scaleto{\textrm{pix}}{4pt}}^{m,j}$\texttildelow1648--flagged pixels). Deviant pixels flagged during data reduction or forward modeling are not included in the sum.

Nightly error bars are computed via an unbiased weighted standard deviation, divided by the square root of the number of spectra used for that night, $N_{\textrm{Obs}}^{i}$.

\begin{gather}
    \delta RV_{i,m} = \sqrt{\frac{\sum\limits_{j}^{N_{\textrm{Obs}}^{i}} w_{m,j}}{(\sum\limits_{j}^{N_{\textrm{Obs}}^{i}} w_{m,j})^{2}-\sum\limits_{j}^{N_{\textrm{Obs}}^{i}} w_{m,j}^{2}} \frac{\sum\limits_{j}^{N_{\textrm{Obs}}^{i}}w_{m,j}[RV_{m,j}-RV_{i,m}]^{2}}{\mathrel{\raisebox{-0.75em}{$N_{\textrm{Obs}}^{i}$}}}} \label{eq:rv3}
\end{gather}

Before RVs from different echelle orders are combined, the weighted average RV $\overline{RV_{m}}$ of each order is subtracted off. Combined nightly RVs are then computed through a second weighted average,

\begin{gather}
    \overline{RV_{m}} = \frac{\sum\limits_{i}^{N_{\textrm{nights}}}w_{i,m}RV_{i,m}}{\sum\limits_{i}^{N_{\textrm{nights}}}w_{i,m}} \label{eq:rv4} \\[15pt]
    RV_{i} = \frac{\sum\limits_{m}^{N_{\textrm{Ord}}} w_{i,m}[RV_{i,m}-\overline{RV_{m}}]}{\sum\limits_{m}^{N_\textrm{{Ord}}}w_{i,m}} \label{eq:rv5} \\[15pt]
    \delta RV_{i} = \sqrt{\frac{\sum\limits_{m}^{N_{\textrm{Ord}}} w_{i,m}}{(\sum\limits_{m}^{N_{\textrm{Ord}}} w_{i,m})^{2}-\sum\limits_{m}^{N_{\textrm{Ord}}} w_{i,m}^{2}} \frac{\sum\limits_{m}^{N_{\textrm{Ord}}}w_{i,m}[RV_{i,m}-\overline{RV_{m}}-RV_{i}]^{2}}{\mathrel{\raisebox{-0.75em}{$N_{\textrm{Ord}}$}}}} \label{eq:rv6} \\[15pt]
    \textrm{with } w_{i,m}=1/\delta RV_{i,m}^{2} \label{eq:rv7}
\end{gather}

where $N_{\textrm{Ord}}$ is the number of echelle orders used.

\subsection{Detrending Minimization} \label{sec:rv_calc2}
Second, to better constrain the intrinsic order dependent characteristic RVs (assumed to be $\overline{RV_{m}}$ above), we utilize a version of the Trend Filtering Algorithm (TFA) \citep{2005MNRAS.356..557K} which is frequently used to remove systematics and detrend \kepler\ light curves \citep{2017MNRAS.471..759A,2012ascl.soft08004S}. We implement and minimize a modified weighted formula from \cite{2019arXiv190100503B} akin to the weighted implementation of TFA in \cite{2016PASP..128h4504G}:

% nights, orders
\begin{gather}
    \sum\limits_{i,m} w_{i,m} [RV_{i,m}-\overline{RV_{m}}-RV_{i}'-\overline{RV_{m}'}]^{2} \textrm{ where } w_{i,m}=1/\delta RV_{i,m}^{2} \label{eq:rv8}
\end{gather}

\noindent $RV_{i,m}$ are the nightly RVs from Eq. \ref{eq:rv1}, $\overline{RV_{m}'}$ are the new order offsets, and $RV_{i}'$ are the ``detrended'' RVs for the $i^{\textrm{th}}$ night. $\overline{RV_{m}'}$ and $RV_{i}'$ are sets of free parameters with lengths $N_{\textrm{Ord}}$ and $N_{\textrm{nights}}$, respectively. The weighted average $\overline{RV_{m}}$ of each order from Eq. \ref{eq:rv4} is still subtracted from $RV_{i,m}$ before optimizing. Values of $\overline{RV_{m}'}$ and $RV_{i}'$ are set to zero as an initial guess with bounds $\pm$ 5 ms$^{-1}$ and $\pm$ 50 ms$^{-1}$, respectively. Final error bars are computed using Eq. \ref{eq:rv6} with the detrended orders, $RV_{i,m}-\overline{RV_{m}}-\overline{RV_{m}'}$. The parameters are optimized using the same Nelder-Mead algorithm described in Section \ref{sec:numerical_solver}.

%%%% RESULTS %%%%%%%%%%%%%%%%%%%%%%%%%%%%%%%%%%%%%%%%%%%%%%%%%%%%%%%%%%%%%%
% Section 5
\section{Results} \label{sec:results}

For each of the three stars, we run orders $5-26$ $(m=216-237)$ through 41 iterations. Order numbers $1-4$ $(m=212-215)$ contain sufficient stellar and gas cell RV information content, but are also higher in water absorption and haven't yielded comparable RV precision ($>$30 ms$^{-1}$ long-term). We aim to explore a more sophisticated telluric model for these orders in future work. Higher order numbers shortward of the CO band ($< 2.29\ \micron,\ m>229$) are relatively low in stellar RV content and have fewer gas cell lines for a precise wavelength calibration (see Section \ref{sec:err_analysis}).

\begin{figure}[H]
    \centering
    \includegraphics[width=0.75\textwidth]{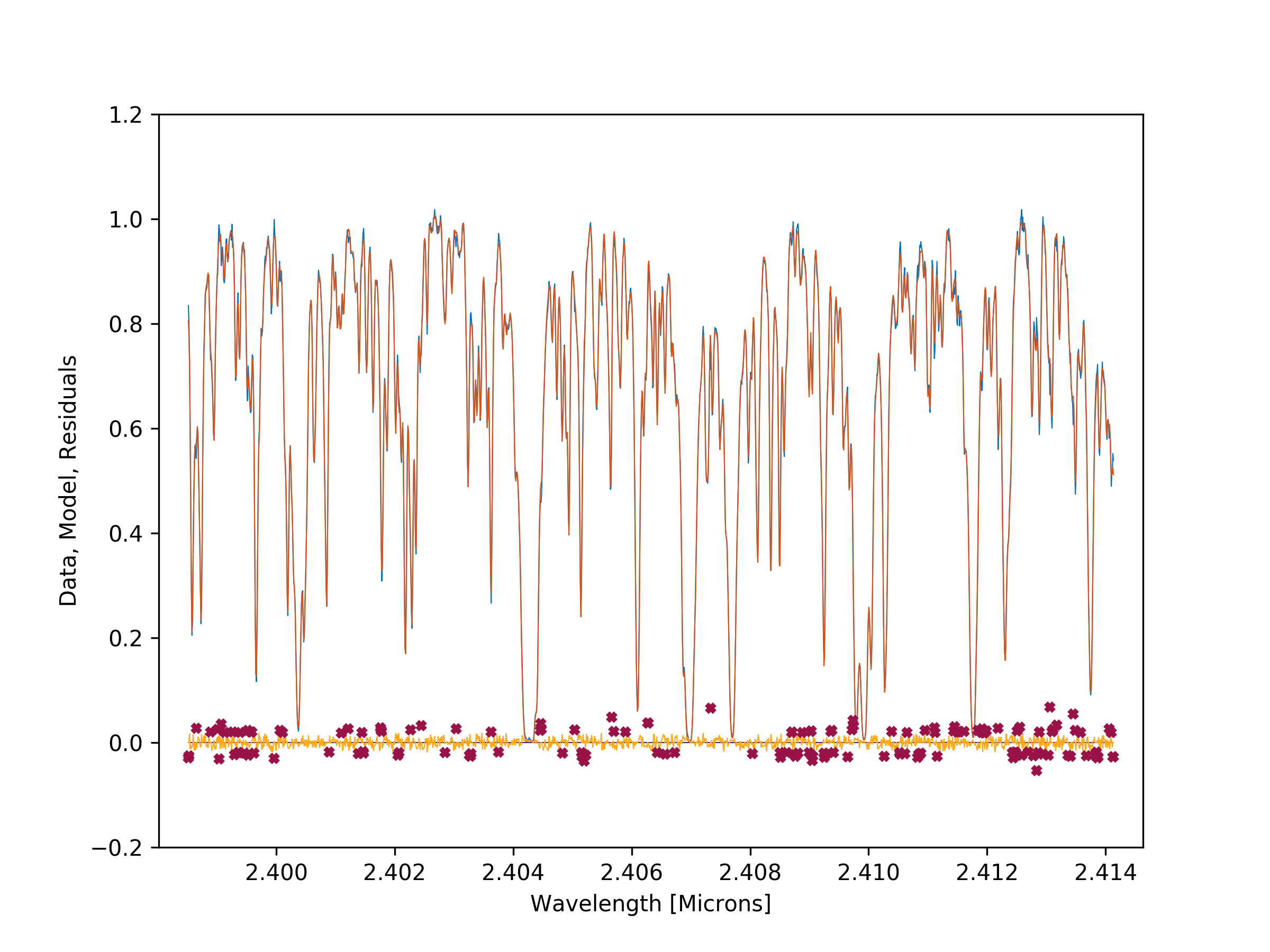}
    \includegraphics[width=0.75\textwidth]{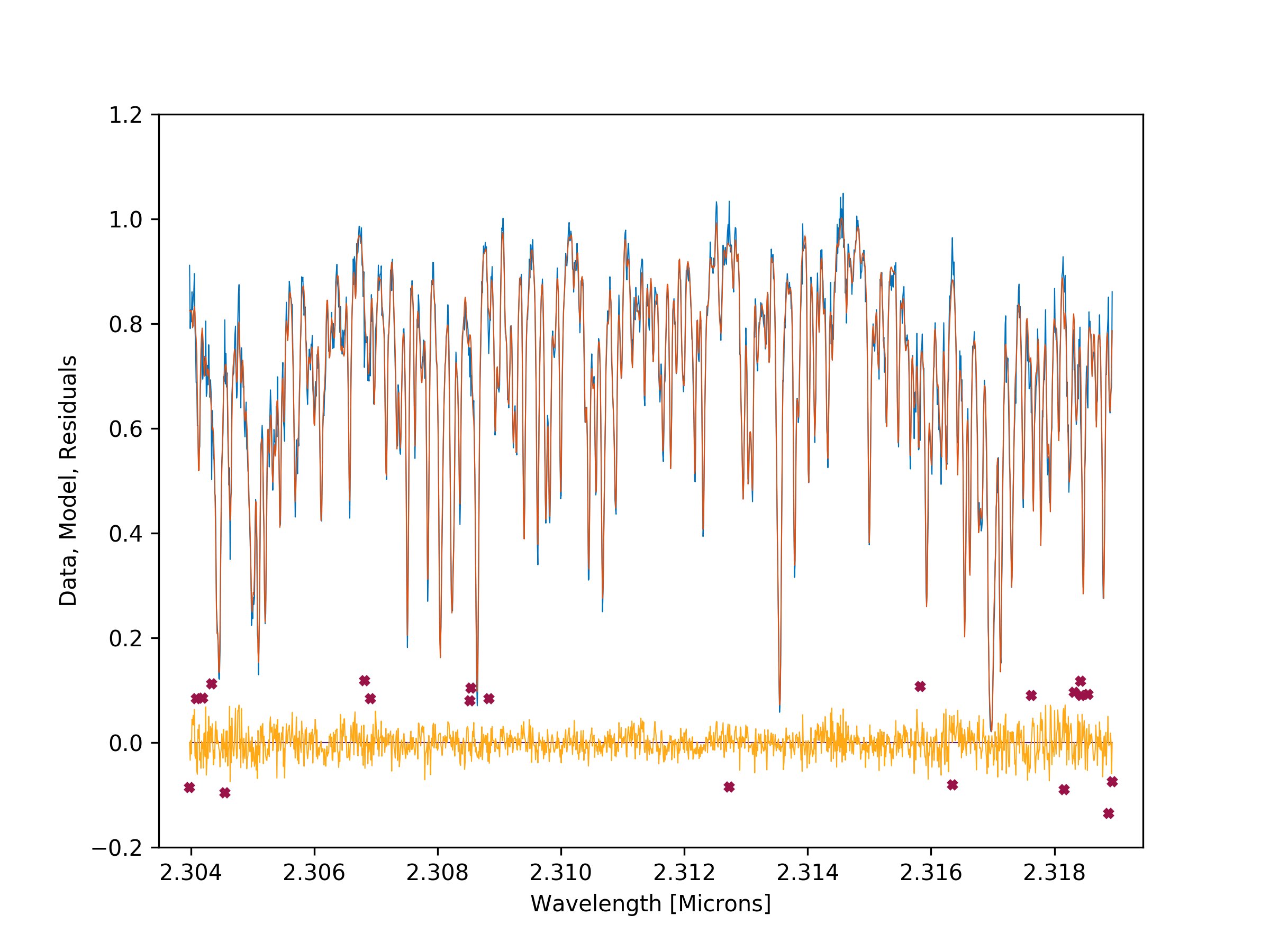}
    \caption{\textit{Top}: An example fit to a spectrum of Barnard's Star from July 29, 2017, for order 6 $(m=217)$ from iteration 41 (last) from the high SNR run. The data is in blue and the model in red. The deep and wide absorption lines with near zero transmission correspond to water in Earth's atmosphere. The worst pixels flagged between iterations 10-41 are marked as red X's. \textit{Bottom}: A lower SNR example fit from Oct. 20, 2017, to a spectrum of Barnard's Star for order 15 $(m=226)$ and iteration 15. The data is in blue and the model in red. Any major stellar features will have visually converged at this point.}
    \label{fig:data_model}
\end{figure}

For Barnard's Star, we also compute RVs separately from the first nine nights for orders $6-17$ $(m=217-228)$, which we refer to as the ``high SNR run'' in the rest of this paper. Barnard's Star has historically been shown to have the highest long-term RV stability with precisions below our expected noise floor, so we use it to assess multi-order RV precision (Section \ref{sec:rvs}), forward model parameter distributions (Section \ref{sec:model_params}) and alternative forward model implementations (Section \ref{sec:discussion}). This also shows the impact of including lower SNR observations in the stellar template generation. Fig. \ref{fig:data_model} shows example fits of the model spectrum to a high and low SNR observation of Barnard's Star. The residuals (and thus RMS) for the low SNR observations are typically twice as large compared to the high SNR observations (\texttildelow 2\% vs. 1\%), and are therefore weighted less in generating the stellar template.

% Section 5.1
\subsection{RVs} \label{sec:rvs}

\begin{figure}
    \center
    \includegraphics[width=0.4\textwidth]{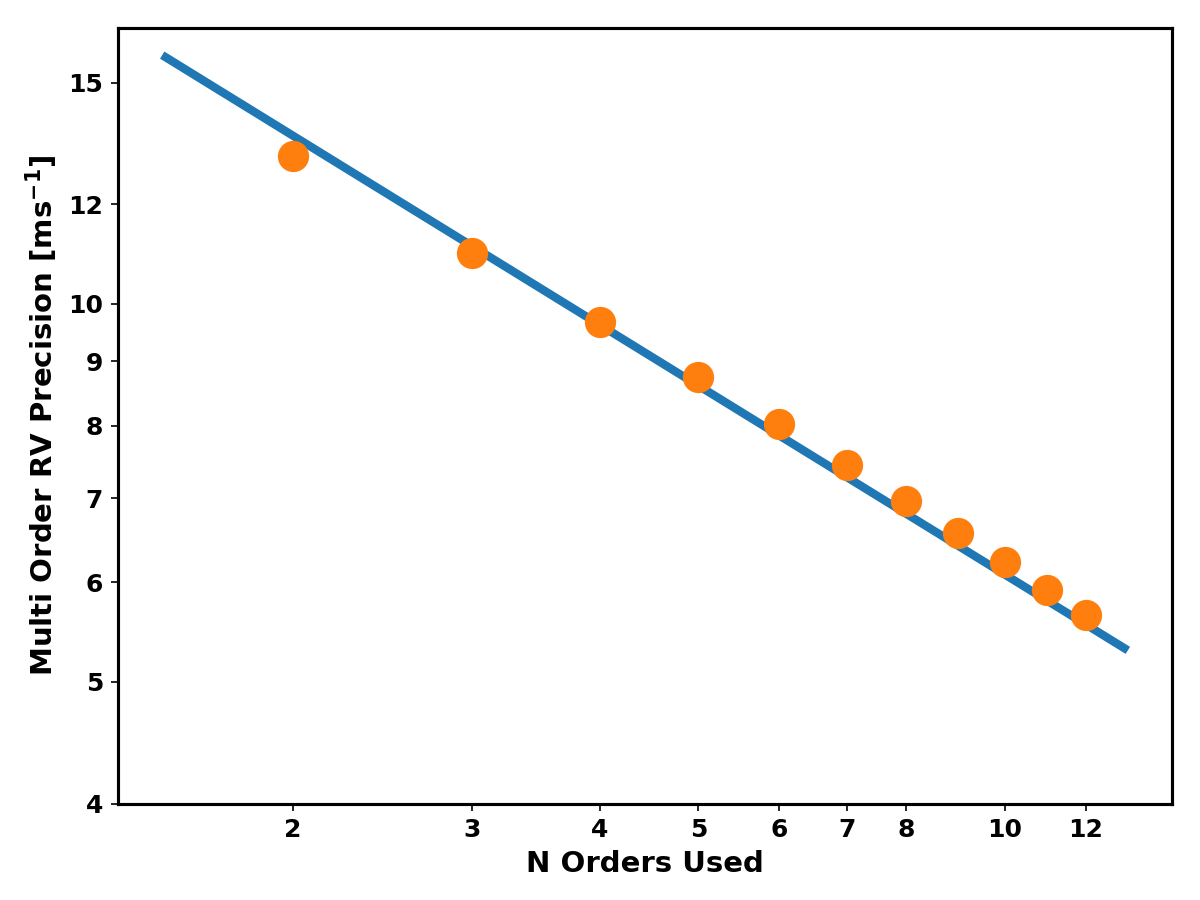}
    \includegraphics[width=0.4\textwidth]{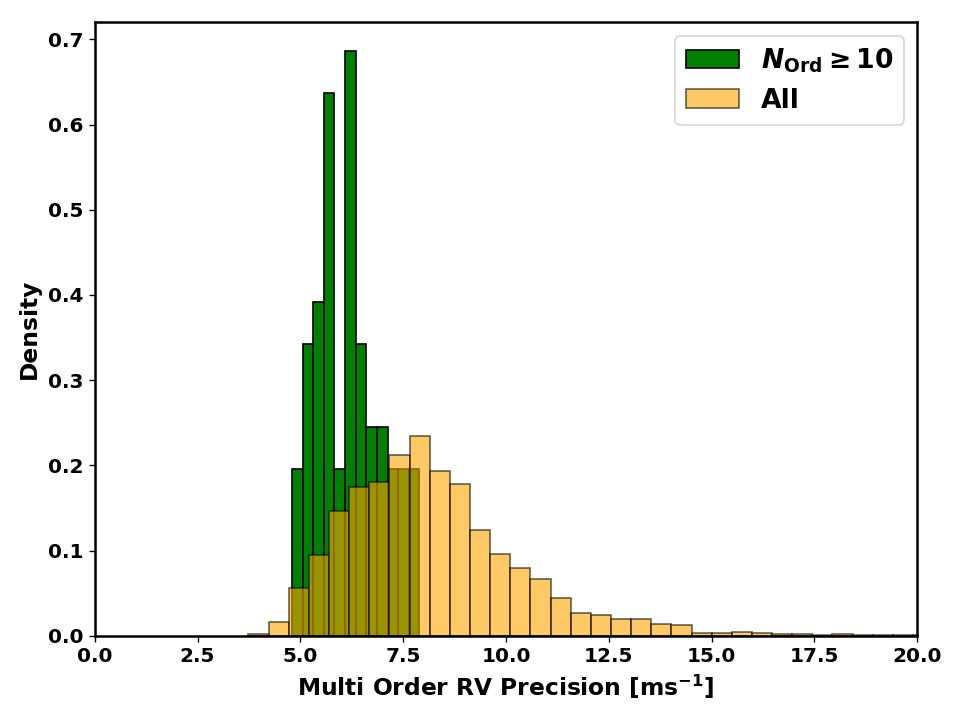}
    \caption{\textit{Left}: The orange circles correspond to the average long-term RV RMS obtained for all possible combinations of $N$ orders. The trend is obtained by fitting a function $A N_{\textrm{Ord}}^{-1/2}$ where A is a constant parameter. On average our multi-order velocities are consistent with averaging out random noise. \textit{Right}: A histogram of long-term RV precisions obtained trying all possible order combinations for the 12 orders (6--17, $m=217-228$) for the high SNR Barnard's run. In yellow, we show all order combinations of 2--12 orders (e.g. there is only one 12-order combination, 12 11-order combinations, etc.), and in green order combinations with 10--12 combined orders. The total number of order combinations are 4083 and 79, respectively. For the latter green histogram with 10 or more combined orders, the 5th percentile is 5.0 ms$^{-1}$, while the 10th percentile is 5.2 ms$^{-1}$, and the 20th is 5.4 ms$^{-1}$.}
    \label{fig:rv_prec1}
\end{figure}

\begin{figure}[H]
    \center
    \includegraphics[width=0.65\textwidth]{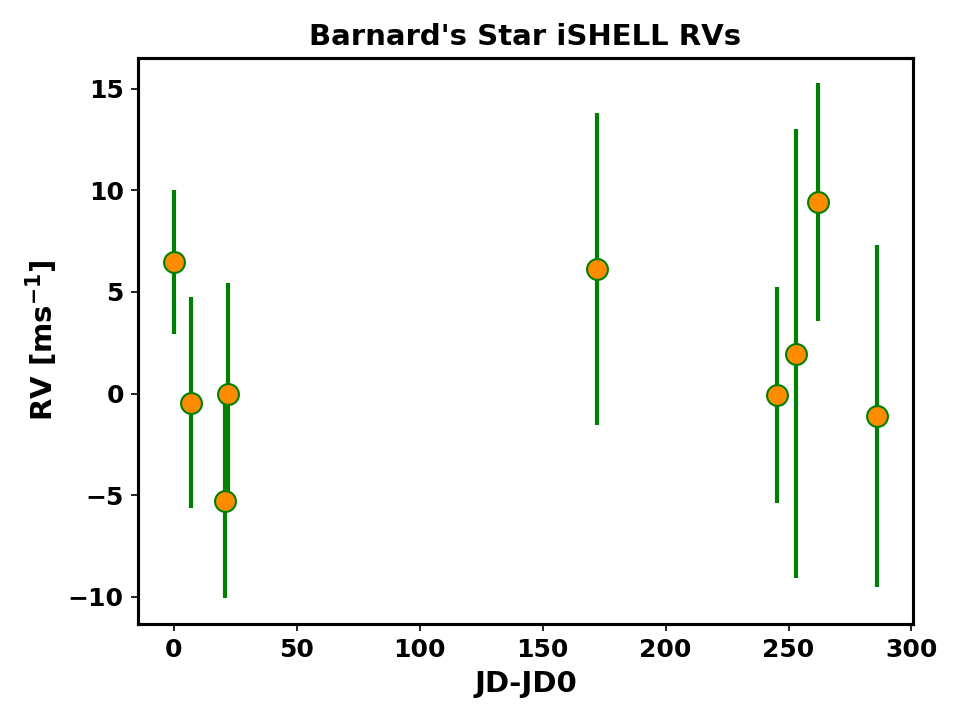}
    \caption{The best case multi-order RV combination that yielded the lowest RMS for Barnard's Star from the high SNR run. The unweighted standard deviation is 4.33 ms$^{-1}$. JD0 corresponds to the first nightly JD for each target given in Table \ref{tab:rvs2}.}
    \label{fig:gj699_rvs_highsnr}
\end{figure}

\begin{figure}[H]
    \center
    \includegraphics[width=0.65\textwidth]{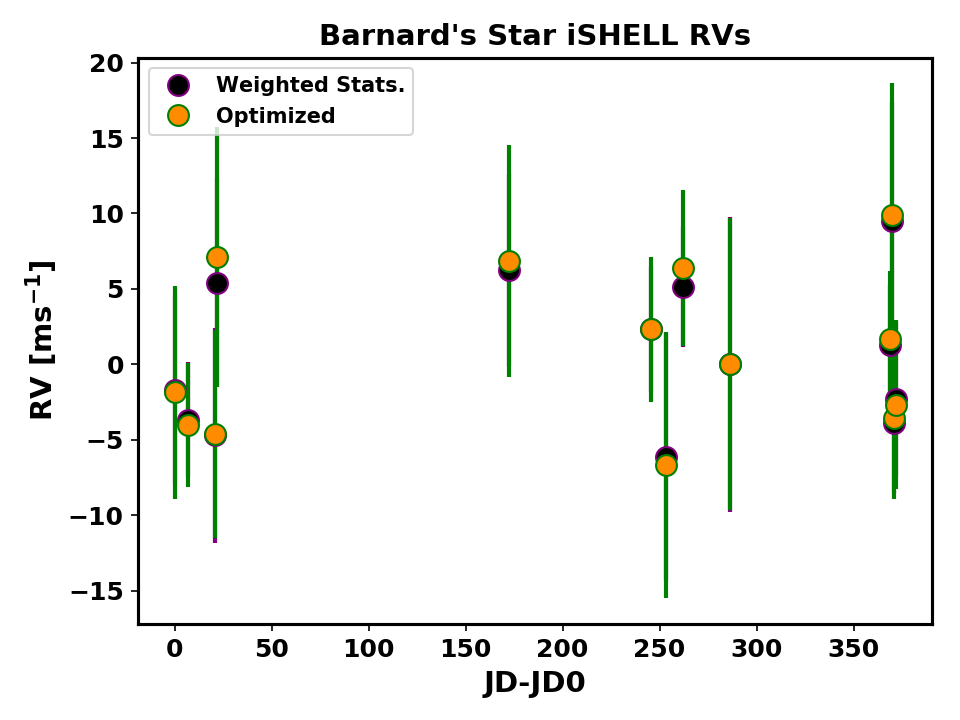}
    \caption{The best case multi-order RV combination that yielded the lowest RMS for Barnard's Star for the full dataset. The unweighted standard deviation is 5.13 ms$^{-1}$ for the optimized set. The weighted statistics formulation (Section \ref{sec:rv_calc1}) agrees well with the optimized RVs (Section \ref{sec:rv_calc2}). Hidden error bars are of similar size.}
    \label{fig:gj699_rvs_all}
\end{figure}

\begin{figure}[H]
    \center
    \includegraphics[width=0.65\textwidth]{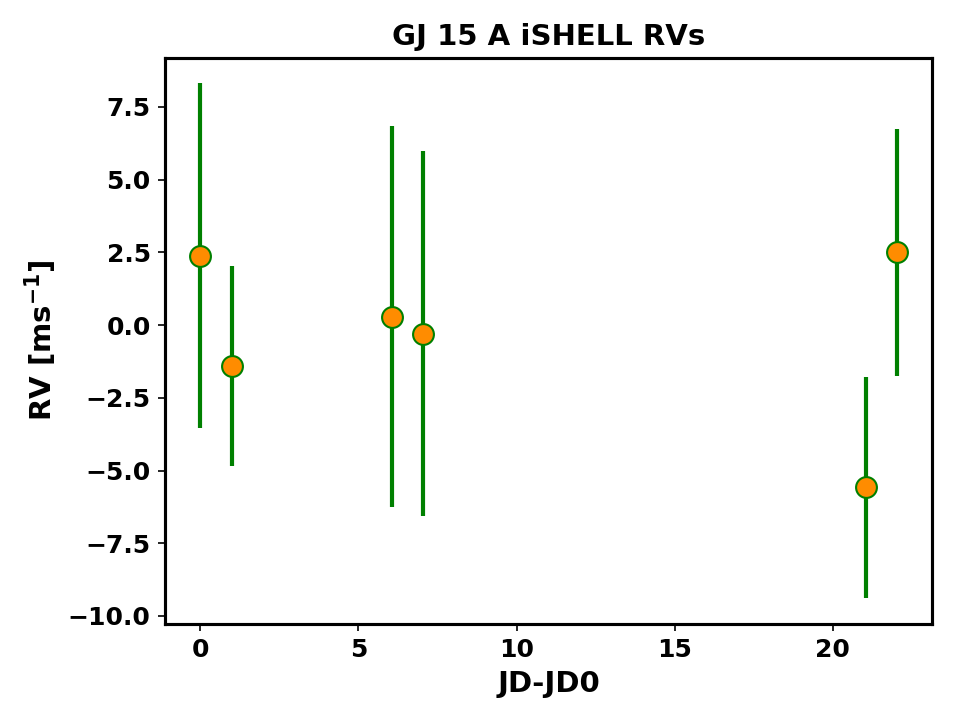}
    \caption{The best case multi-order RV combination that yielded the lowest RMS for GJ 15 A. The unweighted standard deviation is 2.72 ms$^{-1}$.}
    \label{fig:gj15a_rvs}
\end{figure}

\begin{figure}[H]
    \center
    \includegraphics[width=0.65\textwidth]{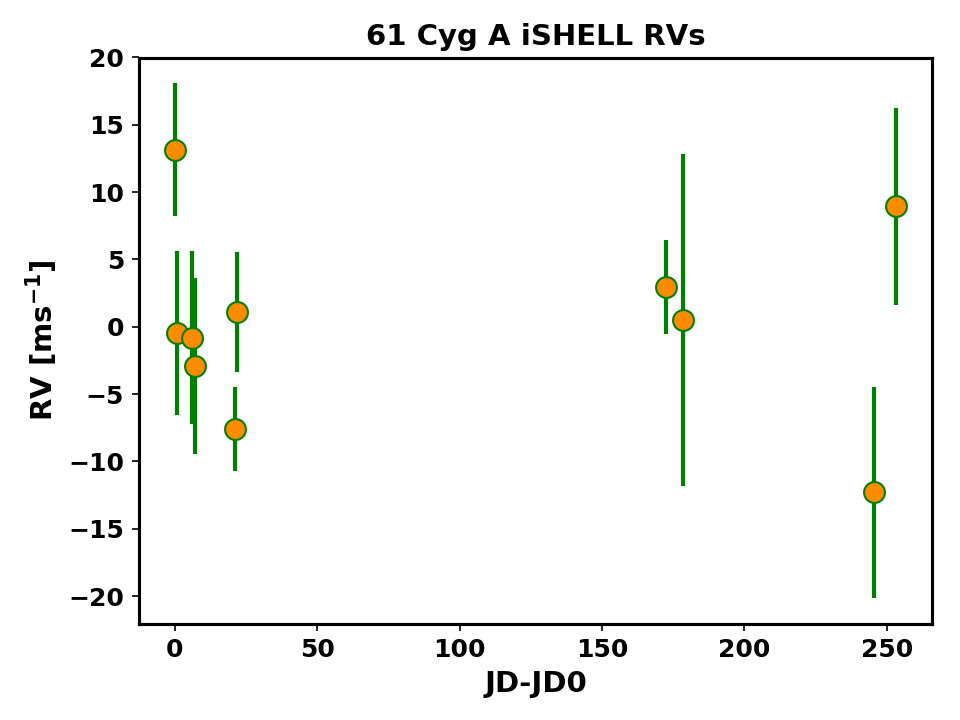}
    \caption{The best case multi-order RV combination that yielded the lowest RMS for 61 Cyg A. The unweighted standard deviation is 3.77 ms$^{-1}$. The last data point is not shown.}
    \label{fig:61cyga_rvs}
\end{figure}

To assess our combined order precision, we utilize a powerset (all possible subsets of a given set) to analyze the RV precision as a function of orders used and look for orders that statistically yield lower combined RV precision. We do so using the weighted statistical approach given by Eqs. \ref{eq:rv1}-\ref{eq:rv7}. Using Eq. \ref{eq:fischer} (see Section \ref{sec:err_analysis}), we take our RV precision $\sigma_{\textrm{RV}}\propto N_{\textrm{Ord}}^{-1/2}$, and find the long-term RV precision follows this relationship (Fig. \ref{fig:rv_prec1}). Lastly, for Barnard's Star, we subtract off from each individual (single spectrum) RV the secular acceleration of 4.515 ms$^{-1}$yr$^{-1}$ given by \cite{2012AAS...21924504C} before any multi-order or nightly RVs are computed. We don't perform this for other targets as their relative spatial motion is not significant enough to produce a detectable acceleration. Orders that yielded the lowest long-term precision are then optimized through Eq. \ref{eq:rv4}, and typically reproduce RVs from the weighted formulation (Fig. \ref{fig:gj699_rvs_all}). The long-term RV precisions for each individual order are presented in Table \ref{tab:rvs1}. We present the best combined order precision in Table \ref{tab:rvs2} and corresponding Figures \ref{fig:gj699_rvs_highsnr}-\ref{fig:61cyga_rvs}. We obtain best case long-term RV precisions of 4.3 ms$^{-1}$ for Barnard's Star, 2.7 ms$^{-1}$ for GJ 15 A, and 3.8 ms$^{-1}$ for 61 Cyg A. For 61 Cyg A, the RV is a large outlier for the last night (+ 1 kms$^{-1}$). This outlier has a typical multi-order RV uncertainty, and survived numerous modifications to the code during development. We suspect this is an observational error where we mistakenly observed 61 Cygni B, or a flare event on the surface of the star. We therefore disregard this night from any long-term RV calculations.

% Table for RV precision for each order (all targets)
\begin{table}[H]
\caption{The best single-order long-term RV precisions (unweighted standard deviation) for each of the four runs, and the corresponding best iteration. We only include the first 10 nights for 61 Cyg A in the calculation.}
\begin{center}
    \begin{tabular}{ | m{8mm} | m{8mm} | m{18mm} | m{4mm} | m{18mm} | m{4mm} | m{18mm} | m{4mm} | m{24mm} | m{4mm} |}
    \hline
    Image\newline Order & Echelle\newline Order & Barnard's Star\newline (High SNR) [ms$^{-1}$] & Iter & Barnard's Star \newline (All) [ms$^{-1}$] & Iter & GJ 15 A [ms$^{-1}$] & Iter & 61 Cyg A [ms$^{-1}$] & Iter\\
    \hline
    5  & 216 & -     & -  & 19.02 & 10 & 50.07  & 32 & 16.68  & 31\\
    6  & 217 & 15.87 & 28 & 13.68 & 33 & 32.39  & 15 & 32.92  & 40\\
    7  & 218 & 13.95 & 9  & 16.25 & 12 & 3.61   & 32 & 13.17  & 15\\
    8  & 219 & 11.66 & 20 & 10.68 & 23 & 7.39   & 32 & 9.48   & 16\\
    9  & 220 & 18.05 & 7  & 14.50 & 40 & 6.59   & 13 & 15.14  & 39\\
    10 & 221 & 16.53 & 27 & 17.15 & 21 & 8.36   & 18 & 19.32  & 32\\
    11 & 222 & 15.10 & 23 & 14.93 & 24 & 4.81   & 40 & 10.95   & 26\\
    12 & 223 & 21.11 & 40 & 16.59 & 6  & 17.38  & 34 & 11.99  & 12\\
    13 & 224 & 12.99 & 40 & 27.16 & 25 & 9.39   & 12 & 11.88  & 20\\
    14 & 225 & 29.20 & 19 & 33.23 & 33 & 13.38  & 6 & 24.46  & 18\\
    15 & 226 & 16.17 & 17 & 11.19 & 21 & 9.07   & 15  & 30.22  & 19\\
    16 & 227 & 16.22 & 40 & 15.44 & 16 & 20.88  & 6  & 27.24  & 6\\
    17 & 228 & 31.07 & 14 & 28.14 & 15 & 7.34   & 26 & 26.85  & 34\\
    18 & 229 & -     & -  & 28.91 & 40 & 127.77 & 18 & 488.75 & 6\\
    19 & 230 & -     & -  & 25.14 & 23 & 49.54  & 17 & 475.78 & 6\\
    20 & 231 & -     & -  & 27.34 & 24 & 48.82  & 11  & 96.62  & 7\\
    21 & 232 & -     & -  & 20.17 & 22 & 67.97  & 14 & 895.76 & 6\\
    22 & 233 & -     & -  & 49.28 & 40 & 89.55  & 40 & 240.85 & 6\\
    23 & 234 & -     & -  & 46.82 & 40 & 90.12  & 18 & 124.50 & 9\\
    24 & 235 & -     & -  & 41.13 & 15 & 42.95  & 6 & 60.39  & 14\\
    25 & 236 & -     & -  & 68.48 & 27 & 132.62 & 40 & 195.59 & 6\\
    26 & 237 & -     & -  & 45.51 & 40 & 95.76  & 16 & 67.75  & 21\\
    \hline
    \end{tabular}
\end{center}
\label{tab:rvs1}
\end{table}

% Table for Combined RVs (all targets)
\begin{table}[H]
\centering
\begin{threeparttable}
\caption{The best multi-order RVs for each target obtained through a powerset. The unweighted standard deviation $\sigma$ and value of $\chi_{red}^{2}$ of the measurements is noted.} \label{tab:rvs2}
    \begin{tabular}{ | m{24mm} | m{28mm} | m{24mm} | }
    \hline
    JD-2457677 & Nightly RV [ms$^{-1}$] & Unc. [ms$^{-1}$] \\
    \hline
    \multicolumn{3}{|c|}{|Barnard's Star (high SNR) (Orders 7-9, 11, 13), $\sigma=4.33$ ms$^{-1}$, $\chi_{red}^{2}=0.81$|} \\
    \hline
    0.76914091 & 5.26 & 3.55 \\
    7.72960279 & -1.70 & 5.17 \\
    21.71375211 & -6.50 & 4.78 \\
    22.69412594 & -1.20 & 5.43 \\
    173.07680052 & 4.89 & 7.68 \\
    246.08399455 & -1.35 & 5.32 \\
    253.97949511 & 0.69 & 11.04 \\
    262.9147927 & 8.16 & 5.84 \\
    286.90298491 & -2.39 & 8.42 \\
    \hline
    \multicolumn{3}{|c|}{|Barnard's Star (All) (Orders 6-10, 14, 17, 20), $\sigma=5.13$ ms$^{-1}$, $\chi_{red}^{2}=0.61$|} \\
    \hline
    0.76914091 & -1.94 & 7.06 \\
    7.72960279 & -4.02 & 4.11 \\
    21.71375211 & -4.68 & 6.92 \\
    22.69412594 & 7.09 & 8.61 \\
    173.07680052 & 6.86 & 7.70 \\
    246.08399455 & 2.30 &  4.81 \\
    253.97949511 & -6.65 & 8.81 \\
    262.9147927 & 6.35 & 5.19 \\
    286.90298491 & -0.04 & 9.68 \\
    369.6983861 & 1.65 & 4.47 \\
    370.68776439 & 9.96 & 8.76 \\
    371.69478571 & -3.60 & 5.38 \\
    372.68679116 & -2.62 & 5.60 \\
    \hline
    \multicolumn{3}{|c|}{|GJ 15 A (Orders 8, 9, 10), $\sigma=2.72$ ms$^{-1}$, $\chi_{red}^{2}=0.50$|} \\
    \hline
    0.82538026 & 2.81 & 5.93 \\
    1.83099163 & -0.980 & 3.43 \\
    6.88132949 & 0.71 & 6.57 \\
    7.88033167 & 0.13 & 6.29 \\
    21.8549003 & -5.16 & 3.80 \\
    22.86261265 & 2.92 & 4.24 \\
    \hline
    \multicolumn{3}{|c|}{|61 Cyg A\tnote{1} (Orders 8-9, 11-12, 17), $\sigma=3.77$ ms$^{-1}$, $\chi_{red}^{2}=0.71$|} \\
    \hline
    0.79182255 & -3.28 & 10.66 \\
    1.7336577 & -0.65 & 11.27 \\
    6.8404918 & -0.570 & 2.34 \\
    7.80991716 & 6.21 & 2.38 \\
    21.79603189 & 3.44 & 5.67 \\
    22.7468759 & 3.97 & 8.79 \\
    173.15354752 & 1.02 & 15.35 \\
    179.14649648 & -0.10 & 23.62 \\
    246.12869699 & -7.58 & 12.27 \\
    254.07051542 & -2.45 & 4.30 \\
    \hline
    263.01044249 & 1403.12 & 17.50 \\
    \hline
    \end{tabular}
    \begin{tablenotes}
    \item [1] Only the first ten nights are considered in any calculations for 61 Cyg A.
    \end{tablenotes}
\end{threeparttable}
\end{table}

\pagebreak
\subsection{Error Analysis} \label{sec:err_analysis}

We compare our obtained RV precisions with the expected analytic precision in the optimistic photon noise limit. Following \cite{2001A&A...374..733B}, we compute a photon noise model precision:

\begin{gather}
    \sigma_{\textrm{RV}} = c \bigg[\sum_{i} \frac{(\lambda_{i} \textrm{d}A_{i}/\textrm{d}\lambda_{i})^{2}}{A_{i}} \bigg]^{-1/2} \label{eq:fischer}
\end{gather}

\noindent for both the convolved stellar template and gas cell used in our forward model, which we then add in quadrature to obtain a photon noise estimated RV precision. We do this for each order. $A_{i}$ is the signal at pixel $i$ given in photo-electrons (PEs). We adopt a peak SNR of 300 (per detector pixel) and gain of 1.8 to convert SNR to PEs\footnote{\url{http://irtfweb.ifa.hawaii.edu/\~ishell/iSHELL_observing_manual.pdf}}. This is performed on the data grid, ignoring cropped pixels. A $\sinc^{\sim 1.6}$ models the observed blaze function prior to flat-fielding sufficiently well, so we modulate the templates to approximately account for the lower SNR near the edges of the orders. We also convolve both templates with a Gaussian \textit{LSF} with $a_{0}=8$, which is a representative \textit{LSF} width in our model grid, and is roughly equal to one data pixel. 

\begin{figure}[H]
    \center
    \includegraphics[width=0.75\textwidth]{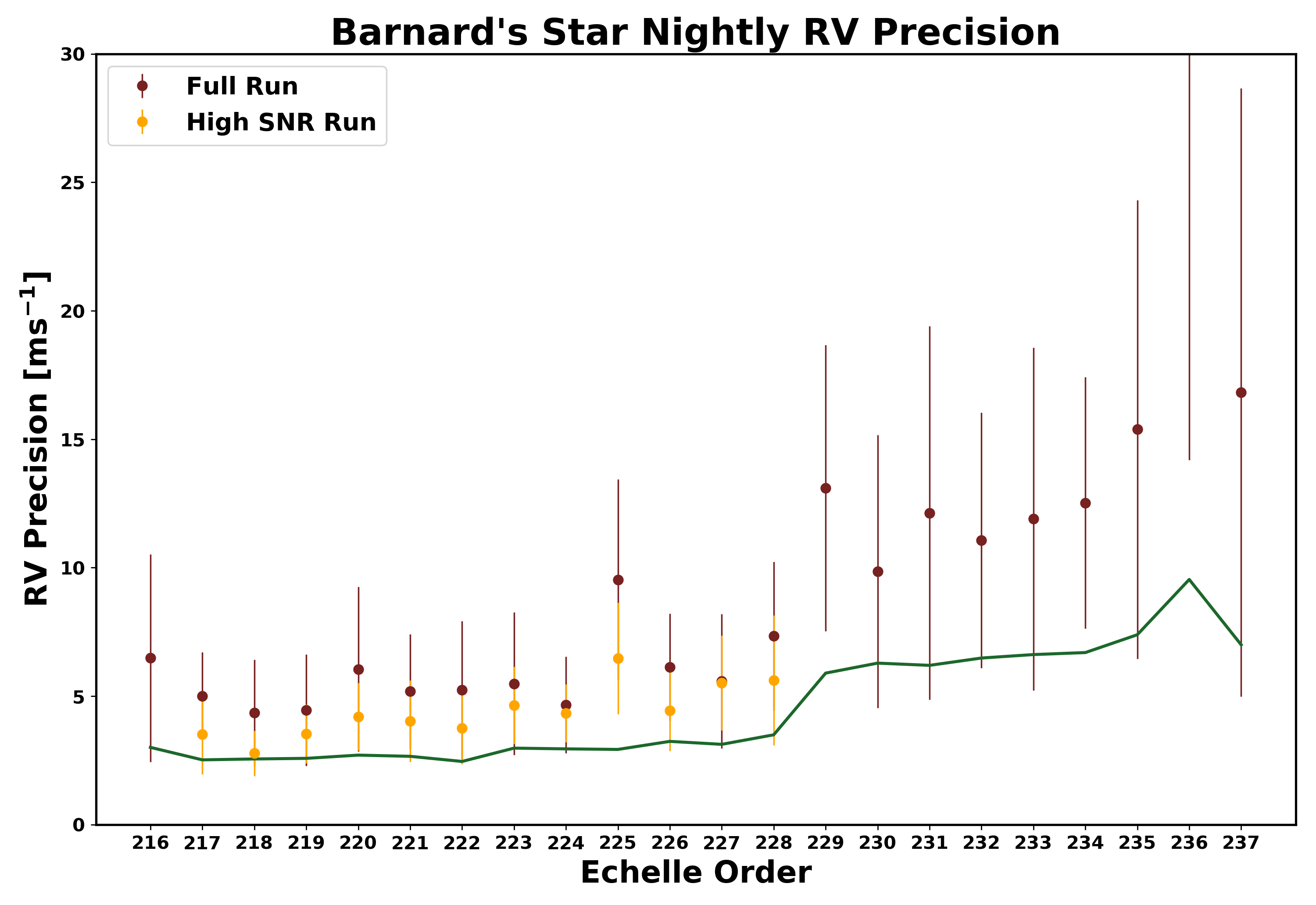}
    \caption{The nightly Barnard's Star RV uncertainties for each order (markers), averaged over nights, alongside the estimated photon noise limit (solid line). Nights from the full data set are in red, and the high SNR run are shown in orange. The lower SNR data (last 4 nights) are ignored in generating this plot. Error bars represent a $1\sigma$ spread of the uncertainties in the nightly RVs. Barnard's Star nightly RV uncertainties are above the noise floor, unlike GJ 15 A and 61 Cyg A. Including lower SNR measurements can still impact nights at higher SNRs due to the common stellar template generation. The CO bandhead for cool stars starts at $2.29\ \micron$ $(m\leq228)$.}
    \label{fig:photon_gj699}
\end{figure}

\begin{figure}[H]
    \center
    \includegraphics[width=0.75\textwidth]{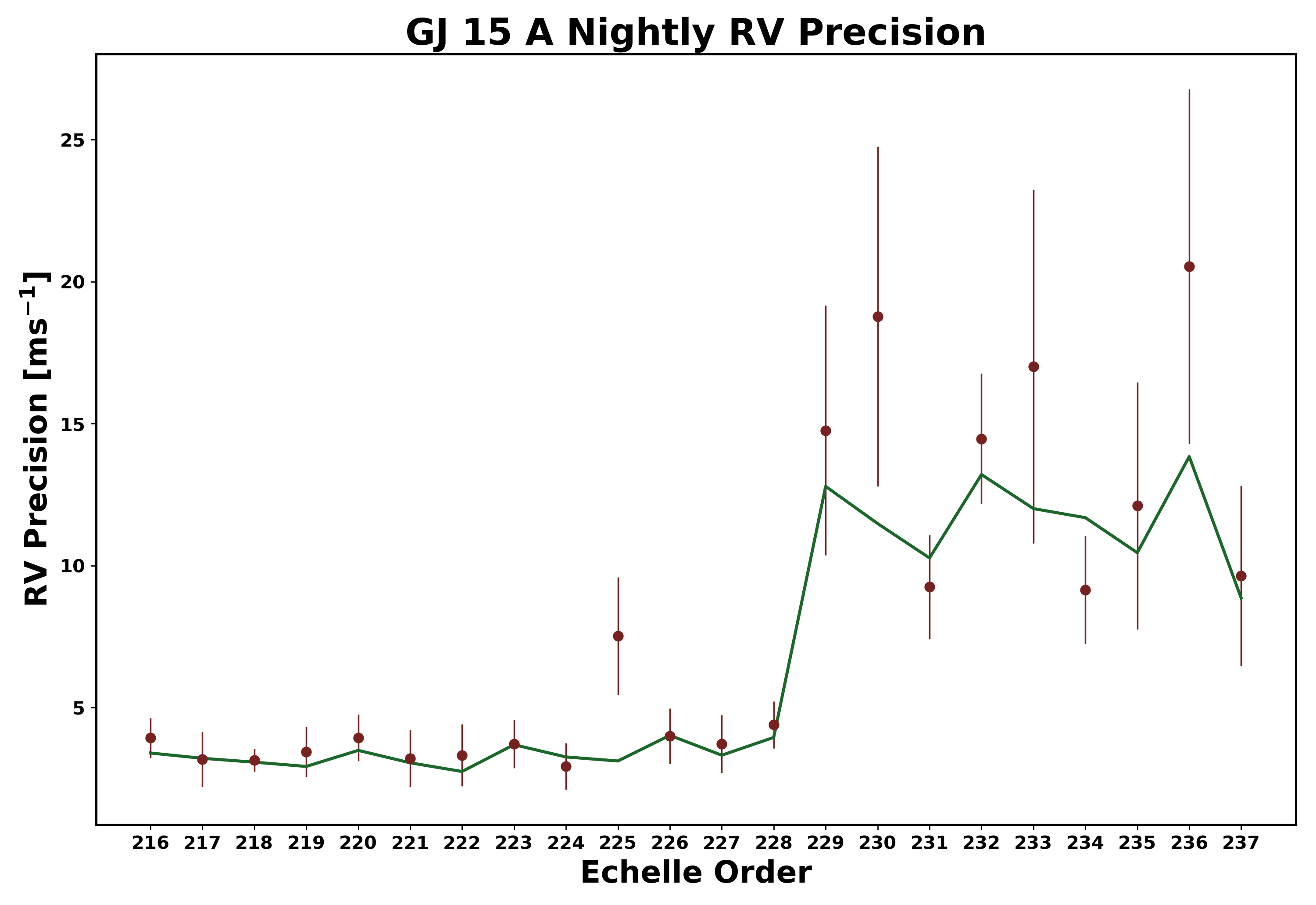}
    \caption{Same as Fig. \ref{fig:photon_gj699}, but for GJ 15 A.}
    \label{fig:photon_gj15a}
\end{figure}

\begin{figure}[H]
    \center
    \includegraphics[width=0.75\textwidth]{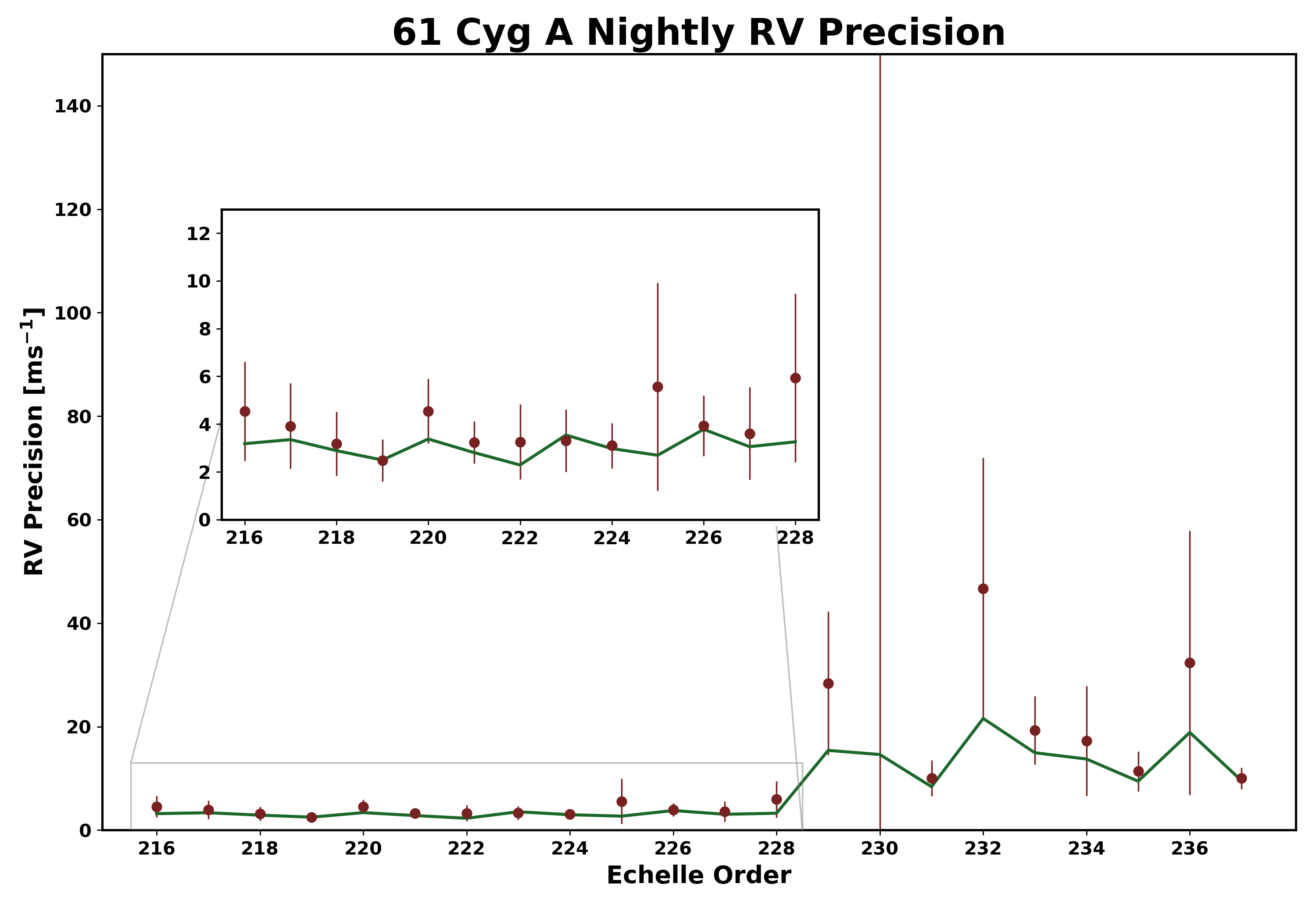}
    \caption{Same as Fig. \ref{fig:photon_gj699}, but for 61 Cyg A.}
    \label{fig:photon_61cyga}
\end{figure}

For GJ 15 A and 61 Cyg A, our nightly RV precision, $\delta RV_{i,m}$, is comparable to the photon noise estimate (Figs. \ref{fig:photon_gj15a}, \ref{fig:photon_61cyga}). Nightly scatter in RVs for Barnard's Star are a few ms$^{-1}$ above the photon noise estimate, even when ignoring the lower SNR data (Fig. \ref{fig:photon_gj699}).

Achieving this precision over long timescales is challenging due to other standard and non-standard sources in the RV error budget unaccounted for in the photon noise approximation. Known sources of external error arise from the fact that iSHELL is mounted at Cassegrain focus, and thus mechanically flexes as the telescope moves. Finally, iSHELL has both standard and non-standard fringing sources that will induce errors of $>10-20$ ms$^{-1}$ if not modeled sufficiently, and $>$50 ms$^{-1}$ if not modeled at all \citep{2016PASP..128j4501G}. Determining telluric induced error on RVs is the subject of a future investigation, but regions of large residuals are not found to be correlated with regions of high telluric absorption. We find that order 14 ($m=225$) is an outlier for all three targets, and suggests that the gas cell spectrum or telluric template is in error for this order.

\begin{figure}
    \center
    \includegraphics[width=0.8\textwidth]{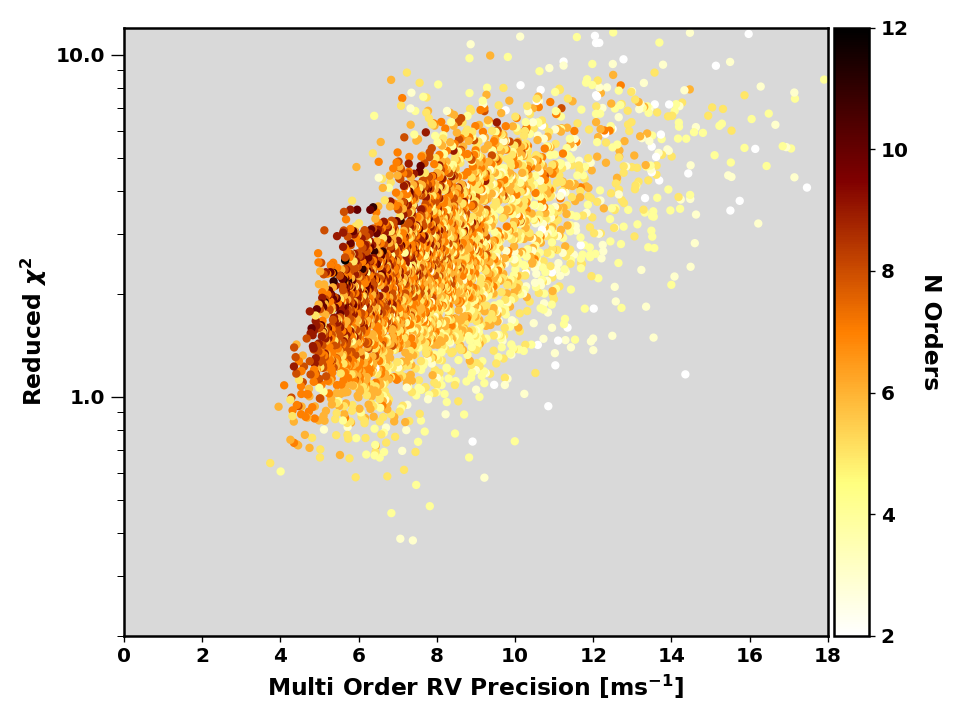}
    \caption{The corresponding values of $\chi_{red}^{2}$ from all possible combinations of multi-order RVs. Points are colored according to the number of orders used for that combination, showing the expected improvement in RV precision by using increasing numbers of orders.}
    \label{fig:rv_prec2}
\end{figure}

\pagebreak
For our optimized multi-order RVs, we also compute the reduced chi-squared statistic given by:

\begin{gather}
    \chi_{red}^{2} = \frac{1}{\nu} \sum\limits_{i}^{N_{\textrm{nights}}} \bigg(\frac{RV_{i}-\overline{RV_{i}}}{\delta RV_{i}}\bigg)^{2}
\end{gather}

\noindent where $\nu=N_{\textrm{nights}}-1$ corresponds to the largest possible degrees of freedom \citep{2010arXiv1012.3754A}, $\overline{RV_{i}}$ is the average RV of all nights weighted by $1/\delta RV_{i}^{2}$, and $\delta RV_{i}$ is the uncertainty given by Eq. \ref{eq:rv6}. By looking at all possible values of $\sigma_{\textrm{RV}}$ from the powerset, we find for the high SNR Barnard's run, $\chi_{red}^{2}=1$ corresponds to approximately 4--6 ms$^{-1}$ (Fig. \ref{fig:rv_prec2}). When observing stars with unknown RVs, we do not have this freedom of picking the orders that lead to the lowest long-term $\sigma_{\textrm{RV}}$. However, when using at least 8 orders, less than 1 percent of $\chi_{red}^{2}$ are less than 1. So, we can be confident in obtaining long-term multi-order precision of 5--7 ms$^{-1}$, so long as we are using a sufficient number of orders and if the RV content allows for it, which will be the case for most K \& M dwarfs, and late G dwarfs as well.

\section{Stellar Template Generation} \label{sec:template_2}

\begin{figure}[H]
    \center
    \includegraphics[width=0.95\textwidth]{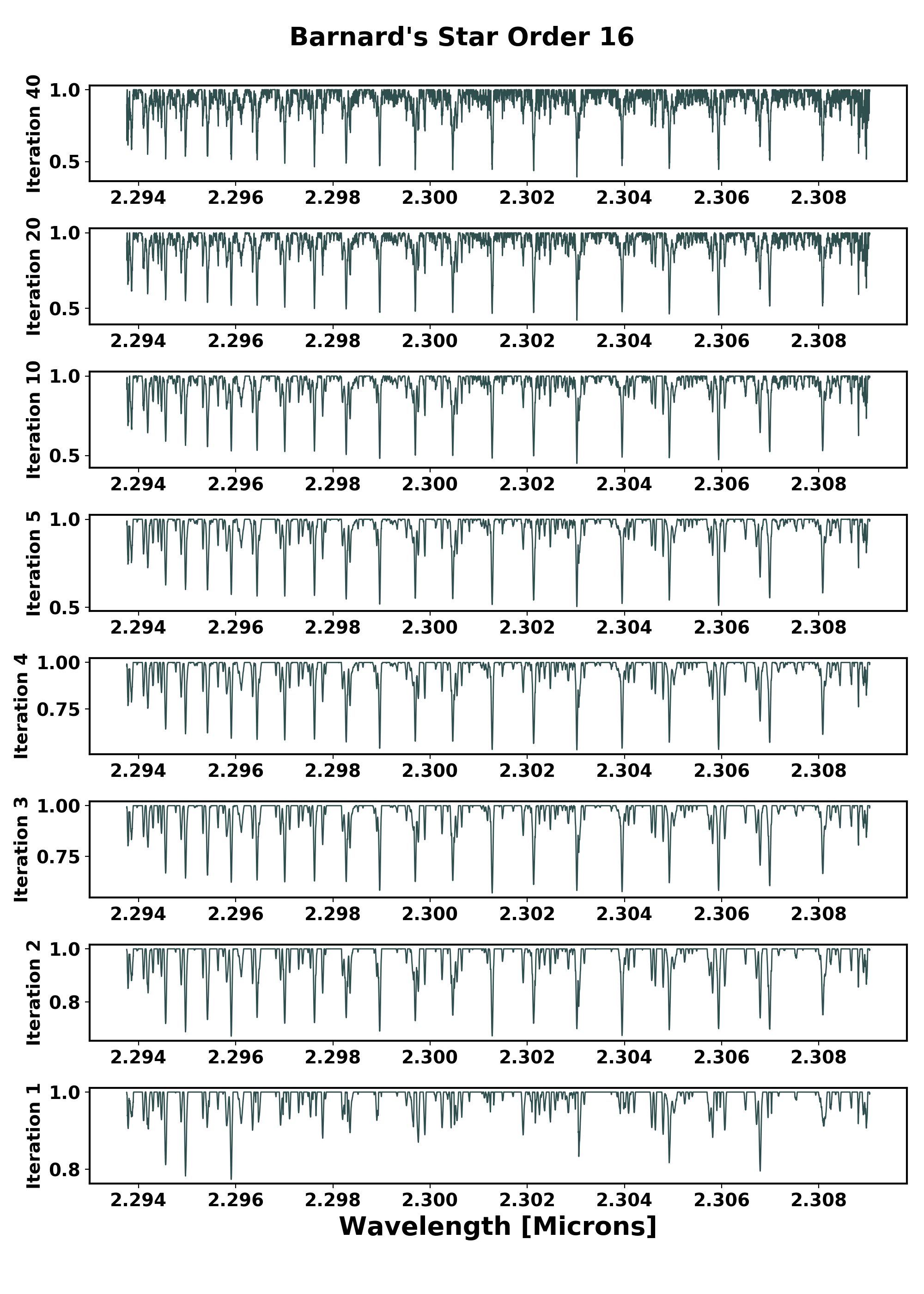}
    \caption{The generation of the stellar template for Barnard's Star for order 16 ($m=227$) for the high SNR run. Stellar features continue to get added to the template through early iterations, but a noisy continuum develops at later iterations, although RVs continue to improve.}
    \label{fig:template_gj699_all}
\end{figure}

\begin{figure}[H]
    \center
    \includegraphics[width=0.95\textwidth]{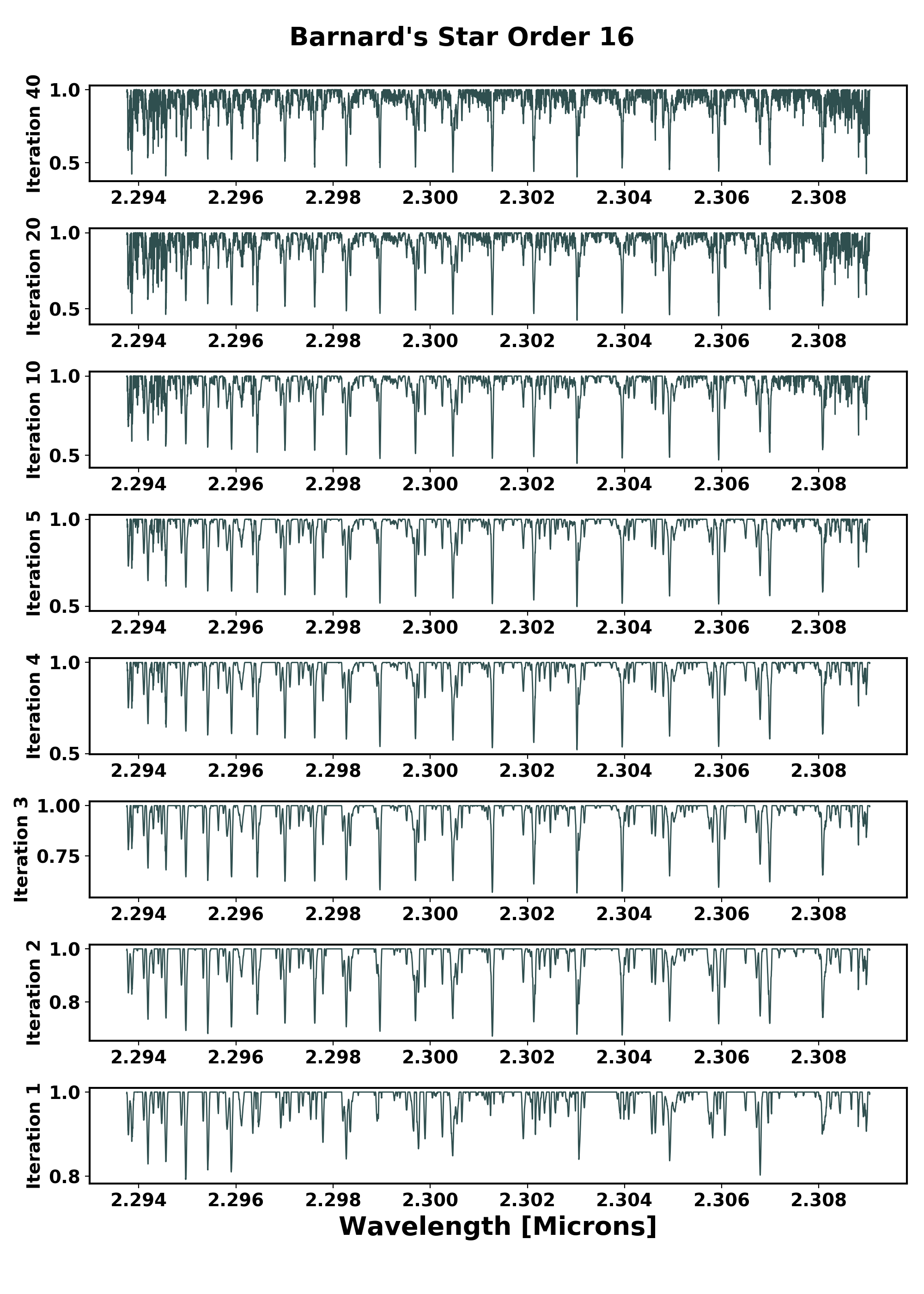}
    \caption{Same as Fig. \ref{fig:template_gj699_all}, but using the full data set. The noisy continuum that develops at later iterations is worse at the edges compared to the high SNR run, because relatively lower SNR data is being used to generate the template, even though they are down-weighted.}
    \label{fig:template_gj699_highsnr}
\end{figure}

\begin{figure}[H]
    \center
    \includegraphics[width=0.95\textwidth]{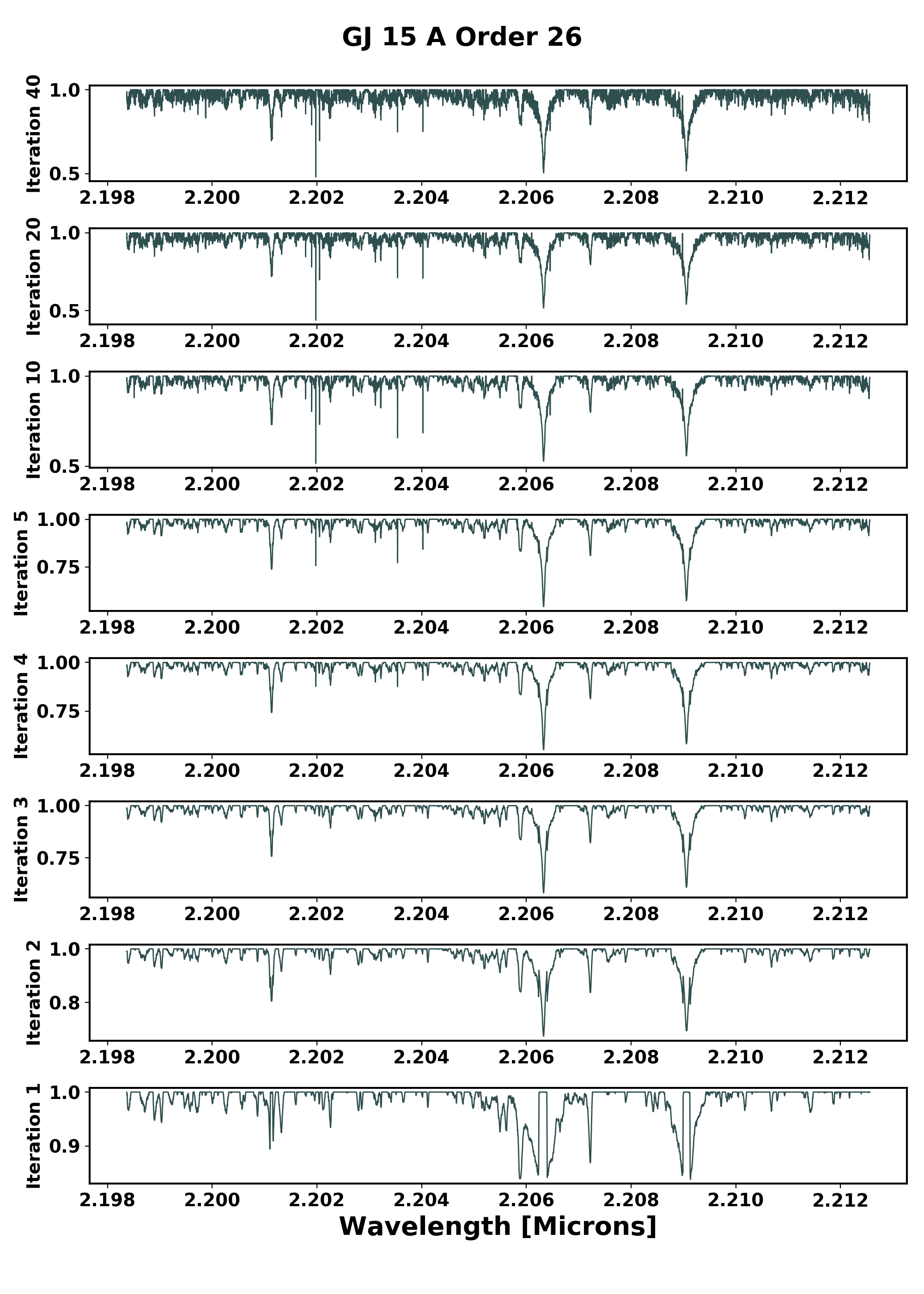}
    \caption{The generation of the stellar template for GJ 15 A for order 26 (m=237). The stellar RV information is less shortward the CO bandhead ($<2.29\ \micron$), but there are still broad lines from other molecules that can provide nightly RV precisions of 10--20 ms$^{-1}$ (see Fig. \ref{fig:photon_gj15a}). Sharp lines like those found at $2.202\ \micron$ are bad pixels.}
    \label{fig:template_gj15a}
\end{figure}

\begin{figure}[H]
    \center
    \includegraphics[width=0.95\textwidth]{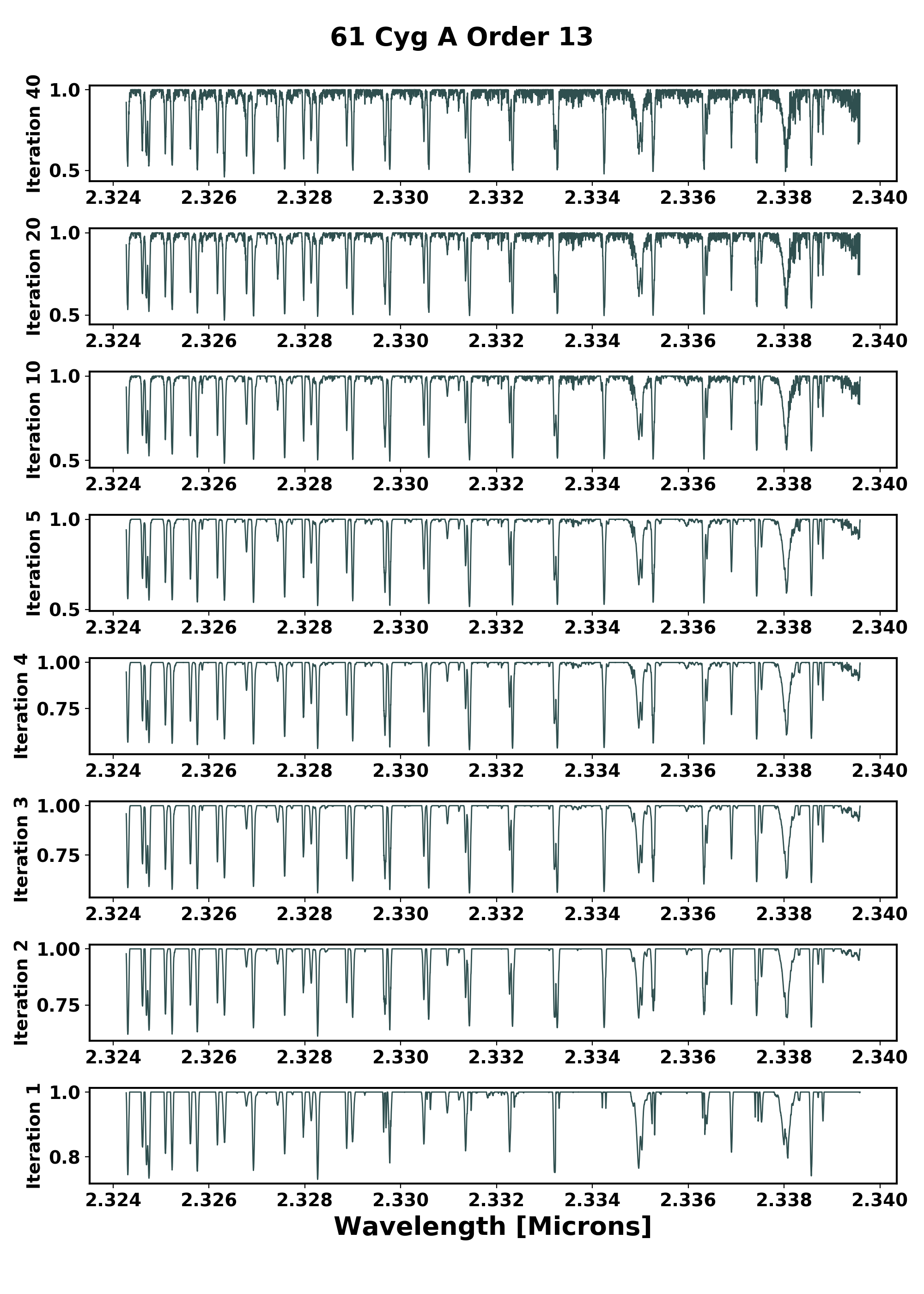}
    \caption{The generation of the stellar template for 61 Cyg A for order 16. K dwarfs also exhibit a strong CO bandhead past $2.29\ \micron$.}
    \label{fig:template_61cyga}
\end{figure}

For each star and each order, a high resolution (8 times the data) deconvolved stellar template is obtained. For all orders, after a large number of iterations, randomly coherent noise eventually begins to accumulate in the stellar template, particularly for values near the continuum where the RV content is less, and especially near the edges where the SNR is relatively low. Additionally, the empirically derived template wavelength grid is still Doppler shifted by the unknown absolute RV of the star relative to the Solar system barycenter. This can be estimated by cross-correlating our empirically-derived template to a synthetic template. Examples of retrieved stellar templates are shown in Figs. \ref{fig:template_gj699_all}-\ref{fig:template_61cyga}.

In our approach to extract heliocentric RVs, all spectra are compared to a common empirically-derived stellar template, and therefore we must be concerned whether or not our RV errors are caused by inherent astrophysical RV variability or internal errors in the stellar template spectrum itself. We do not quantitatively investigate the RV precision as a function of the number of epochs used in the analysis to identify a minimum number of epochs required for adequate barycenter velocity sampling in the stellar template derivation. Instead, in order to test how robust our stellar template retrieval is, we run two seasonal data sets of Barnard's Star and compare the generated templates. We choose only the high SNR data set taken in October - November 2016 and the following high SNR data set taken from April - July 2017. We do this for order 13 which is high in stellar and gas RV content. 

\begin{figure}[H]
    \center
    \includegraphics[width=0.85\textwidth]{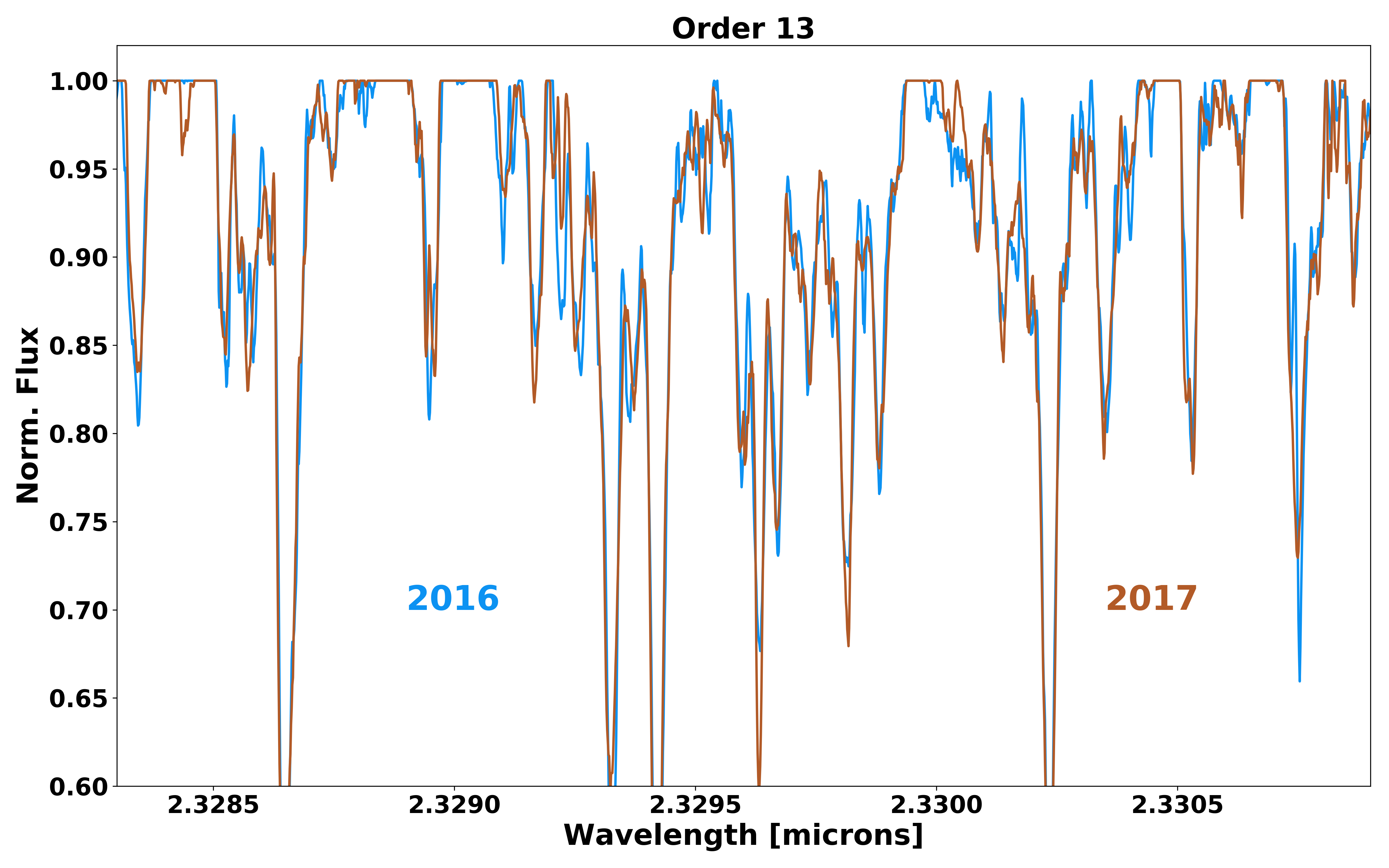}
    \caption{Two separately retrieved stellar templates for Barnard's Star (fall 2016, spring-summer 2017).}
    \label{fig:template_compare_2}
\end{figure}

\begin{figure}[H]
    \center
    \includegraphics[width=0.5\textwidth]{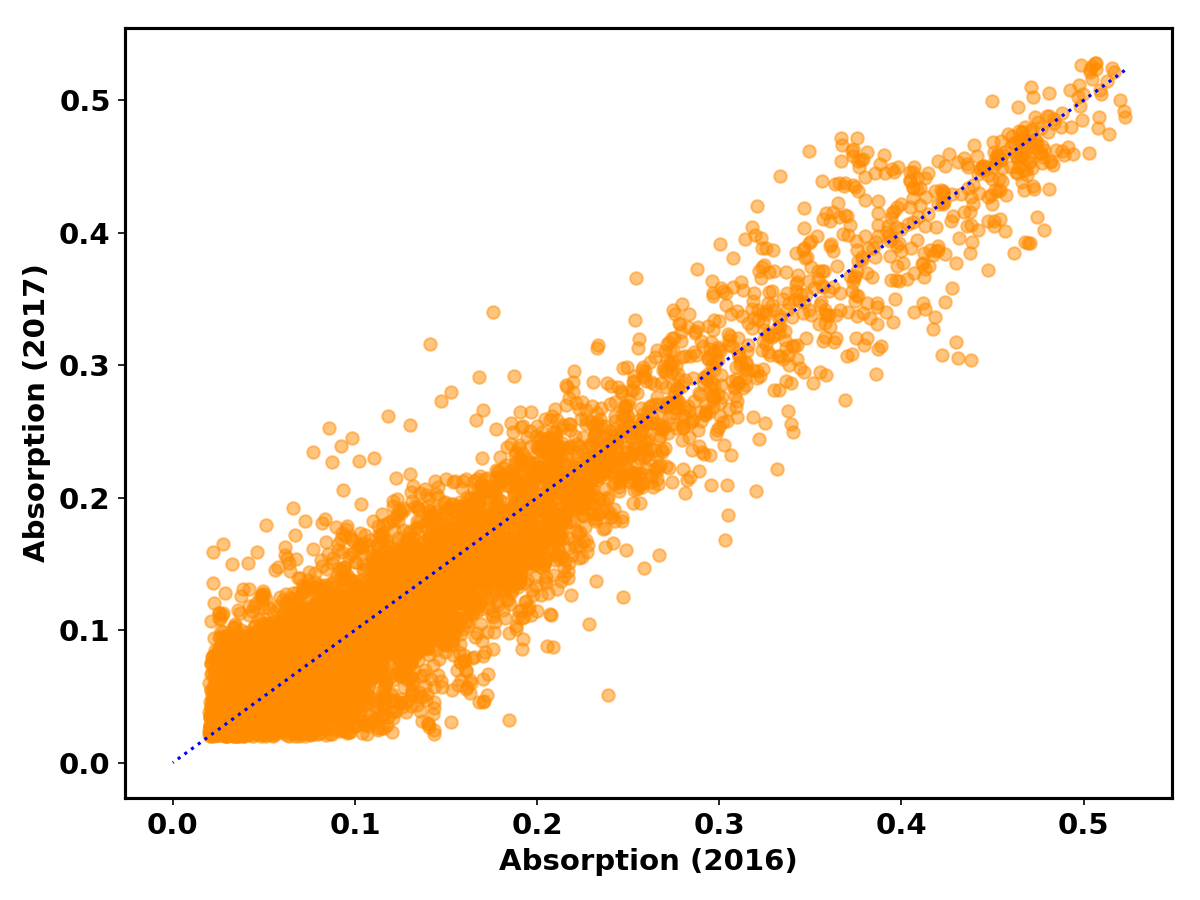}
    \caption{An absorption vs. absorption plot for two separately retrieved stellar templates for Barnard's Star (fall 2016, spring-summer 2017). Only features deeper than 2\% in both templates are shown. A one-to-one line corresponding to perfect agreement between the separately retrieved templates is shown in blue. The disagreement is slightly larger for values with less absorption (near the continuum).}
    \label{fig:template_compare_1}
\end{figure}

Qualitatively, we find that using fewer spectra in the analysis allow bad pixels to increasingly affect the template (Fig. \ref{fig:template_compare_2}). We find that it is critical in our analysis to flag bad pixels in the data or in the residuals on the data wavelength grid, because a single bad pixel gets spread out into many on the template grid due to the high resolution of the model. Additionally, we find that deep lines with high RV content are fairly consistent between the two seasons and the mismatches are typically found for values near the continuum (Fig. \ref{fig:template_compare_1}).

\section{Model Parameters} \label{sec:model_params}

\subsection{Multi-Order Consistencies}

\begin{figure}[H]
    \center
    \includegraphics[width=0.45\textwidth]{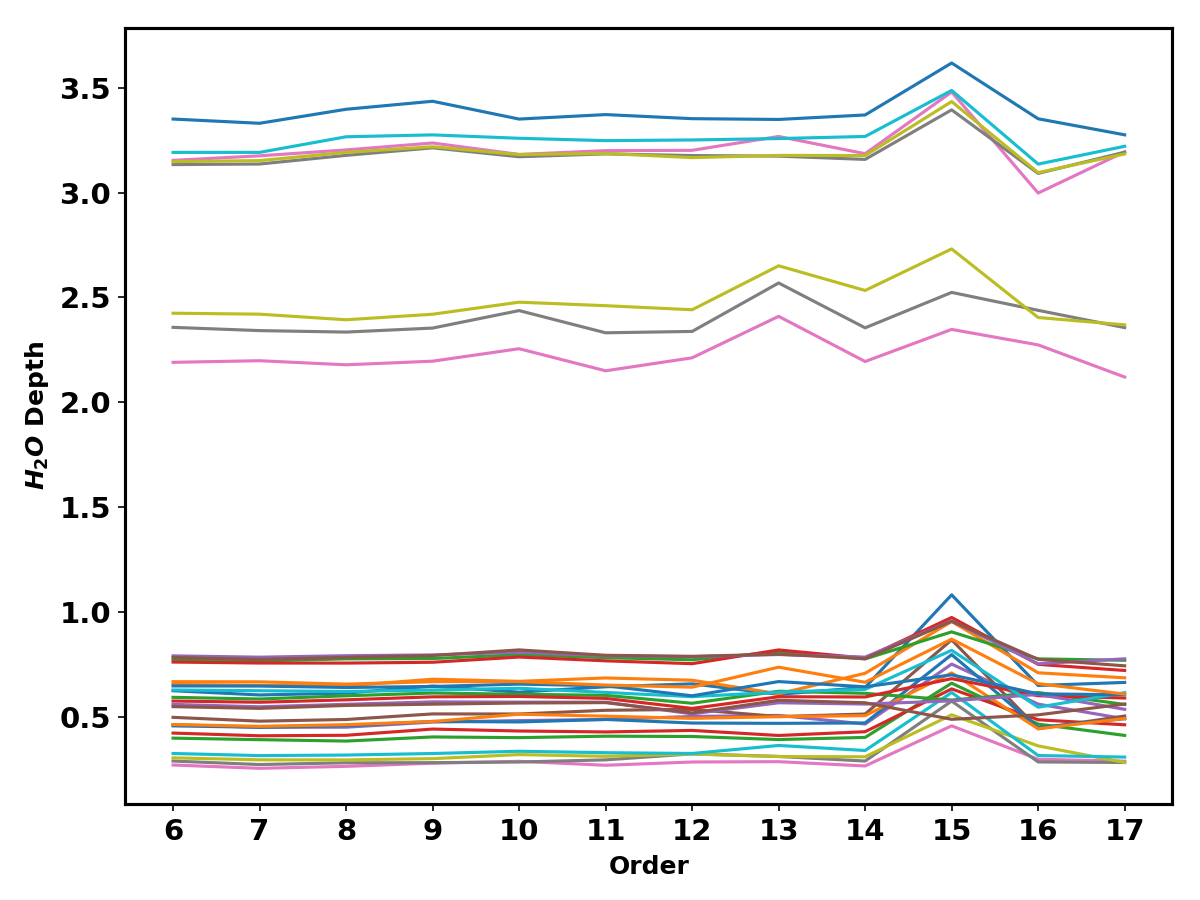}
    \includegraphics[width=0.45\textwidth]{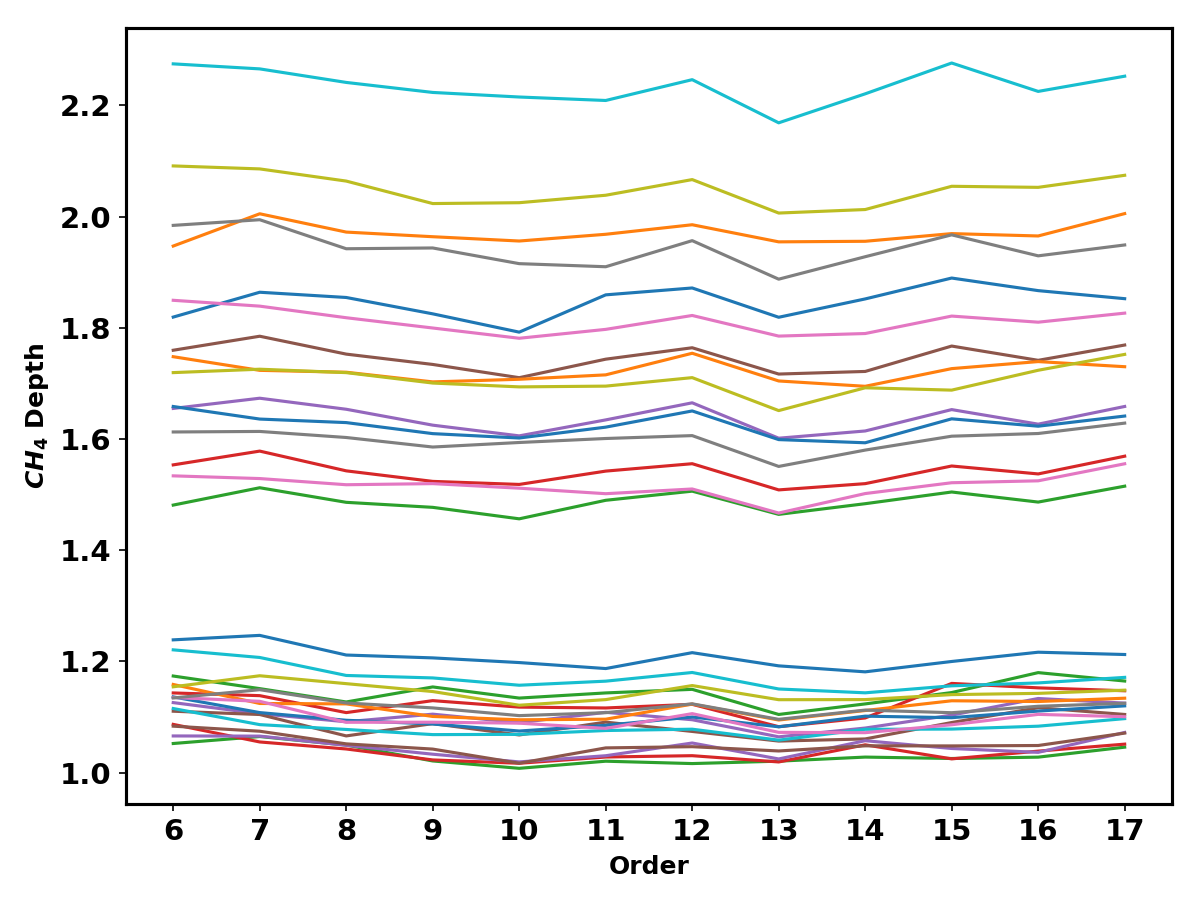}
    \caption{\textit{Left}: The water optical depth for multiple orders from the high SNR run. \textit{Right}: Same, but for telluric methane. Only every other observation is plotted. The water and methane depths are also unique supporting our hypothesis of variable atmospheric content.}
    \label{fig:depths}
\end{figure}

The same set of forward model parameters are used for all orders (see Table \ref{tab:pars}). We forward model all orders independently - e.g. the parameters derived from one order are not used to constrain the parameters for other neighboring orders, when in principle some parameters should be identical across orders or related by simple analytic approximations. Thus, we can investigate parameters that are consistent across orders as a sanity check on our analysis. For this Section, we use the high SNR Barnard's Star run results. We find the telluric water and methane optical depths are consistent across orders (Fig. \ref{fig:depths}). Order 15 tends to require a systematically higher water optical depth compared to the other orders, indicating error in the synthetic telluric template at that wavelength.

\begin{figure}[H]
    \center
    \includegraphics[width=0.45\textwidth]{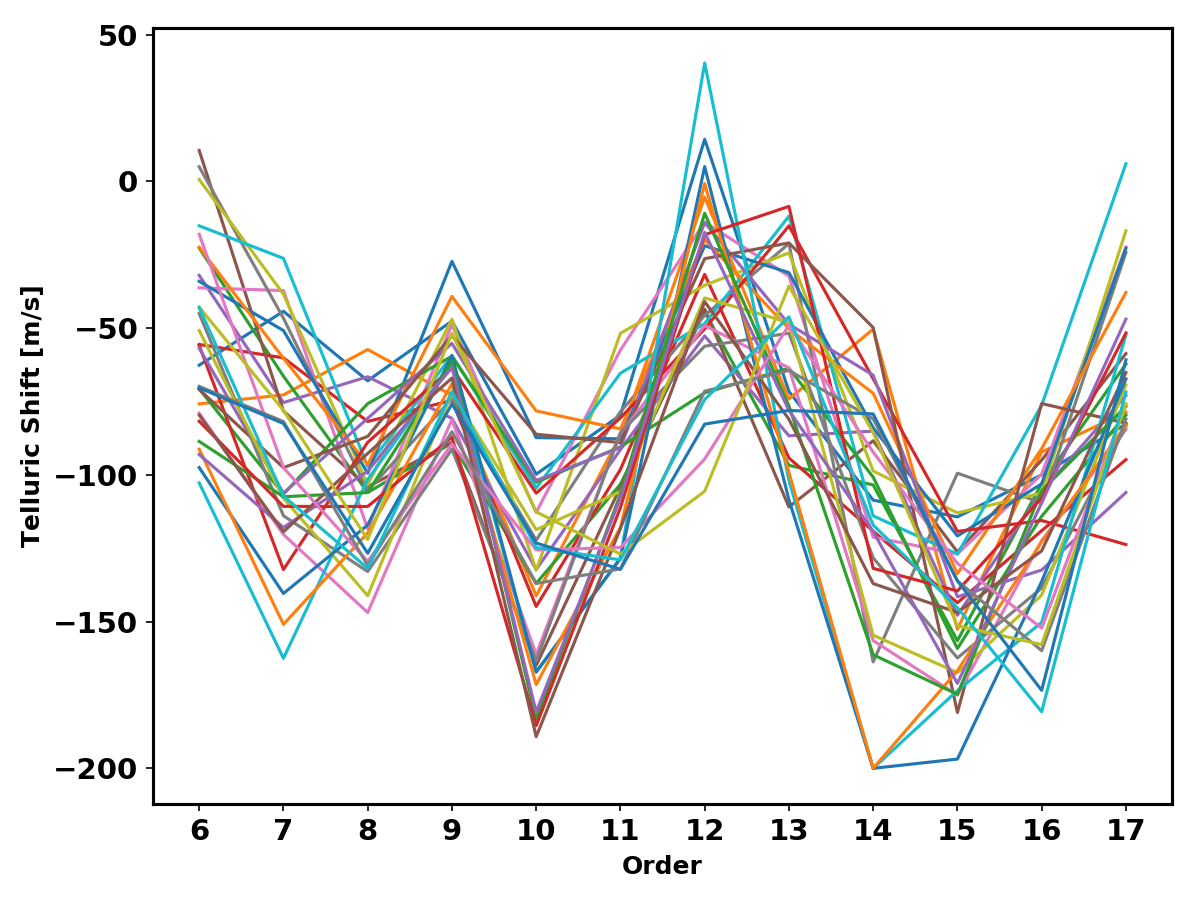}
    \includegraphics[width=0.45\textwidth]{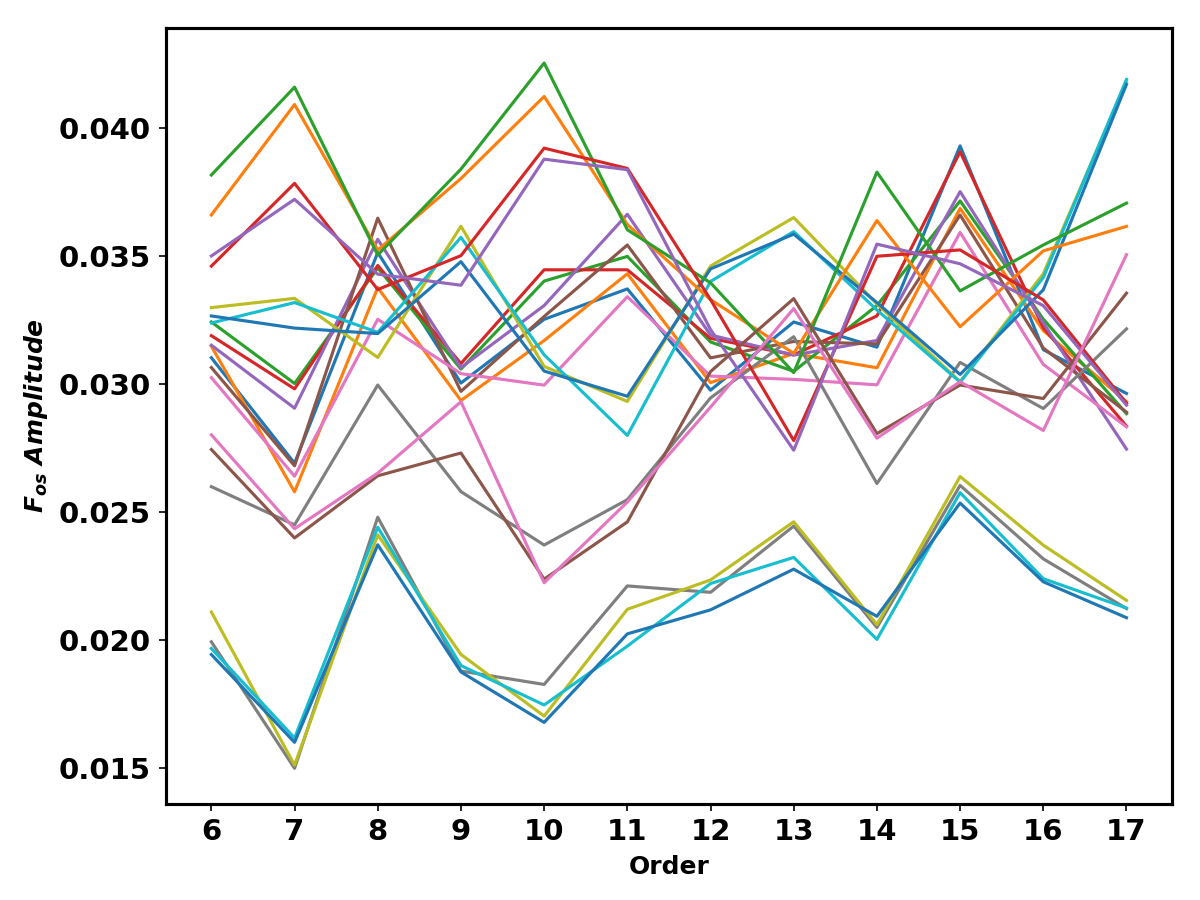}
    \caption{\textit{Left}: Same as Fig. \ref{fig:depths} but for the telluric shift. Nights within an order show less scatter in the fit telluric shifts than the scatter between orders. This implies that there is room for improvement in our telluric model components. \textit{Right}: Same, but for the OS filter fringing amplitude. There is little inter-order agreement, but intra-night stability still allows for nights to cluster together.}
    \label{fig:multi_2}
\end{figure}

The fringing parameters are not well-behaved across orders, but show clear nightly consistency (Fig. \ref{fig:multi_2}). The telluric shift shows a large scatter order to order relative to our RV precision, and is relatively more consistent intra-order across all nights (Fig. \ref{fig:multi_2}). This could be used in future work to refine our telluric template.

The quadratic wavelength solution and AR fringing component (Eq. \ref{eq:fringing_ar}) wavelength set points will influence multi-order consistencies, but we still find the former ($\lambda_{i}$) are fairly consistent across orders, especially $\lambda_{2}$ (not shown).

\subsection{Correlations} \label{sec:corrs}

\begin{figure}[H]
    \center
    \includegraphics[width=0.95\textwidth]{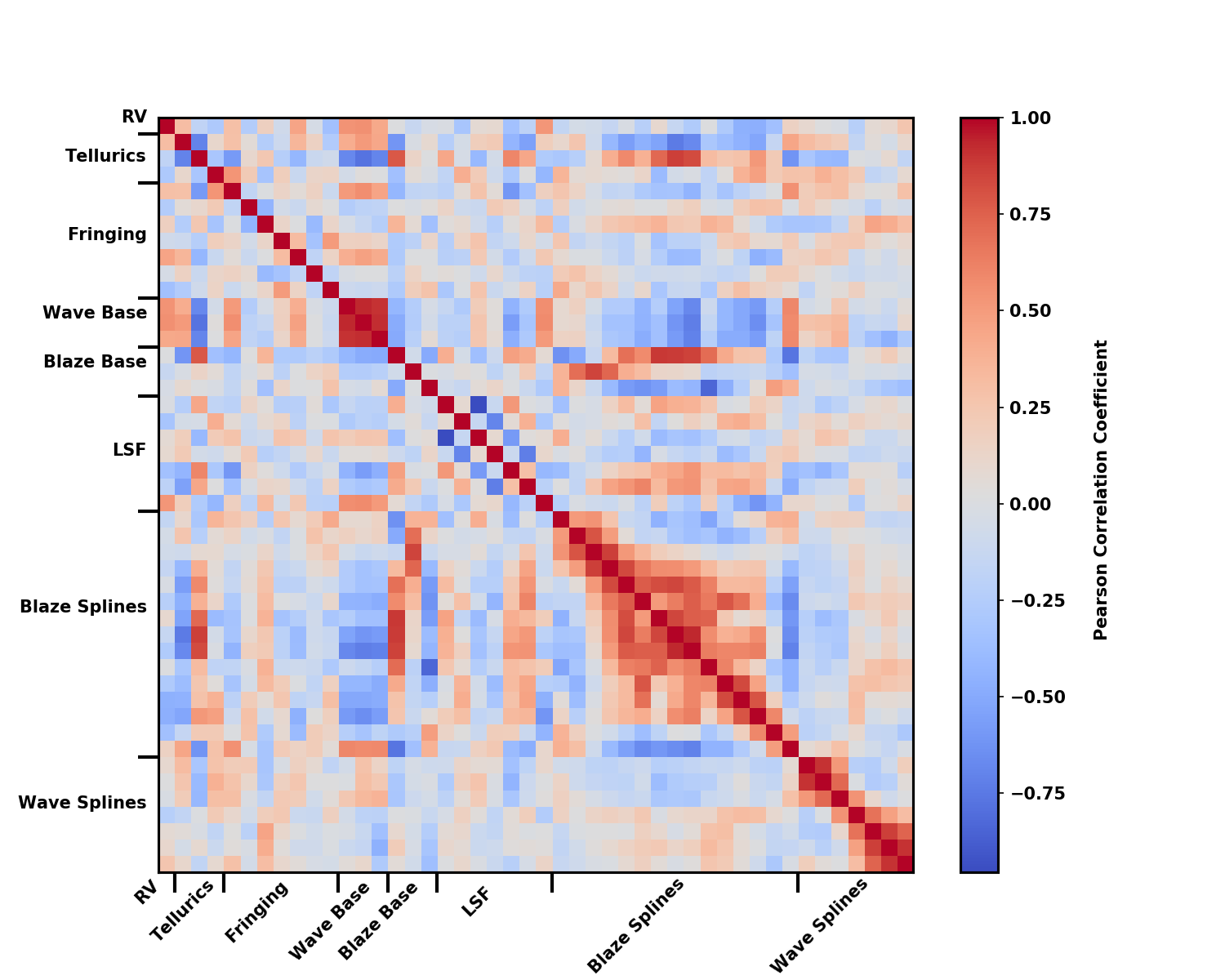}
    \caption{A correlation plot for all forward model parameters from order 8 ($\textrm{CO}_{2}$ and $\textrm{N}_{2}\textrm{O}$ are not considered here). Parameters are in the same order as given in Table \ref{tab:pars}. Neighboring spline points for the blaze and wavelength solution are heavily correlated. Other orders exhibit qualitatively similar correlation plots.}
    \label{fig:pcc_pars1}
\end{figure}

\begin{figure}[H]
    \center
    \includegraphics[width=0.95\textwidth]{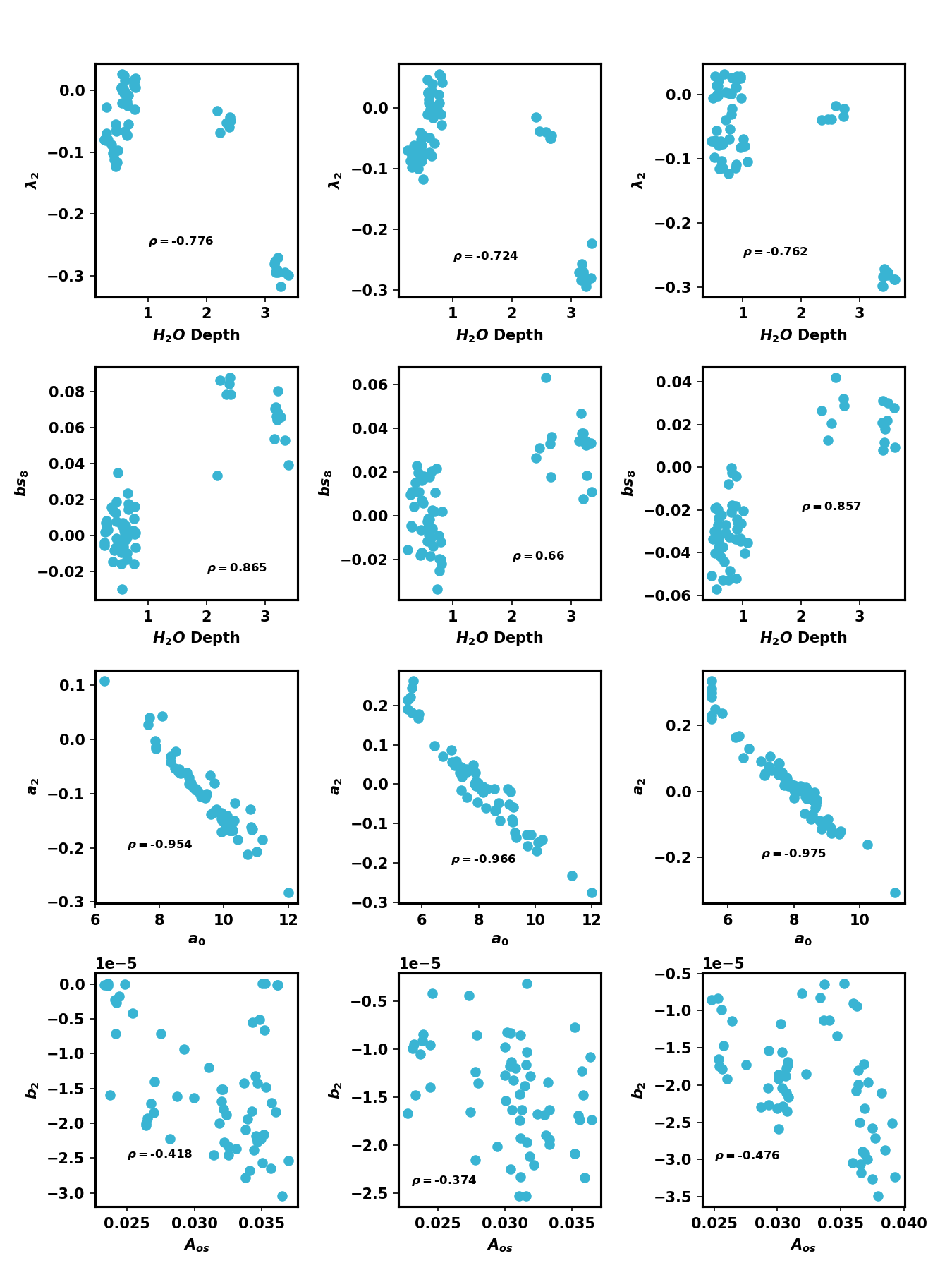}
    \caption{A series of correlation plots for orders 8, 13, \& 15 (from left to right) highlighting strongly correlated parameters. Parameter symbols are defined in Table \ref{tab:pars}.}
    \label{fig:pcc_pars2}
\end{figure}

\begin{figure}[H]
    \center
    \includegraphics[width=0.6\textwidth]{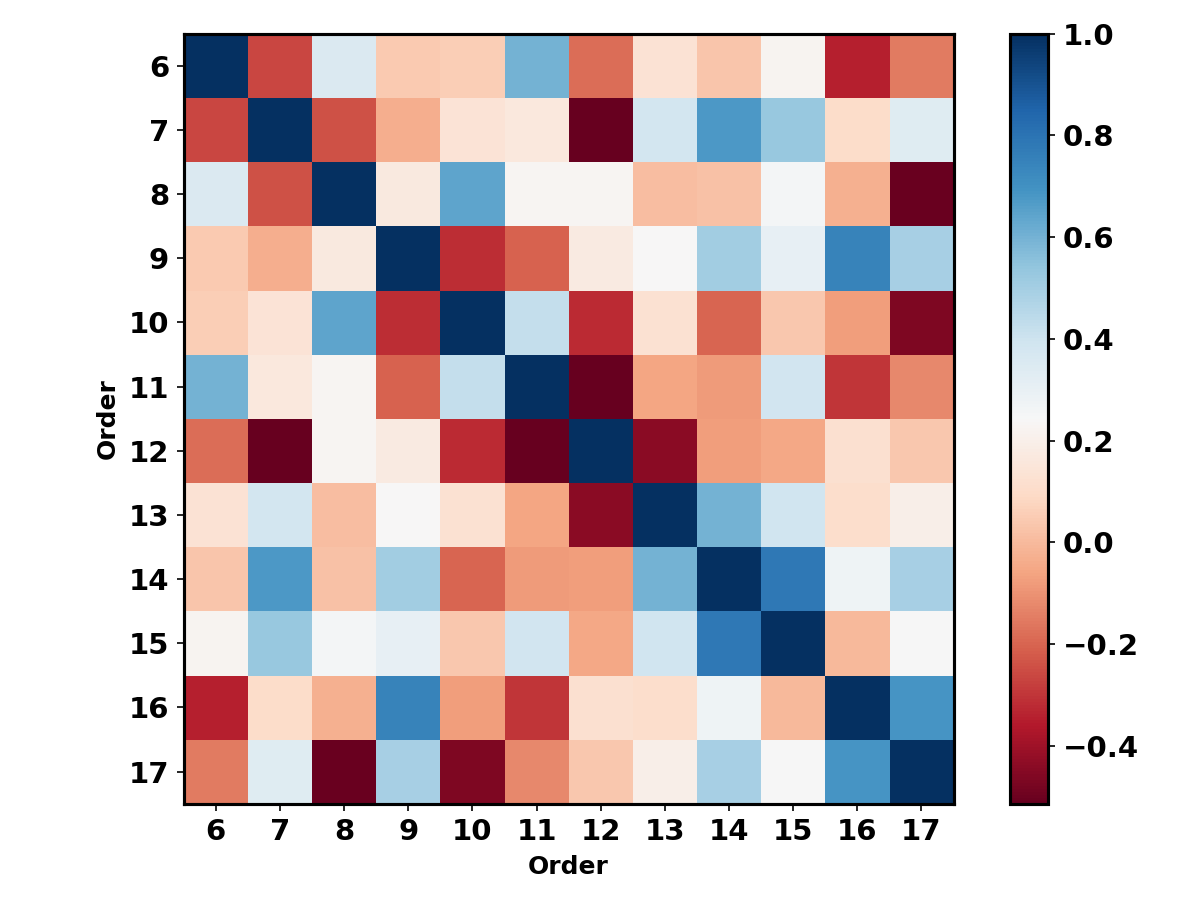}
    \caption{A correlation plot for the single-order nightly RVs from the Barnard's star high SNR run. Each block is colored according to the value of the Pearson Correlation Coefficient. Neighboring orders (near the diagonal) are typically more correlated than orders further away, perhaps because they overlap in wavelength.}
    \label{fig:rv_pcc}
\end{figure}

Our choice of forward model implementation for the work presented here uses 48 parameters. We investigate parameters that are highly correlated with RVs or other parameters. Some parameters are expected to be correlated without concern. The quadratic wavelength solution Lagrange points are not orthogonal parameters, and are indeed strongly correlated with one another. We also find neighboring spline points to be correlated for the blaze and wavelength corrections and are not further discussed as they are also not orthogonal. Other correlated parameters are found through computing the Pearson linear correlation coefficient $\rho$ defined in \citet{1895RSPS...58..240P} for all pairs of parameters and for each order. Significant linear correlation or anti-correlation corresponds to $\rho \rightarrow \pm1$. We calculate $\rho$ for all pairs of parameters, including $RV_{\star}$ for each spectrum. We flag all pairs of parameters such that $\mid\rho\mid>0.5$ for all orders (6-17) using the high SNR Barnard's Star results. We find that the \textit{LSF} width $a_{0}$ is degenerate with even \textit{LSF} Hermite terms $a_{j}$ (odd terms are usually zero), despite being an orthogonal basis. We also find that the water optical depth is correlated with several parameters, but only consistently with the base (quadratic) wavelength solution points $\lambda_{i}$ across multiple orders (Figs. \ref{fig:pcc_pars1}, \ref{fig:pcc_pars2}). Most of the water depth correlations are due to two nights with relatively high water vapor content/airmass. Otherwise, we find no other parameters with $\mid\rho\mid>0.5$ consistently across orders. A full correlation plot is shown in Fig. \ref{fig:pcc_pars1}, and several examples of correlated parameters are shown in Fig. \ref{fig:pcc_pars2}. We also check for correlation in the single-order nightly (co-added) RVs. We find neighboring orders are moderately correlated or anti-correlated, which is expected with the large spectral region of overlap, but find no other strong correlation (Fig. \ref{fig:rv_pcc}).

\section{Discussion} \label{sec:discussion}

With a complex forward model of 48 parameters, we investigate the benefits and drawbacks of our choice of parameter space. Proper analysis of our forward model requires a thorough analysis for each component, but here we only focus on the wavelength solution and \textit{LSF}, as we identify they significantly impact the derivation of our RVs. Without a robust \textit{LSF} and wavelength solution, the model breaks down and remaining parameters will fail to converge.  We conclude the discussion with a comparison to other NIR RV spectrographs, methodologies, and prospects for planet confirmation.

% Section 5.4.1
\subsection{Wavelength Solution} \label{sec:wave_sol}

\begin{figure}[H]
    \center
    \includegraphics[width=0.95\textwidth]{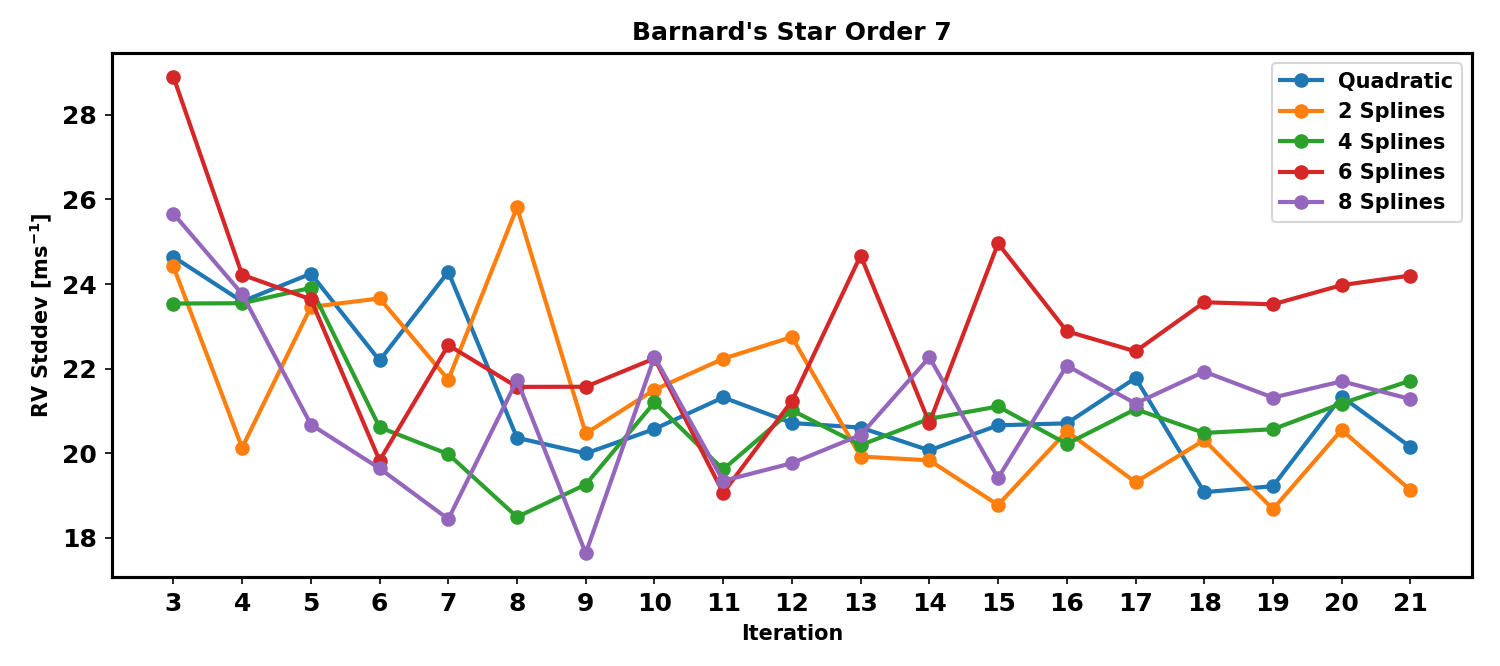}
    \includegraphics[width=0.95\textwidth]{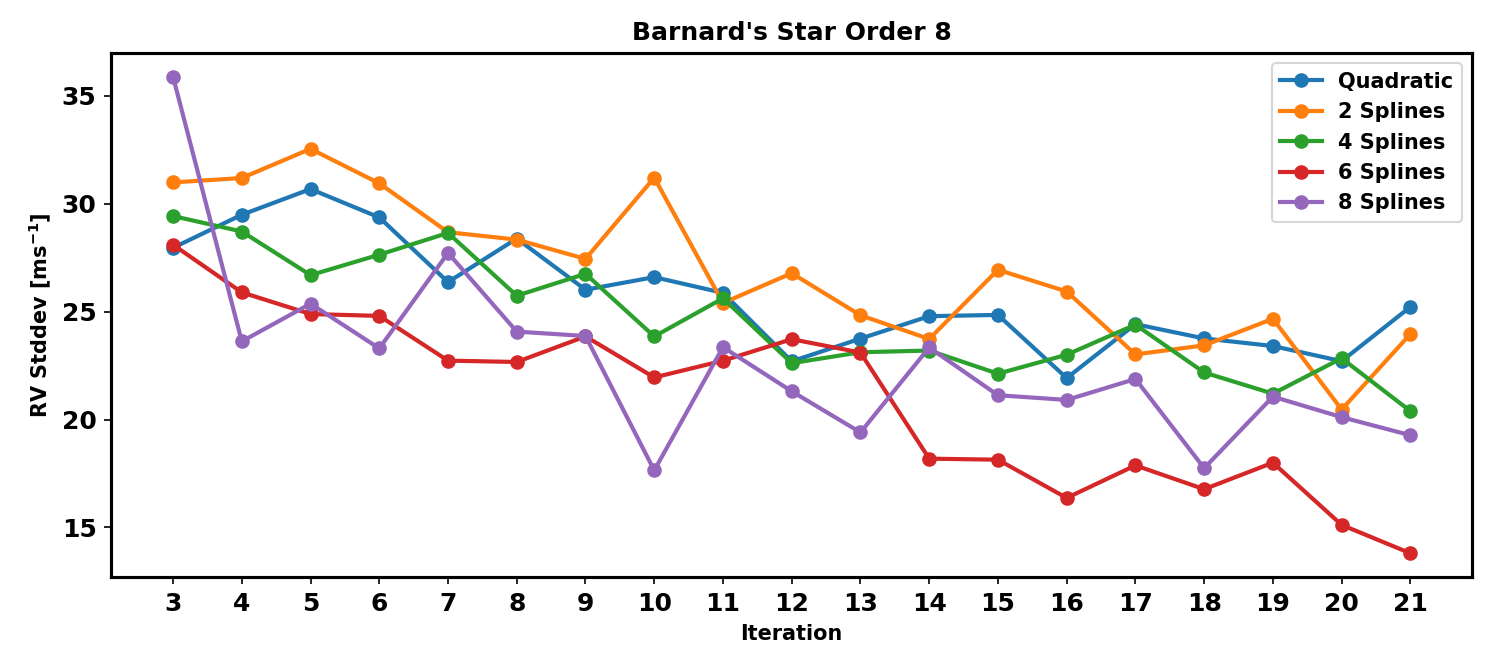}
    \includegraphics[width=0.95\textwidth]{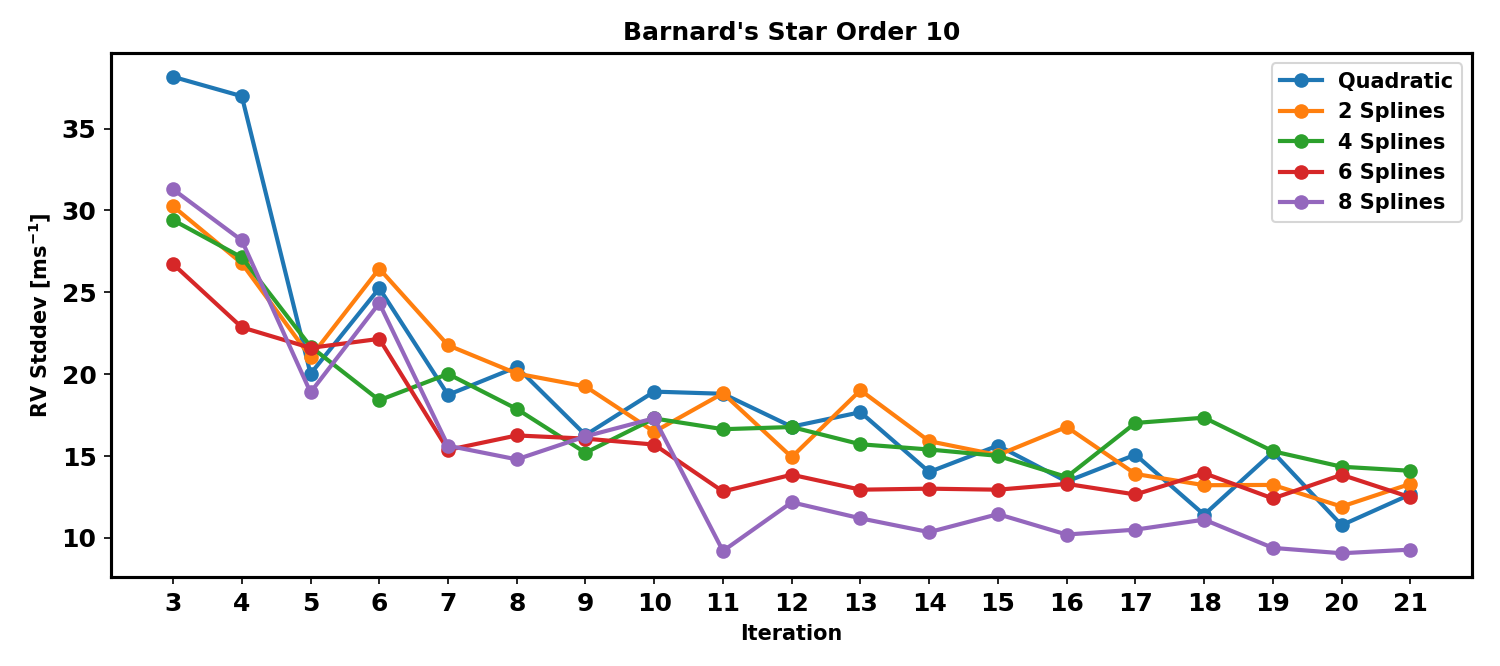}
    \caption{The obtained RV precision for Barnard's Star using different spline implementations for the wavelength solution for orders 7, 8, \& 10 using the high SNR data set. Most orders show improvement when using splines, but the number of splines is inconsistent and can in some cases make RV precisions larger.}
    \label{fig:wave_splines_1}
\end{figure}

\begin{figure}[H]
    \center
    \includegraphics[width=0.95\textwidth]{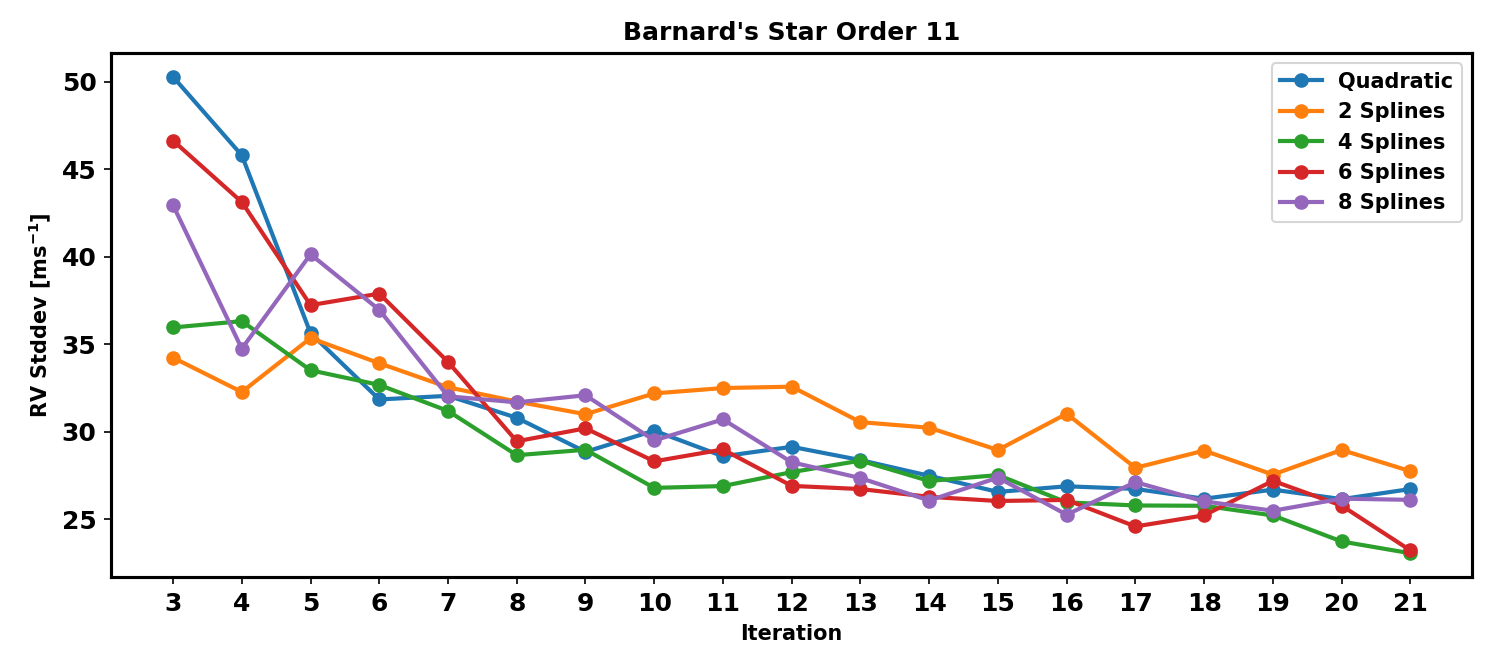}
    \includegraphics[width=0.95\textwidth]{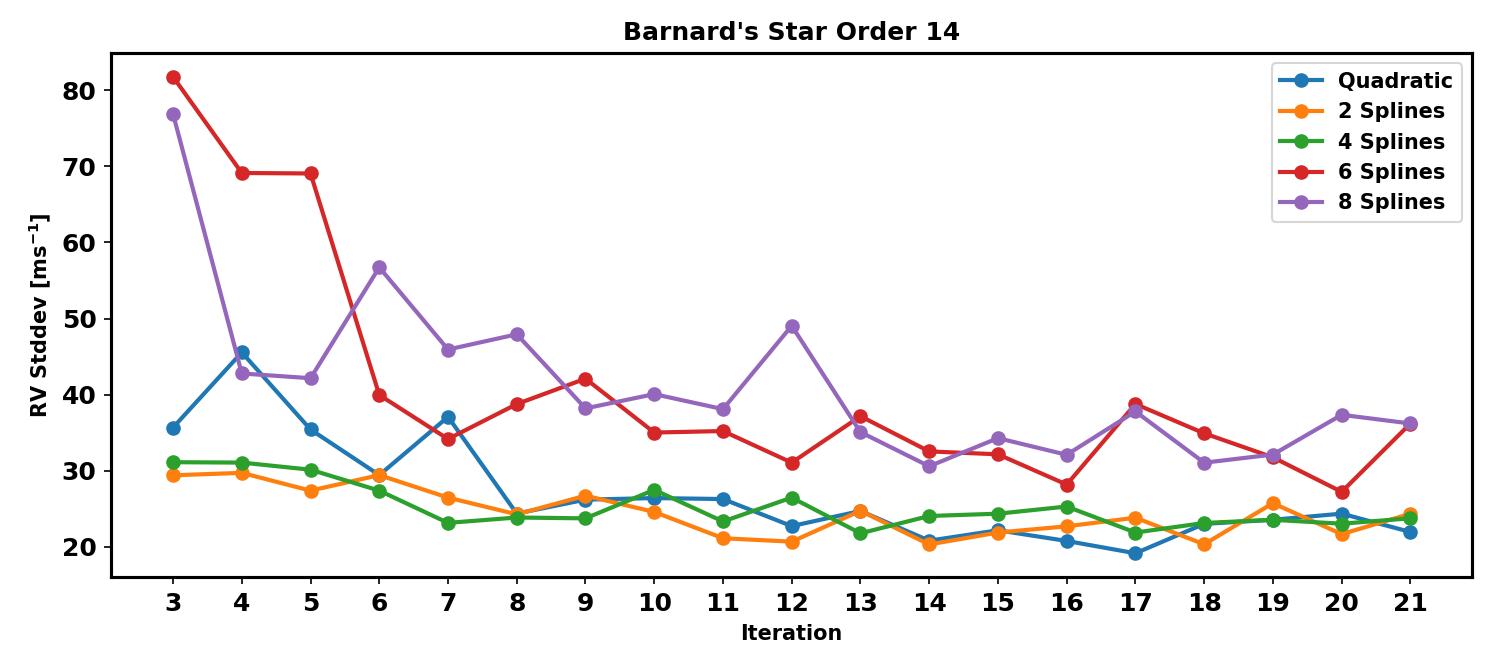}
    \includegraphics[width=0.95\textwidth]{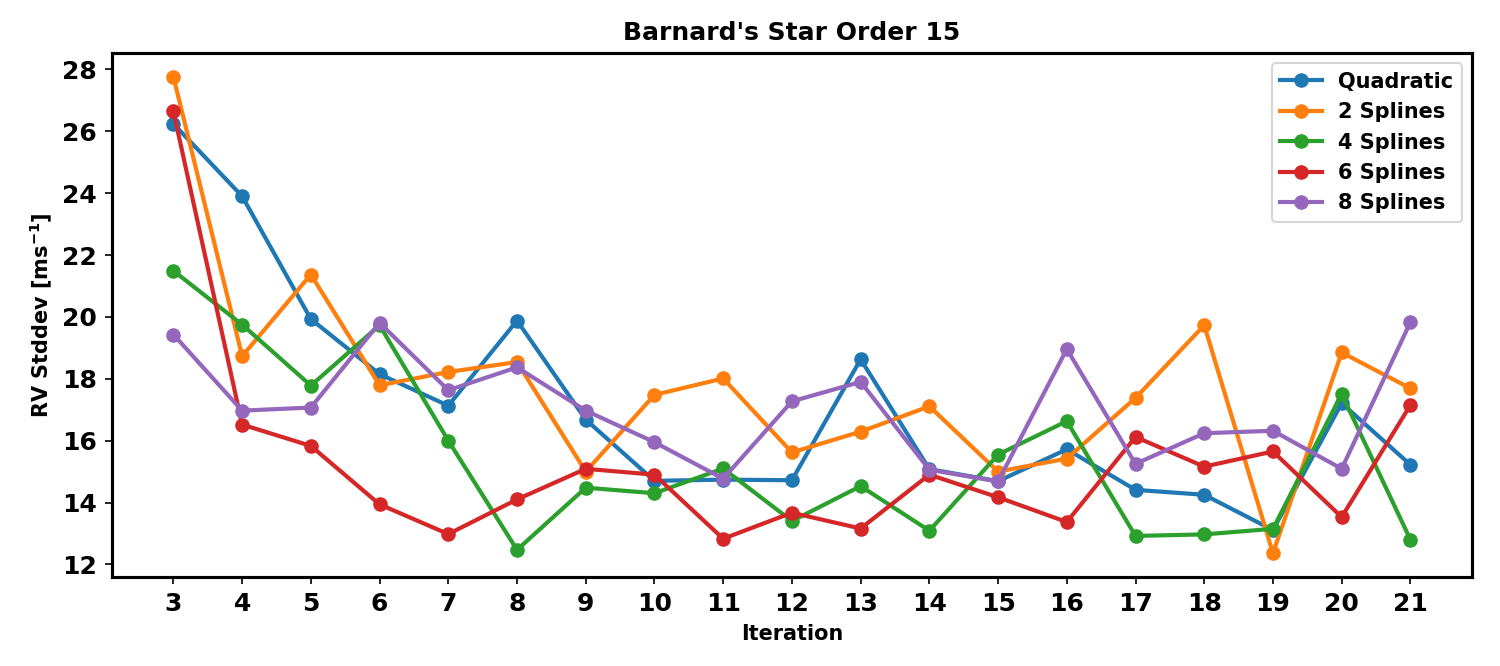}
    \caption{Same as Fig. \ref{fig:wave_splines_1} but for orders 11, 14, \& 15.}
    \label{fig:wave_splines_2}
\end{figure}

\begin{figure}[H]
    \center
    \includegraphics[width=0.95\textwidth]{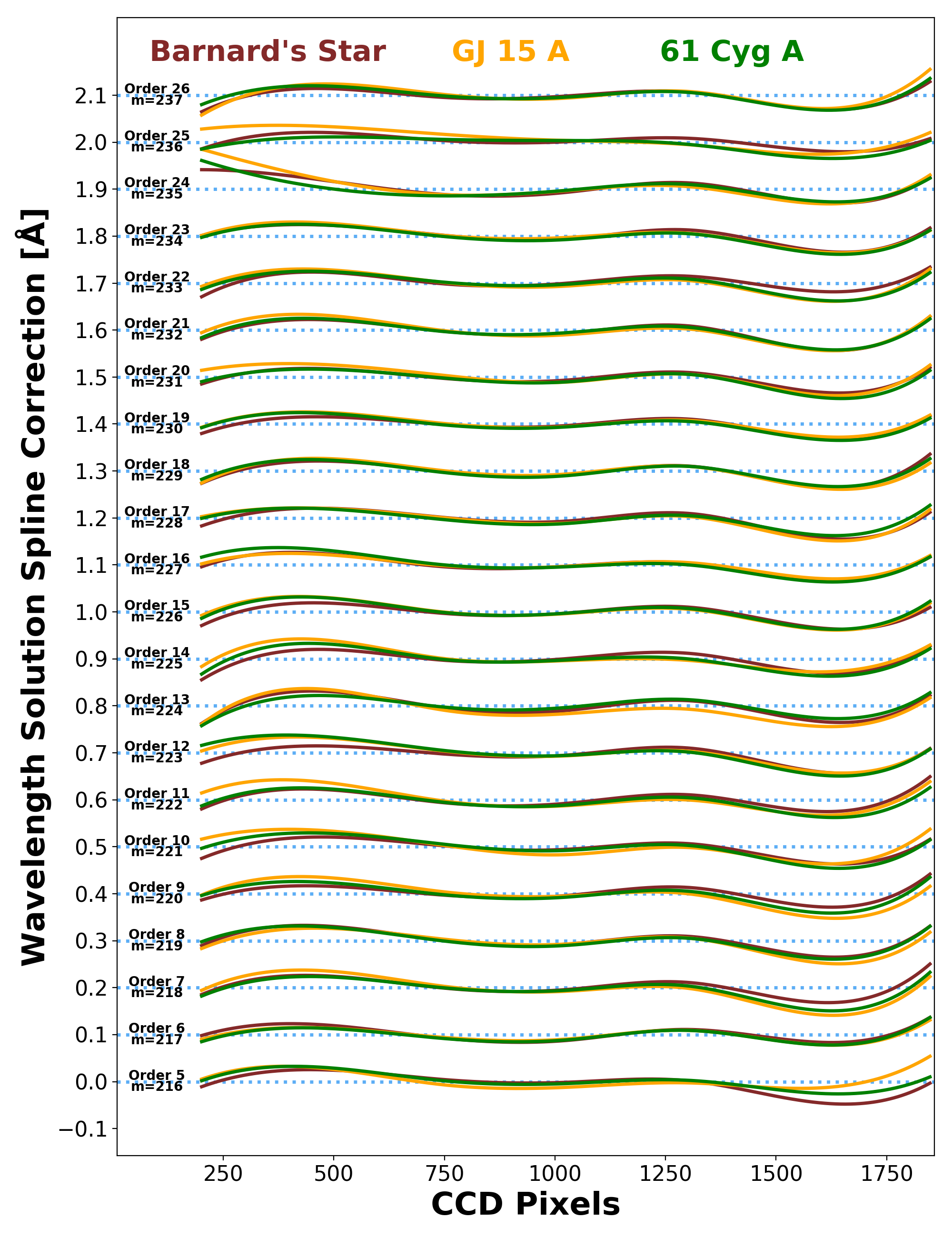}
    \caption{The average spline correction that gets added to a quadratic in the wavelength solution for all three runs (using the full data set for Barnard's Star). The average correction is approximately the same for all orders and targets, strengthening the case for including the correction.}
    \label{fig:wave_spline_maps}
\end{figure}

We expect the wavelength solution to be well-modeled by a quadratic, but considering both the non-ideal stability conditions for iSHELL and extremely fine RV measurements being performed, there are good motives to try a wavelength solution that allows for local perturbations. To test this, we run several orders of Barnard's Star from the high SNR data set with considerable stellar RV content using a various number of splines for the wavelength solution. For orders near the middle of the detector, the addition of splines can yield worse RV precision, but in most cases the RVs are improved (Figs. \ref{fig:wave_splines_1}-\ref{fig:wave_splines_2}). Unfortunately, there is little agreement on the number of splines. However, we find the average spline corrections for all targets and orders are similar (order-to-order consistency), with most deviations occurring at the end points, justifying the spline correction (Fig. \ref{fig:wave_spline_maps}).

\begin{figure}[H]
    \center
    \includegraphics[width=0.75\textwidth]{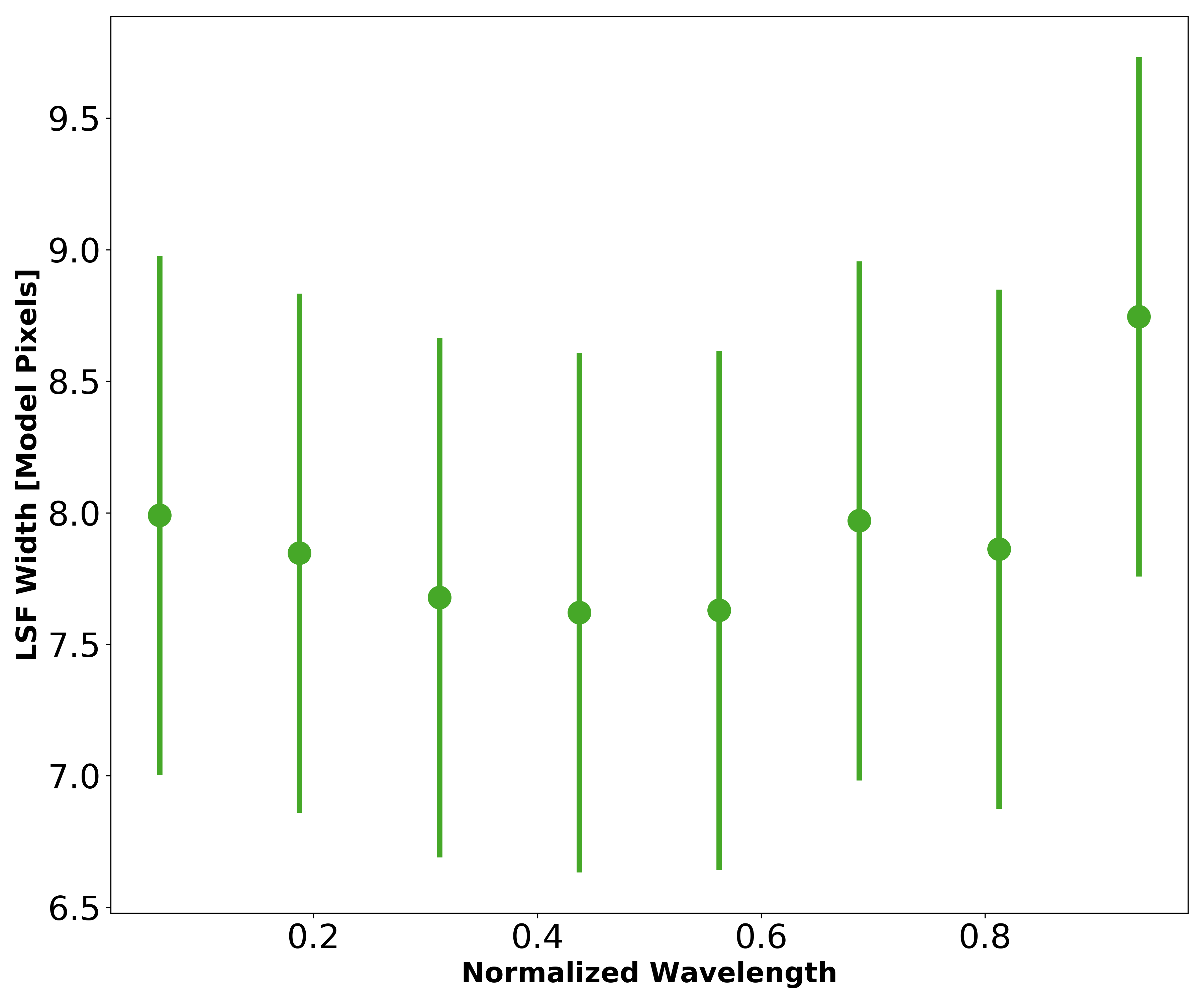}
    \caption{The \textit{LSF} width across the detector for the cropped portion of the data (spectral pixels 200-1848). This data set is an average of the high SNR Barnard's Star observations (61 spectra) and 6 echelle orders high in gas cell RV content (Eq. \ref{eq:fischer}). A higher order \textit{LSF} model may over-compensate for a larger width on the ends of the detector (further from the blaze angle), so only a 3 Hermite term model was used. The error bars represent a $1\sigma$ spread. The \textit{LSF} width tends to be higher on the ends (especially the red end), but is generally consistent in the middle of the detector.}
    \label{fig:lsf_width}
\end{figure}

\subsection{LSF Model} \label{sec:lsf_model}

Like \cite{2016PASP..128j4501G} with CSHELL, we assess different parameterizations of our \textit{LSF} model. With iSHELL's larger spectral grasp, we find that a varying \textit{LSF} model within the order can improve the RMS in fitting. If the \textit{LSF} truly is dynamic across a single-order, then it would be advantageous to allow for a unique \textit{LSF} model at each model pixel using spline continuity relations, similar to the wavelength solution and blaze function. Unfortunately this would be too computationally expensive having to compute over 16,000 \textit{LSF}'s for a single model. Further, there is no reason to use a finer \textit{LSF} model than a single resolution element (\texttildelow 0.03 nm, or 3 detector pixels). The downside of a binned \textit{LSF} is it drastically increases the number of model parameters and therefore runtime. Further, from the limited number of cases performed with a dynamic \textit{LSF}, we find that this degrades RV precision. A 3 Hermite term model with 8 equally sized bins across the detector (Fig. \ref{fig:lsf_width}) typically yields \texttildelow 5 ms$^{-1}$ higher single-order RV precision, while a 7 Hermite term model (8 bins) is anywhere from 0--10 ms$^{-1}$ worse on average. Since it is possible to over-fit the data, a lower RMS from a more complex LSF model does not necessarily lead to the lowest RV precision.

\pagebreak
\subsection{Other NIR Precise RV Instruments \& Methodologies}

We compare our results with other on-sky NIR precise RV spectrographs. Early instruments like NIRSPEC \citep{1995SPIE.2475..350M} on Keck were capable of 40--50 ms$^{-1}$ precisions using tellurics as a wavelength reference and were mostly limited by the smaller spectral resolution of R\texttildelow25,000 \citep{2012ApJS..203...10T}. The CRIRES \citep{2004SPIE.5492.1218K} spectrograph on the VLT obtained 5 ms$^{-1}$ long-term RVs at K-band using an ammonia gas cell for wavelength calibration, and was primarily limited by imperfect modeling of telluric lines \citep{2010ApJ...713..410B}. The Habitable Zone Planet Finder (HPF) spectrograph (Y- \& J-band) on the 10 meter Hobby-Eberly telescope has reached $<$ 3 ms$^{-1}$ long-term precisions on Barnard's Star \citep{2019arXiv190200500M}. Unlike iSHELL which uses a gas cell to serve as a common optical path wavelength reference, HPF uses a laser frequency comb providing a series of evenly spaced emission lines to serve as a wavelength reference \citep{2019arXiv190202817M}. The CARMENES instrument utilizes two spectrographs (visible and J-, Y-band) with the goal of characterizing stellar activity through analyzing the color (wavelength) dependence on RVs. The visible arm has shown 1--5 ms$^{-1}$ is possible \citep{2018A&A...612A..49R}, but the NIR is still impacted by the mitigation of tellurics using the CCF method and the lower than expected (from synthetic spectra) RV information content in the Y- and J-bands \citep{2018A&A...614A.122T}.

While not used here, the \textit{wobble} pipeline \citep{2019arXiv190100503B} is a second data-driven technique to retrieve $I_{\star}$ and has shown notable precision at optical wavelengths further validating our approach. In their work, an initial template is determined in the same method we outline above, but is then treated as a high resolution grid of values to be optimized. The grid must be the same for all spectra, but each is then Doppler shifted with a unique $\Delta v$. This implies all spectra are optimized simultaneously with a single likelihood function, although the temporal variations are fit separately.

\subsection{Prospects for iSHELL Planet Confirmation}

With the launch of the NASA \tess\ (Transiting Exoplanet Survey Satellite) mission, there will be a plethora of planet candidates needing RV follow-up to constrain the mass, and therefore density of the planets. Given our demonstrated precision, many of these candidates orbiting K and M dwarfs brighter than $K_{\textrm{mag}}=9$ and with velocity semi-amplitudes $>$3 ms$^{-1}$ will be detectable with iSHELL. From the existing list of \tess\ objects of interest\footnote{https://tess.mit.edu/toi-releases/} that meet this brightness and predicted semi-amplitude criteria, and the total estimated yield from \cite{2018ApJS..239....2B}, we estimate \texttildelow100 candidates will be amenable to follow-up with iSHELL. We have already demonstrated iSHELL's planet detection capabilities with the discoveries of a Jovian planet system (Plavchan et al. 2019, submitted).

With its unique wavelength coverage, iSHELL measurements will provide a valuable window to confirm planets around K and M dwarfs, particularly those that are more magnetically active and less amenable to confirmation at visible wavelengths. To first order, we expect RV variations induced by stellar activity from stellar rotation modulated spots and plages to be reduced in amplitude in the NIR w/r/t to the visible by a factor proportional to the frequency ratio \citep{2010ApJ...710..432R}. For example, a star with 5 ms$^{-1}$ stellar activity in the visible may be reduced to $<$1.5 ms$^{-1}$ in the NIR, improving sensitivity to planets with velocity semi-amplitudes of \texttildelow1--10 ms$^{-1}$.

% Section 9
\section{Summary \& Future Improvements} \label{sec:summary}

% Summary & Pipeline improvements
We have developed a data analysis pipeline that can robustly extract RVs from K-Band spectra taken with the iSHELL spectrograph on the NASA IRTF using a $^{13}\textrm{CH}_{4}$ gas cell as a wavelength reference. By iteratively minimizing the RMS between the model and observed spectrum, we retrieve both the best-fit RVs as well as a deconvolved high-resolution spectrum of the star. The model uses 48 parameters and accounts for our gas cell, tellurics, fringing, blaze, \textit{LSF}, and wavelength solution. Our initial efforts have shown 5 ms$^{-1}$ precision for Barnard's Star and 61 Cyg A over a \texttildelow 1 year baseline, and 3 ms$^{-1}$ for GJ 15 A over one month. We note a summary of accomplishments shown in this work below.

\begin{enumerate}
  \item Achieve 5 ms$^{-1}$ RV precision over 1 year with
   \begin{itemize}
        \item A unique calibration source at an unfrequented wavelength range for precise RV work,
        \item A spectrograph that slews with the telescope at Cassegrain focus,
    \end{itemize}
  \item In the presence of
  \begin{itemize}
        \item Deep and dynamic telluric lines across entire spectral orders,
        \item Non-standard fringing greatly sophisticating an already high-dimensional forward model,
        \item Starting from the assumption of an unknown stellar template.
  \end{itemize}
\end{enumerate}

Future improvements to our analysis may come with a more sophisticated RV forward model. We have yet to utilize the large overlap of the echelle orders and order-to-order parameter consistency. Low SNR regions (from blaze modulation) near the ends of order $m$ are closer to the blaze wavelength (and thus a higher SNR) in orders $m\pm1$ which will help yield better estimates of the proper line shapes across the entire order, as well as bad pixel flagging. Lastly, we may want to introduce a time-varying telluric or stellar component to account for dynamic features (in a similar fashion to the \textit{wobble} pipeline).

\section{Acknowledgments} \label{sec:ack}

We acknowledge support from the the National Science Foundation (Astronomy and Astrophysics grant 1716202) and George Mason University start-up funds. All RVs extracted through \textit{PySHELL} were run on ARGO, a research computing cluster provided by the Office of Research Computing, and the exo computer cluster, both at George Mason University, VA. We thank John Rayner and the IRTF and iSHELL support astronomers, telescope operators, and engineers for the dedicated efforts in helping enable the collection of the data presented in this paper. We acknowledge Guillem Anglada-Escud\'{e}, Russel White, Todd Henry, and Bernie Walp for their feedback. The authors wish to recognize and acknowledge the very significant cultural role and reverence that the summit of Maunakea has always had within the indigenous Hawaiian community. We are most fortunate to have the opportunity to conduct observations from this mountain.

% Bibliography
\bibliographystyle{apj}
\bibliography{bib_master}

\software{\textbf{ishell\_reduction} (https://github.com/jgagneastro/ishell\_reduction), PySHELL (Available upon request), Scipy \citep{jones2001scipy}, Matplotlib, \citep{Hunter:2007}, barycentric vel.pro \citep{2014PASP..126..838W}}

\end{document}